\documentclass[11pt]{article}
\pdfoutput=1
\usepackage{jcappub,natbib,float,caption,comment,bm}
\usepackage{graphicx,epsfig}
\usepackage[dvipsnames]{xcolor}
\usepackage[utf8]{inputenc}

\usepackage{slashed}

\newcommand{\be}{\begin{equation}}         
\newcommand{\ee}{\end{equation}}
\newcommand{\ba}{\begin{eqnarray}}
\newcommand{\ea}{\end{eqnarray}}

\newcommand{\ncdm}{\texttt{ncdm}}

\newcommand\lsim{\mathrel{\rlap{\lower4pt\hbox{\hskip1pt$\sim$}}
        \raise1pt\hbox{$<$}}}
\newcommand\gsim{\mathrel{\rlap{\lower4pt\hbox{\hskip1pt$\sim$}}
        \raise1pt\hbox{$>$}}}

\title{Probing Decoupling in Dark Sectors with the Cosmic Microwave Background}
\author{Gongjun Choi,}
\author{Chi-Ting Chiang,}
\author{and Marilena LoVerde}
\affiliation{C.N. Yang Institute for Theoretical Physics and Department of Physics \& Astronomy,
Stony Brook University, Stony Brook, NY 11794}

\emailAdd{gongjun.choi@stonybrook.edu}
\emailAdd{chi-ting.chiang@stonybrook.edu}
\emailAdd{marilena.loverde@stonybrook.edu}

\abstract{The acoustic peaks in the angular power spectrum of cosmic microwave background (CMB) temperature and polarization anisotropies play an important role as a probe of the nature of new relativistic particles contributing to the radiation density in the early universe, parametrized by $\Delta N_{eff}$. The amplitude and phase of the acoustic oscillations provide information about whether the extra species are free-streaming particles, like neutrinos, or tightly-coupled, like the photons, during eras probed by the CMB. On the other hand, some extensions of the Standard Model produce new relativistic particles that decouple from their own non-gravitational interactions after neutrinos, but prior to photons. We study the signature of new relativistic species that decouple during this intermediate epoch. We argue that the decoupling species will cause a scale-dependent change in the amplitude and phase shift of the acoustic oscillations, different from the usual constant shifts on small scales. For intermediate decoupling times, the phase and amplitude shifts depend not only on $\Delta N_{eff}$ but the redshift $z_{dec,X}$ at which the new species decoupled. For $\Delta N_{eff} >0.334$, a Stage IV CMB experiment could determine $N_{eff}$ at the percent level and $z_{dec,X}$ at the $\sim 10\%$ level. For smaller values, $\Delta N_{eff}\sim 0.1$, constraints on $z_{dec,X}$ weaken but remain $\sim 20-50\%$ for $z_{dec,X} \sim \mathcal{O}(10^3-10^4)$. As an application, we study the contributions to $\Delta N_{eff}$ and determine the $z_{dec,X}$ values for simple implementations of the so-called $N$naturalness model.}

\begin{document}
\subheader{\rm YITP-SB-18-09}

\maketitle
\flushbottom

\section{Introduction}
\label{sec:section1}

The anisotropies in the cosmic microwave background (CMB) temperature and polarization provide information about the energy density in relativistic particles, or radiation, in the early universe. In the Standard Model and at temperatures below $\sim1$ MeV, this radiation is comprised of photons and relativistic neutrinos. The photon contribution to the radiation energy is accurately determined by the temperature of the CMB today. The remaining contribution is characterized by $N_{eff}$, a parameter defined through
\be
\rho_{rad}(T \lesssim 1{\rm MeV}) = \rho_{\gamma}\left[1+ \frac{7}{8}\left(\frac{4}{11}\right)^{4/3}N_{eff}\right]\,,
\ee
where $\rho_\gamma = (\pi^2/15)T_\gamma^4$ is the energy density in CMB photons and $T_\gamma = 2.725K$ today (see e.g. \cite{Ade:2015xua}). With this definition, $N_{eff} = 3$ corresponds to the radiation energy density expected from three Standard Model neutrinos that decouple instantaneously. In the standard cosmology $N_{eff} \approx 3.046$, due to residual heating of neutrinos from electron-positron annihilation \cite{Mangano:2001iu, Mangano:2005cc, Gnedin:1997vn, Hannestad:1995rs, Heckler:1994tv, Dolgov:1992qg, Dolgov:2002wy, Dolgov:1997mb, Dodelson:1992km, Rana:1991xk, Esposito:2000hi}.\footnote{Taking neutrino oscillation effects into account, \cite{deSalas:2016ztq} found $N_{eff}=3.045$ for both the normal and inverted neutrino mass hierarchies.} 

Any deviation from the Standard Model prediction for $N_{eff}$ would indicate the presence of new relativistic particles, dark radiation, or a change to the standard cosmological history (for a recent review, see \cite{Abazajian:2016yjj}). For instance, a new relativistic species $X$ with a thermal Fermi-Dirac or Bose-Einstein distribution will change the inferred value of $N_{eff}$ by an amount 
\begin{equation}
\Delta N_{eff,X}=
\begin{cases}
a_{\nu}\frac{\mathcal{N}_{X}}{2}(\frac{T_{X}}{T_{\gamma}})^{4}& ({\rm bosonic} X)\\
a_{\nu}\frac{\mathcal{N}_{X}}{2}\frac{7}{8}(\frac{T_{X}}{T_{\gamma}})^{4}& ({\rm fermionic} X)
\end{cases}
\label{eq:defneff}
\end{equation}
where $a_{\nu}\equiv\frac{8}{7}(\frac{11}{4})^{4/3}$, $\mathcal{N}_{X}$ is the internal degrees of freedom of $X$ and $T_{X}$ is the temperature of $X$ at the time of interest. The current constraint on $N_{eff}$ from a combination of CMB and baryon acoustic oscillation (BAO) data is $N_{eff}=3.15\pm0.23$ (68\% C.L.)  \cite{Ade:2015xua}. Separately, Big Bang nucleosynthesis (BBN) puts another constraint $\Delta N_{eff}\leq1.0$ (95\% C.L.) \cite{2011PhLB..701..296M}. These measurements leave room for the existence of additional beyond the Standard Model (BSM) contributions to $N_{eff}$. Increasing the precision of the $N_{eff}$ constraints is a major goal for the current and near-term high-resolution CMB experiments such as Advanced ACT, SPT-3G, the Simons Observatory, and a Stage IV CMB experiment \cite{Benson:2014qhw,Henderson:2015nzj,Abazajian:2016yjj}. From here on, whenever we discuss $\Delta N_{eff}$, we implicitly mean $\Delta N_{eff}$ determined from the CMB, i.e. $\Delta N_{eff}=\Delta N_{eff}^{CMB}$ unless specified otherwise.

Cosmological measurements of $\Delta N_{eff}$ are a particularly interesting test of BSM physics because they are sensitive to new particles even if the new particles do not have any non-gravitational interactions with Standard Model particles. In fact, there are many possible BSM scenarios that generate some new contribution to the radiation density of the early universe \cite{Jungman:1995bz, Kojima:2009gw, Cadamuro:2010cz, Menestrina:2011mz, Boehm:2012gr, Brust:2013xpv, Weinberg:2013kea, Cyr-Racine:2013fsa, Vogel:2013raa, Millea:2015qra, Chacko:2015noa, Lancaster:2017ksf, Arkani-Hamed:2016rle, Buen-Abad:2015ova, Salvio:2013iaa, Kawasaki:2015ofa, Baumann:2016wac, Abazajian:2001nj, Strumia:2006db, Boyarsky:2009ix, Boyle:2007zx, Stewart:2007fu, Meerburg:2015zua, Kaplan:2011yj, Ackerman:mha, CyrRacine:2012fz}. The parameter $\Delta N_{eff}$ could also be negative if photons get heated after the decoupling of neutrinos \cite{Steigman:2013yua, Boehm:2013jpa}. A high-significance detection of $\Delta N_{eff} > 0$ alone would not, however, be sufficient to determine the nature of the new contribution to the radiation density. In this paper, we will show that if $\Delta N_{eff} > 0$, CMB measurements can place limits on and potentially even detect the epoch at which the particles contributing to $\Delta N_{eff}$ decoupled from interactions in their own dark sector. In the event of a future detection of $\Delta N_{eff}$, this additional information could help to identify the new species. 

The CMB is primarily sensitive to $N_{eff}$ through two distinct physical effects. First, the total radiation energy density sets Hubble rate and therefore the damping scale of CMB anisotropies \cite{Zaldarriaga:1995gi}. Second, the presence of free-streaming relativistic particles, such as neutrinos, induces a shift in the phase and a decrease in the amplitude of the acoustic peaks of the CMB \cite{Bashinsky:2003tk}. The damping tail probes the amount of radiation density, but provides no information about the nature of the particles contributing to it beyond the fact that they are relativistic at CMB times. On the other hand, the changes to the acoustic oscillations are specifically generated by particles that have ceased to scatter frequently by CMB times.\footnote{Note that isocurvature perturbations may also generate a phase shift of the acoustic oscillations \cite{Baumann:2015rya}, but we will not consider that in this paper.} The phase shift generated by the Standard Model neutrinos was first pointed out in \cite{Bashinsky:2003tk} and recently detected in \cite{Follin:2015hya}. Subsequently, Baumann, Green, Meyers, and Wallisch \cite{Baumann:2015rya} pointed out that taken together, the two effects mentioned here allow one to jointly constrain the amount of free-streaming relativistic particles at CMB times, $N_{eff}$, and the amount of relativistic particles that are tightly coupled and {\em fluid-like}, as opposed to free-streaming, parametrized by $N_{\text{fl}}$ \cite{Baumann:2015rya}. Furthermore, the BAO feature in large-scale structure data also provides information about the phase of the acoustic oscillations that can supplement CMB constraints on $N_{eff}$ \cite{Baumann:2017lmt, Baumann:2017gkg, Baumann:2018qnt}.

In this paper we extend the work of \cite{Bashinsky:2003tk, Baumann:2015rya} to study the changes to CMB power spectra caused by relativistic particles that transition from fluid-like to free-streaming. That is, they decouple from scattering interactions in their own sector, at epochs probed by the CMB anisotropies. \footnote{We restrict our analysis to dark radiation, or particles that remain relativistic for the entire cosmic history. For studies of the relevance of various parameters (such as mass, coupling strength, etc) of a dark sector to CMB, see, e.g. \cite{Archidiacono:2017slj, Cyr-Racine:2013jua, Cyr-Racine:2013fsa, Cui:2018imi}.} For a species $X$ that decouples at a redshift $z_{dec,X}$, the Fourier modes that cross the horizon at $z_{dec,X}$ predominantly contribute to the CMB anisotropies at multipoles $\ell_{dec,X} \sim \pi d_{LSS}/c_\gamma\tau_{dec,X}$ where $\tau_{dec,X}$ is the conformal time at $z_{dec,X}$, $d_{LSS}$ is the comoving distance to the surface of last scattering, and $c_\gamma = 1/\sqrt{3}$. The CMB anisotropies at multipoles $\ell \gtrsim \ell_{dec,X}$ are therefore sensitive to modes that entered the horizon at  $z \gtrsim z_{dec,X}$. As we shall see, the decoupling of species $X$ changes the $\ell$-dependence of the phase shift and amplitude suppression from appearing nearly constant \cite{Bashinsky:2003tk, Follin:2015hya} at high $\ell$ to dropping off at $\ell \gtrsim \ell_{dec,X}$. If $\Delta N_{eff}\neq0$ is caused by a species that decouples at $z_{dec,X} \lesssim 25000$, $\ell_{dec,X}$ falls  within the observable range of multipole values of CMB Stage IV (e.g. $\ell \lesssim 5000$). The Standard Model neutrinos were, of course, at one time tightly coupled as well but for neutrinos $\ell_{dec,\nu}\sim 10^8$ so probing the Standard Model neutrino decoupling from these effects seems unlikely. Detecting a feature from dark decoupling therefore requires that $z_{dec,X}$ be not too much earlier than our own decoupling $z\sim 1100$. While the existence of a dark sector with a dark decoupling time so close to our own may seem contrived, we will show that this is precisely what can occur in the $N$naturalness model \cite{Arkani-Hamed:2016rle}. In any case, we advocate using the full set of parameters $N_{eff}$, $N_{\text{fl}}$, and $z_{dec,X}$ in future searches for light relics in the CMB. We note that nonstandard low-energy effective four neutrino interactions can produce neutrinos that start to free-stream at $z \sim10^{4}$, which also give rise to an $\ell$-dependent phase shift.  \cite{Cyr-Racine:2013jua, Lancaster:2017ksf}. 

The outline of this paper is as follows. In Section \ref{sec:cosmologicalPT}, we review the analytic calculation of the phase and amplitude shift in the CMB anisotropy power spectrum and then provide a simple extension to incorporate species that decouple at finite $z_{dec,X}$. In Section \ref{sec:deltal}, we present our modifications to the \texttt{CLASS} Boltzmann code \cite{Blas:2011rf} to model a new species that decouples at finite time and then use the modified code to numerically study the changes to the temperature and polarization power spectra caused by the new species. In Section \ref{sec:Nnaturalness}, we show how the $N$naturalness model \cite{Arkani-Hamed:2016rle} gives rise to new particles with such intermediate decoupling times and show how to map $N$naturalness parameters onto our phenomenological parametrization of $\Delta N_{eff}$ and $z_{dec,X}$.  In Section \ref{sec:forecast}, we perform a Fisher forecast to assess the sensitivity of a Stage IV CMB experiment to $\Delta N_{eff}$ and $z_{dec, X}$, and to explore the degeneracies between these parameters. Our conclusions are presented in Section \ref{sec:conclusion}.

\section{Analytic Computation}
\label{sec:cosmologicalPT}

\subsection{Background}
In this section, we review the cosmological perturbation theory needed to study the effects of free-streaming particles on the acoustic peaks of the CMB anisotropy power spectra. For a more thorough discussion, the reader is referred to the classic paper of Bashinsky \& Seljak \cite{Bashinsky:2003tk} or the more recent paper by Baumann et al \cite{Baumann:2015rya}. The discussion here follows \cite{Bashinsky:2003tk, Baumann:2015rya} closely and readers familiar with those works can skip directly to Section \ref{sec:thdeltal}.  

\subsubsection{Metric and Conventions}

\label{sec:perturbation_metric}
We parametrize the perturbed metric as
\begin{equation}
ds^{2}=a^{2}(\tau)[(-1-2A)d\tau^{2}-2B_{i}d\tau dx^{i}+[(1-2H_{L})\delta_{ij}-2\bar{H}_{ij}]dx^{i}dx^{j}] \,,
\label{eq:metric}
\end{equation}
where $\tau$ is conformal time and $\bar{H}_{ij}$ is traceless and symmetric.
In this paper we only study the scalar modes and so
\begin{equation}
B_{i}=\nabla_{i}b\quad,\quad \bar{H}_{ij}=(\nabla_{i}\nabla_{j}-\frac{1}{3}\delta_{ij}\nabla^{2})\chi \,,
\label{eq:scalarpotential}
\end{equation}
where $b$ and $\chi$ are scalar potentials for perturbations. We can write the diagonal part of the perturbation to the spatial curvature (proportional to $\delta_{ij}$) as $\Psi=H_{L}-\frac{1}{3}\nabla^{2}\chi$. Apart from the discussion of our modification to Boltzmann codes in Section \ref{sec:deltal}, we will work in the Newtonian gauge ($b=\chi=0$) with metric given by \cite{Ma:1995ey}
\begin{equation}
ds^{2}=a^{2}(\tau)[(-1-2\Phi)d\tau^{2}+(1-2\Psi)\delta_{ij}dx^{i}dx^{j}] \,,
\label{eq:metricnewton}
\end{equation}
where $\Phi$ and $\Psi$ are the gravitational potential and spatial curvature perturbation, respectively. In what follows, it is useful to define
\begin{equation}
\Phi_{\pm}\equiv\Phi\pm\Psi \,.
\label{eq:phipsi}
\end{equation}
For a species $a$, we define the perturbation of particle number {\em per proper volume} as
\begin{equation}
\delta_{a}\equiv\frac{\delta n_{a}}{\bar{n}_{a}}=\frac{\delta\rho_{a}}{\bar{\rho}_{a}+\bar{P}_{a}} \,,
\label{eq:overdensity}
\end{equation}
where $\bar{\rho}$ and $\bar{P}$ are mean energy density and pressure. The second equality is due to energy conservation. On the other hand, the perturbation of particle number {\em per coordinate volume} is defined as
\begin{equation}
d_{a}\equiv\delta_{a}-3\Psi \,.
\label{eq:overdensity2}
\end{equation}
In the study of the evolution of the photon density, we employ $d_{\gamma}$ rather than $\delta_{\gamma}$
as it provides a simpler description \cite{Bashinsky:2003tk}. In this paper, we assume adiabatic
initial conditions, i.e. $\delta_{a}=\delta$ is a constant for all species on super-horizon scales $k\tau\ll1$ with $k$ being the Fourier mode.
This implies $d_{a}$ is also a constant on super-horizon scales, and we shall set its initial value
to be $d_{a,in}$.
Following \cite{Bashinsky:2003tk}, we write the initial photon overdensity $d_{\gamma,in}=-3\zeta$
in terms of the gauge-invariant primordial curvature perturbation. 
 Finally, we relate the pressure $P_a$ and energy density $\rho_a$ via
\begin{equation}
\frac{P_{a}}{\rho_{a}}=w_{a} \,,
\quad\frac{dP_{a}}{d\rho_{a}}=w_{a}+\frac{dw_{a}}{d\ln\rho_{a}}=c_{a}^{2} \,,
\label{eq:eqnofstate}
\end{equation}
where $w_{a}$ and $c_{a}$ are the equation of state and the adiabatic sound speed, respectively.
In this paper, we assume that all species satisfy $w_{a}=c_{a}^{2}$, meaning that
$w_{a}$ is independent of $\rho_{a}$ \cite{Ma:1995ey}.

\subsubsection{Perturbed Stress-Energy Tensor}
\label{sec:stressenergy}
The perturbed stress-energy tensor for a species $a$ is
\begin{equation}
T^{0}_{0,a}=-(\bar{\rho}_{a}+\delta\rho_{a}),\quad T^{0}_{i,a}=(\bar{\rho}_{a}+\bar{P}_{a})v_{i,a},\quad T^{i}_{j,a}=(\bar{P}_{a}+\delta P_{a})\delta^{i}_{j}+(\bar{\rho}_{a}+\bar{P}_{a})\Sigma^{i}_{j,a} \,,
\label{eq:emtensorcomponent}
\end{equation}
where $v_{i,a}$ is the velocity and $\Sigma^{i}_{j,a}$ is the anisotropic stress tensor with $\Sigma^{i}_{i,a}=0$. 
The velocity and anisotropic stress tensors can be written in terms of their associated scalar potentials as
\begin{equation}
 v_{i,a}=-\nabla_{i} u_{a} \,, \qquad \Sigma^{i}_{j,a}=\frac{3}{2}(\partial^{i}\partial_{j}-\frac{1}{3}\delta^{i}_{j}\nabla^{2})\pi_{a} \,.
\end{equation}

The conservation of stress-energy tensor, $T^{\mu\nu}_{;\nu~,a}$, for each species leads to
\begin{equation}
\dot{\delta}_{a}=\nabla^{2}u_{a}+3\dot{\Psi}\,,\qquad\dot{u}_{a}=c^{2}_{a}\delta_{a}-\chi_{a}u_{a}+\nabla^{2}\pi_{a}+\Phi \,,
\label{eq:ddotudot}
\end{equation}
where $\chi_{a}\equiv\mathcal{H}(1-3c^{2}_{a})$ is the Hubble drag rate and $\mathcal{H}=aH$
is the expansion rate of the universe with respect to the conformal time. Replacing $\delta_a$
with $d_a$ by Eq.~(\ref{eq:overdensity}) and combining Eq.~(\ref{eq:ddotudot}),
we obtain the evolution of $d_a$ as 
\begin{equation}
\ddot{d}_{a}+\mathcal{H}(1-3c^{2}_{a})\dot{d}_{a}-c^{2}_{a}\nabla^{2}d_{a}=\nabla^{4}\pi_{a}+\nabla^{2}(\Phi+3c^{2}_{a}\Psi) \,.
\label{eq:evolution1}
\end{equation}

\subsubsection{Distribution Function and Boltzmann Equation}
\label{sec:boltzmann}
We use the Boltzmann equation to study the evolution of the phase-space distribution function $f_a$
of a species $a$. The distribution function depends on the comoving coordinates, the comoving momenta $\vec{q}\equiv a\vec{p}$
with $\vec{p}$ being the proper momenta, and the conformal time. The full time derivative of $f_a$ is
\begin{equation}
\dot{f}_{a}+\dot{\vec{r}}\cdot\frac{\partial f_{a}}{\partial\vec{r}}+\dot{q}\frac{\partial f_{a}}{\partial q}+\hat{n}\cdot\frac{\partial f_{a}}{\partial\hat{n}}=
\left(\frac{\partial f_{a}}{\partial\tau}\right)_{C} \,,
\label{eq:boltzmanneqn}
\end{equation}
where $\hat{n}=\vec{q}/q$ is the unit direction vector of the momentum.
The right-hand side of Eq.~(\ref{eq:boltzmanneqn}) accounts for the relevant collisions.
For a collisionless species (i.e. free-streaming particles), the collision term vanishes.

Decomposing $f_a$ into background $\bar{f}_a$ and perturbation $\delta f_a$,
the first-order Boltzmann equation in Newtonian gauge is
\begin{equation}
\delta\dot{f}_{a}+\frac{\vec{q}}{\epsilon}\cdot\vec{\nabla}(\delta f_{a})
+q\frac{\partial\bar{f_{a}}}{\partial q}\left(\dot{\Psi}-\frac{\epsilon}{q}\hat{n}\cdot\vec{\nabla}\Phi\right)=(\dot{f}_{a})_{C}-(\dot{\bar{f}}_{a})_{C} \,,
\label{eq:boltzmanneqn2}
\end{equation}
where $\epsilon\equiv ap^{0}=\sqrt{q^{2}+a^{2}m^{2}}$ is the proper energy and $m$ is the mass of the species.
To simplify the notation, we define $df(\tau,\vec{r},q,\hat{n})$ as
\begin{equation}
df_{a}\equiv\delta f_{a}+q\frac{\partial\bar{f}_{a}}{\partial q}\Psi \,.
\label{eq:defdf}
\end{equation}
It will be useful to define
\begin{equation}
D_{a}(\tau,\vec{r},\hat{n})\equiv\frac{3}{4}\frac{\int q^{3}dq\,df_{a}(\tau,\vec{r},q,\hat{n})}{\int q^{3}dq\bar{f}_{a}(q)} \,,
\label{eq:mtmintegrateddf1}
\end{equation}
and the multipole moments of the Fourier transformed $D_a(\tau, k, \mu)$ with $\mu = \hat{n}\cdot \hat{k}$ via
\be
D_{a}(\tau, k, \mu) = \sum_{\ell =0}^\infty (-i)^\ell(2\ell+1)D_{a, \ell}(\tau, k)P_\ell(\mu)\,,
\label{eq:mtmintegrateddf2}
\ee
where the expansion coefficients $D_{a,\ell}$ form the set of multipole moments.

The general expression for the stress-energy tensor of the species $a$ is \cite{Ma:1995ey}
\begin{equation}
T^{\mu}_{\nu,a}=\int d^{3}p\frac{p^{\mu}p_{\nu}}{p^{0}}f_{a} \,,
\label{eq:emtensor}
\end{equation}
and from this one finds the perturbed stress-energy tensor 
\begin{equation}
\delta T^{\mu}_{\nu,a}=a^{-4}\int d\Omega_{\hat{n}}n^{\mu}n_{\nu}\int dqq^{3}\delta f_{a} \,.
\label{eq:deltaemtensor}
\end{equation}
Comparing with Eq.~(\ref{eq:emtensorcomponent}) and using the moments of $D_{a}$, we have
\begin{equation}
D_{a,0}=d_{a},\quad D_{a,1}=ku_{a},\quad D_{a,2}=\frac{3}{2}k^{2}\pi_{a} \,.
\label{eq:multipole}
\end{equation}
In this paper, we take initial conditions for $D_{a,\ell}$ such that $D_{a,\ell}$ is finite in the limit $\tau\rightarrow0$ and grows with time. This can be arranged with setting $D_{a,\ell}\sim(k\tau)^{\ell}$ for $k\tau\ll1$.

\subsubsection{Evolution of a Fluid-like Species}
\label{ssec:fluid}
A tightly-coupled relativistic fluid, such as the photon-baryon plasma prior to decoupling, has a sound speed $c_{\gamma}^{2}\approx \frac{1}{3}$ and no anisotropic stress potential
($\pi_{\gamma}=0$). Thus relying on Eq.~(\ref{eq:evolution1}) rather than the full Boltzmann Eq.~(\ref{eq:boltzmanneqn}), which requires the knowledge of the collision term, we can write the evolution equation for the tightly-coupled photon-baryon plasma as
\begin{equation}
\ddot{d}_{{\rm {\tiny fl}}}-c^{2}_{\gamma}\nabla^{2}d_{{\rm {\tiny fl}}}=\nabla^{2}\Phi_{+}\,.
\label{eq:evolution2}
\end{equation}
Moving to Fourier space and introducing the new variable $y=c_{\gamma}k\tau$ gives
\begin{equation}
d''_{{\rm {\tiny fl}}}+d_{{\rm {\tiny fl}}}=-c_{\gamma}^{-2}\Phi_{+} \,,
\label{eq:evolution3}
\end{equation}
where $'\equiv d/dy$. For adiabatic initial conditions, $d'_{{\rm {\tiny fl}}}(y\rightarrow 0)\rightarrow 0$ so we have $d_{{\rm {\tiny fl}}}=d_{{\rm {\tiny fl}},in}\cos y$ in the absence of $\Phi_{+}$.
The general solution for the non-vanishing $\Phi_{+}$ is then
\be
d_{{\rm {\tiny fl}}}(y) =d_{{\rm {\tiny fl}},in}\cos y-c_{\gamma}^{-2}\int_{y_{in}}^{y}dy'\Phi_{+}(y')\sin(y-y')\,.
\ee

\subsubsection{Evolution of a Free-streaming Species}
\label{ssec:fs}
The distribution function of a free-streaming particle species, $f_{{\rm {\tiny fs}}}$, satisfying the collisionless Boltzmann equation evolves as
\begin{equation}
d\dot{f}_{{\rm {\tiny fs}}}+\frac{q}{\epsilon}\hat{n}\cdot\vec{\nabla}(df_{{\rm {\tiny fs}}})
=q\frac{\partial\bar{f}_{{\rm {\tiny fs}}}}{\partial q}\hat{n}\cdot\vec{\nabla}\left(\frac{q}{\epsilon}\Psi+\frac{\epsilon}{q}\Phi\right) \,.
\label{eq:boltzmanneqn3}
\end{equation}
Since we are interested in massless species, we set $\epsilon\simeq q$. Integrating Eq.~(\ref{eq:boltzmanneqn3}) over $q^{3}dq$, we have
\begin{equation}
\dot{D}_{{\rm {\tiny fs}}}+\hat{n}\cdot\vec{\nabla} D_{{\rm {\tiny fs}}}=-3\hat{n}\cdot\vec{\nabla}\Phi_{+} \,.
\label{eq:boltzmanneqn4}
\end{equation}
Fourier transforming Eq.~(\ref{eq:boltzmanneqn4}) and letting $\mu = \hat{n}\cdot \hat{k}$ gives
\be
\dot{D}_{{\rm {\tiny fs}}}+ik\mu D_{{\rm {\tiny fs}}} = -3ik\mu \Phi_+\,,
\ee
which has solution
\be
\label{eq:Datau}
D_{{\rm {\tiny fs}}}(\tau, k, \mu) = D_{{\rm {\tiny fs}}}(\tau_{in})e^{-ik\mu(\tau-\tau_{in})} -3ik\mu \int_{\tau_{in}}^\tau d\tau' e^{-ik\mu(\tau-\tau')} \Phi_+(\tau')\,.
\ee

\subsection{Analytic Estimate of the Phase and Amplitude Shift for $z_{dec,X}\rightarrow \infty$}
\label{sec:dgammaevolution}
In this section we will use our results from the last few sections to repeat the calculation of the changes to $d_\gamma$, in particular the phase and amplitude shift, induced by a species that is free streaming at the initial time. 

Since CMB photons are tightly coupled with electrons they are fluid-like before the last scattering. From Section \ref{ssec:fluid} we have
\be
d_{\gamma}(y) =d_{\gamma,in}\cos y-c_{\gamma}^{-2}\int_{y_{in}}^{y}dy'\Phi_{+}(y')\sin(y-y') \,,
\ee
which can be rewritten as 
\ba
d_{\gamma}(y) &=&[d_{\gamma,in}+c_{\gamma}^{-2}A(y)]\cos y-c_{\gamma}^{-2}B(y)\sin y\\
&\equiv& C(y) \cos[y+\theta(y)] \,,
\label{eq:dgamma}
\ea
where $A(y)$ and $B(y)$ are defined as
\begin{equation}
A(y)\equiv\int_{y_{in}}^{y}dy'\Phi_{+}(y')\sin y'\quad,\quad B(y)\equiv\int_{y_{in}}^{y}dy'\Phi_{+}(y')\cos y' \,,
\label{eq:AandB1}
\end{equation}
and $C(y)$, $\theta(y)$ are defined through
\ba
C(y) &=& \sqrt{[d_{\gamma,in}+c_{\gamma}^{-2}A(y)]^{2}+[c_{\gamma}^{-2}B(y)]^{2}}\,,\\
\sin[\theta(y)] &=& \frac{B(y)}{\sqrt{[A(y)+c_{\gamma}^{2}d_{\gamma,in}]^{2}+B(y)^{2}}} \,.
\label{eq:dgammaphaseshift1}
\ea

From the perturbed Einstein equations, the equations of motion for $\Phi_+$  and $\Phi_{-}$ read
\begin{equation}
 \nabla^{2}\Phi_{+}-3\mathcal{H}(\dot{\Phi}_{+}+\mathcal{H}\Phi_{+}) =8\pi Ga^{2}\sum\limits_{a}\delta\rho_{a}+S_{1}[\Phi_{-}]\,,
\label{eq:linearEE1}
\end{equation}
\begin{equation}
 \ddot{\Phi}_{+}+3\mathcal{H}\dot{\Phi}_{+}+(2\dot{\mathcal{H}}+\mathcal{H}^{2})\Phi_{+} =8\pi Ga^{2}\sum_{a}\delta P_{a}+S_2[\Phi_{-}] \,,
\label{eq:linearEE2}
\end{equation}
and 
\begin{equation}
 \Phi_{-}=-12\pi Ga^{2}\sum\limits_{a}(\bar{\rho}_{a}+\bar{P}_{a})\pi_{a} \,,
\label{eq:phiminus}
\end{equation}
where
\begin{equation}
 S_{1}[\Phi_{-}]=\nabla^{2}\Phi_{-}-3\mathcal{H}(\dot{\Phi}_{-}-\mathcal{H}\Phi_{-}) \,, \quad
 S_2[\Phi_{-}]=\ddot{\Phi}_{-}+\mathcal{H}\dot{\Phi}_{-}-(2\dot{\mathcal{H}}+\mathcal{H}^{2}+\frac{2}{3}\nabla^{2})\Phi_{-} \,.
\end{equation}
In the radiation dominated era $a\propto \tau$ so Eq.~(\ref{eq:linearEE2}) simplifies to
\be
\label{eq:phiprimeprime}
\Phi_+''+\frac{4}{y}\Phi_+' +\Phi_+ = \frac{8\pi G a^2}{(c_\gamma k)^2}\sum_a(c_a^2 - c_\gamma^2)\delta\rho_a + \tilde{S}[\Phi_-] \,,
\ee
with $\tilde{S}[\Phi_-] = \Phi_-''+(2/y)\Phi_-'+3\Phi_-$. 

In the absence of sources of anisotropic stress, one can see from Eq.~(\ref{eq:phiminus}) that $\Phi_{-}=0$. For a relativistic decoupled species $X$ contributing a fraction $\epsilon_X = \rho_X/\rho_{total}$ to the energy density during the radiation dominated era
\be
\label{eq:phiminusD}
\Phi_-(y) = -\frac{4}{3}\frac{\epsilon_X}{y^2}D_{X,2}(y) \,,
\ee
where we have used Eq.~(\ref{eq:multipole}). Since we are only considering sources with $c_a = c_\gamma$, $\Phi_-$ sourced by $\epsilon_XD_{X,2}(y)$ is the only source for $\Phi_{+}$ in Eq.~(\ref{eq:phiprimeprime}). At zeroth order in $\epsilon_X$ the general solution to 
Eq.~(\ref{eq:phiprimeprime}) is
\be
\Phi^{(0)}_+(y)= C_1\left(\frac{\sin y}{y^3} - \frac{\cos y}{y^2}\right) + C_2 \left(\frac{\sin y}{y^2} +\frac{\cos y}{y^3}\right)\,.
\ee
Only the first term is finite at $y\rightarrow 0$ so we have
\be
\Phi^{(0)}_+(y) = 3\Phi^{(0)}_+(y=0) \frac{j_1(y)}{y} = 4\zeta\frac{j_1(y)}{y}\,,
\ee
where $\Phi^{(0)}_+(y=0)=(4/3)\zeta$ is set by the super-horizon $\Phi_{+}$ solution in \cite{Bashinsky:2003tk}. Plugging this into Eq.~(\ref{eq:AandB1}) one finds that $A^{(0)},B^{(0)}\rightarrow 2\zeta,0$ as $y\rightarrow \infty$. Thus in the large $y$ limit, the phase shift vanishes at zeroth order in $\epsilon_X$. More generally, at zeroth order in $\epsilon_X$ the amplitude and phase of $d_\gamma$ are 
\ba
C^{(0)}(y) &=&  \sqrt{[d_{\gamma,in}+c_{\gamma}^{-2}A^{(0)}(y)]^{2}+[c_{\gamma}^{-2}B^{(0)}(y)]^{2}}\,,\\
 \sin\theta^{(0)}(y) &=& \frac{B^{(0)}(y)}{\sqrt{(A^{(0)}(y)+c_{\gamma}^{2}d_{\gamma,in})^{2}+(B^{(0)}(y))^{2}}}\,.
\ea

To find the first order solutions for $\theta$ and $C$, we use the zeroth order solution for $\Phi_+$ in Eq.~(\ref{eq:Datau}) to determine $D_{2,X}(y)$, which can then be used in Eq.~(\ref{eq:phiminusD}) to solve for $\Phi_-$, and finally $\Phi_+$ at first order in $\epsilon_X$ through Eq.~(\ref{eq:phiprimeprime}). The result is \cite{Baumann:2015rya}
\be
\Phi_+^{(1)}(y) = -\frac{4}{15}\zeta \epsilon_X \frac{\sin y- y \cos y}{y^3}+ \int_{0}^y dy' \tilde{S}[\Phi^{(1)}_-(y')] G_{\Phi_+}(y,y') \,,
\ee
where 
\be
G_{\Phi_+}(y,y') = \Theta(y-y') \frac{y'}{y^3}\left[(y'-y)\cos(y'-y) - (1+ yy')\sin(y'-y)\right]\,,
\ee
and $\Phi^{(1)}_+(y=0)=-(4/45)\zeta \epsilon_X$ is read from the super-horizon $\Phi_{+}$ solution in \cite{Bashinsky:2003tk}. At first order in $\epsilon_X$, the fractional change in the amplitude of $d_\gamma$ is
\be
\frac{C^{(1)}}{C^{(0)}} = \frac{A^{(0)} A^{(1)} + B^{(0)}B^{(1)}+ A^{(1)} c_\gamma^2 d_{\gamma,in}}{(A^{(0)})^2 + (B^{(0)})^2 + 2A^{(0)}c_\gamma^2 d_{\gamma,in} + c_\gamma^4d_{\gamma,in}^2} \,,
\ee
and the first order change in the phase is
\begin{equation}
 \theta^{(1)}(y)=\frac{B^{(1)}(y)[A^{(0)}(y)+c_{\gamma}^{2}d_{\gamma,in}]-A^{(1)}(y)B^{(0)}(y)}{[A^{(0)}(y)+c_{\gamma}^{2}d_{\gamma,in}]^{2}+[B^{(0)}(y)]^{2}} \,.
\label{eq:firstordertheta}
\end{equation}
In the limit of $y\rightarrow \infty$, we have $A^{(1)}\approx -0.27\zeta\epsilon_X$ and $B^{(1)} \approx 0.6\zeta\epsilon_X$ so that $C^{(1)}/C^{(0)}\approx -0.268 \epsilon_X$ and $\theta^{(1)} \approx 0.19 \pi\epsilon_X$ as expected \cite{Bashinsky:2003tk, Baumann:2015rya}. Free-streaming particles therefore suppress the amplitude of $d_\gamma$ and induce shift in the phase of the oscillations. 

Using the
flat sky approximation, we can associate the change of peak locations $\delta\ell$ to
$\theta^{(1)}$ as
\begin{equation}
 \delta\ell\approx\frac{\theta^{(1)}}{\pi}\Delta\ell \,,
\label{eq:thetatol}
\end{equation}
where $\Delta\ell\sim300$ is the averaged separation between two peaks in the CMB power
spectrum \cite{0067-0049-148-1-233}. We will use $\Delta\ell\sim300$ throughout this paper for estimating $\delta\ell$.

\subsection{Analytic Estimate of the Phase and Amplitude Shift for Finite $z_{dec,X}$}
\label{sec:thdeltal}
\begin{figure}[t]
\centering
$\begin{tabular}{cc}
\includegraphics[width=0.495\textwidth]{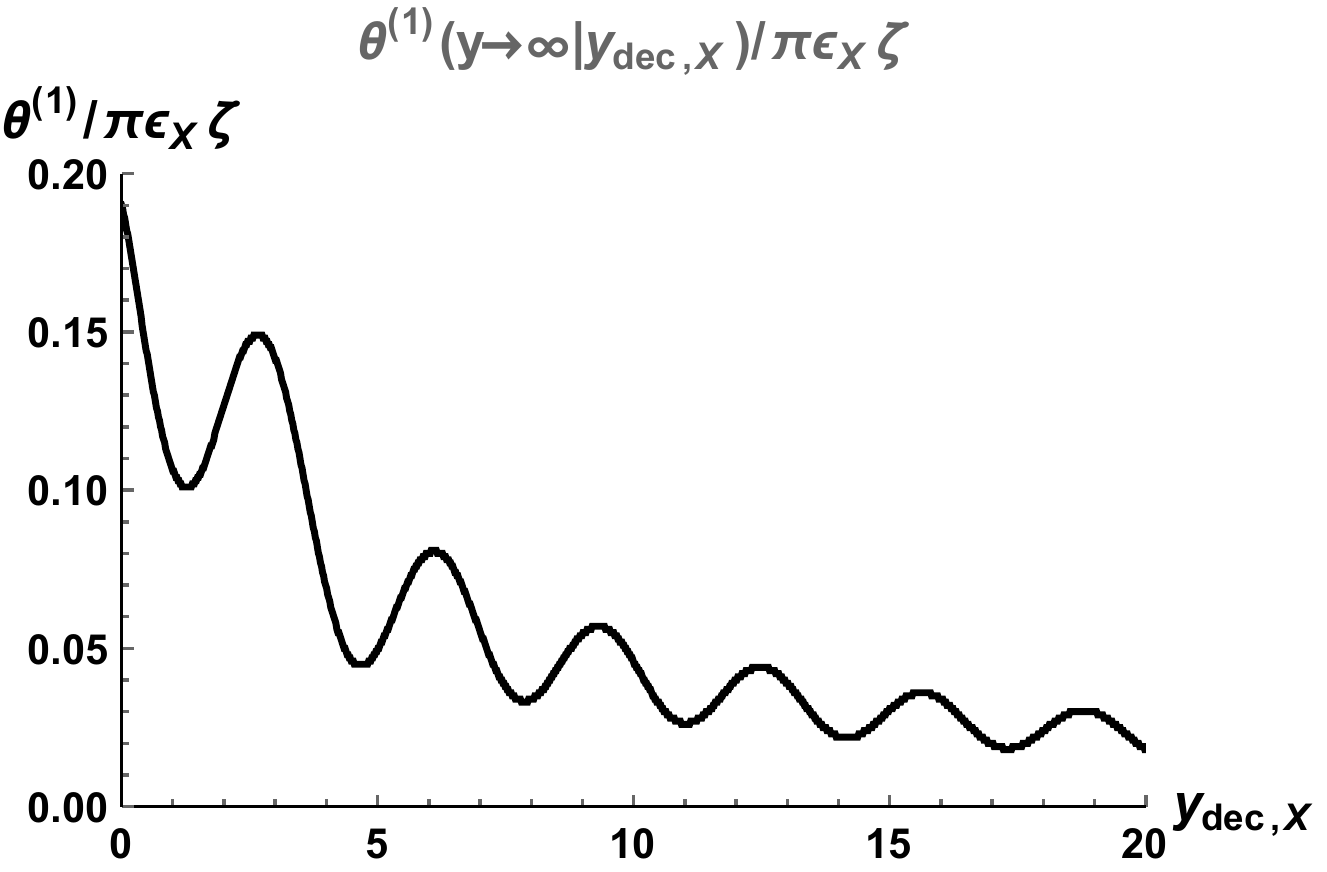}&
\includegraphics[width=0.495\textwidth]{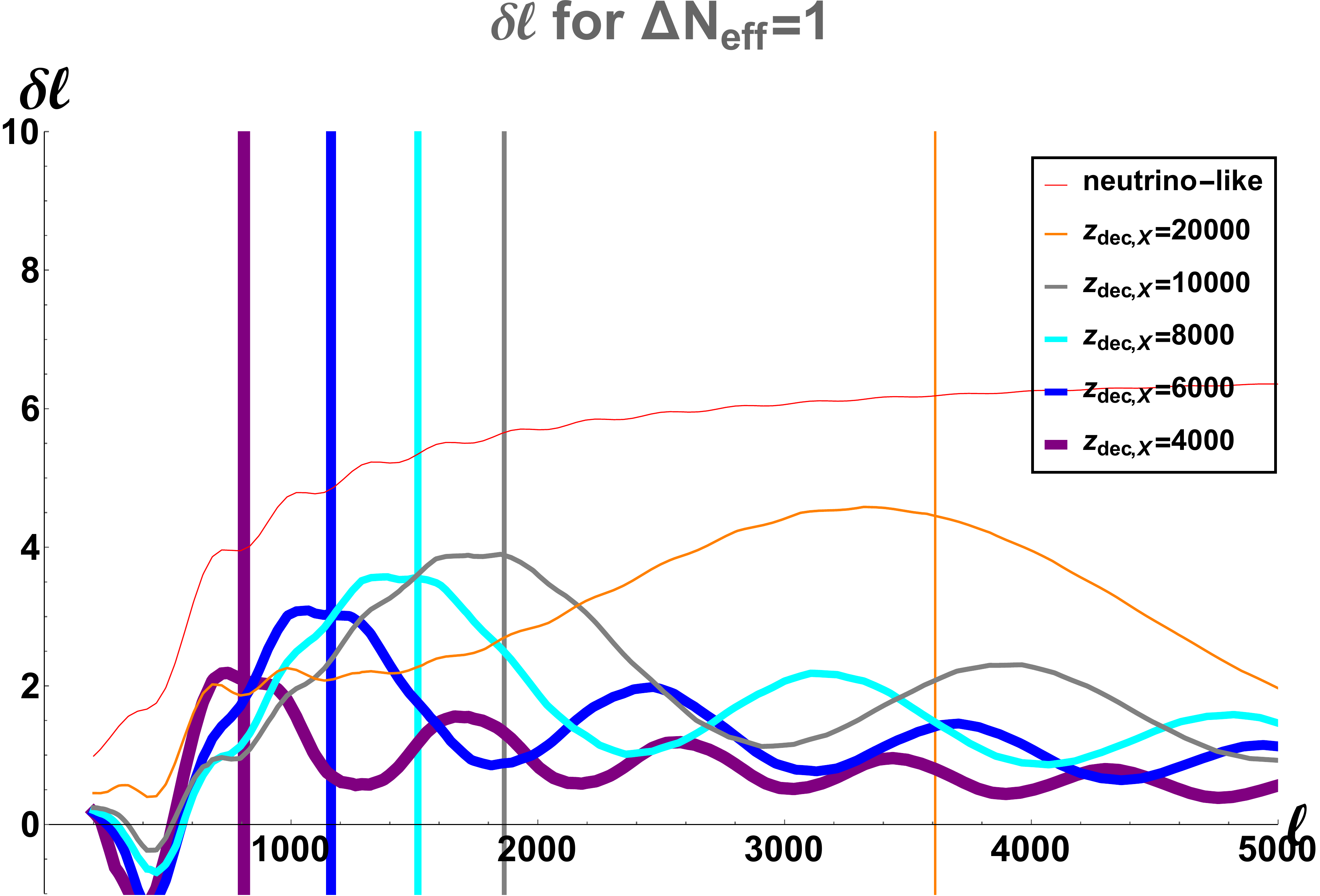}\\
\mbox{(a)} & \mbox{(b)}
\end{tabular}
$
\caption{Analytic calculations of the phase shift for a relativistic species that decouples from dark interactions at finite $z_{dec,X}$. (Left) The phase shift $\theta^{(1)}(y\rightarrow \infty | y_{dec,X})$ in units of the fractional contribution of species $X$ to the energy density $\epsilon_X$ and the primordial spatial curvature perturbation $\zeta$. (Right) Estimate of the phase shift in the CMB power spectra as a function of $\ell$ from Eq.~(\ref{eq:deltaellestimate}) for $\Delta N_{eff} =1$, corresponding to $\epsilon_X = 0.118$ (independent of the values of other cosmological parameters). Different colors and thicknesses represent different decoupling redshifts, $z_{dec,X}$, the thicker lines refers to smaller $z_{dec,X}$. The vertical lines show the location of $\ell_{dec,X}\simeq\sqrt{3}\pi d_{LSS}/\tau_{dec,X}$ of which the associated comoving scale is equal to that of the sound horizon at $z_{dec,X}$. We use the cosmological parameters in Table \ref{table:fiducial} with $\Delta N_{eff}=1$ to evaluate $d_{LSS}\simeq\tau_{0}-\tau_{dec}$ and $\tau_{dec}$.}
\label{fig:analyticphaseshift}
\end{figure}

Now, let us revisit the calculation of the last section allowing for a finite decoupling time for the additional species $X$. We will continue to work in the 
radiation-dominated era and assume that our new species contributes a small fraction $\epsilon_X$ to the total radiation density. 
For simplicity, we assume that a species that decouples at time $\tau_{dec,X}$ instantly switches from satisfying the fluid-like equations
of Section \ref{ssec:fluid} to solving the free-streaming equations of Section \ref{ssec:fs}. We then need to match the free-streaming solution for $\tau \ge \tau_{dec,X}$,
\be
\label{eq:Dxfs}
D_{X}(\tau \ge \tau_{dec,X},k,\mu) = D_{X}(\tau_{dec,X},k,\mu)e^{-ik\mu(\tau-\tau_{dec,X})} -3ik\mu \int_{\tau_{dec,X}}^\tau d\tau' e^{-ik\mu(\tau-\tau')} \Phi_+^{(0)}(\tau')\,,
\ee
to the fluid-like solution for $D_X$ satisfied at $\tau <\tau_{dec,X}$. Prior to $\tau_{dec,X}$, $d_{X}$ will satisfy the fluid-like equations from Section \ref{ssec:fluid}. Defining $y_{dec,X} = c_\gamma k\tau_{dec,X}$
and  using the continuity equation to relate $d_{X}(y<y_{dec,X})$ to $u_{X}(y<y_{dec,X})$ gives
\be
D_{X,0}(y < y_{dec,X}) = d_{X, in} \cos y -c_\gamma^{-2}\int_{y_{in}}^y dy' \Phi^{(0)}_+(y')\sin(y-y')
\ee
and 
\be
D_{X,1}(y < y_{dec,X}) = c_\gamma d_{X, in} \sin y  + c_\gamma^{-1}\int_{y_{in}}^y dy' \Phi^{(0)}_+(y')\cos(y-y')\,.
\ee
Matching these expressions with Eq.~(\ref{eq:Dxfs}) gives
\ba
D_{X}(y >y_{dec,X}) &=& \left[D_{X,0}(y_{dec,X}) -3 i\mu D_{X,1}(y_{dec,X})\right]e^{-ic_{\gamma}^{-1}\mu(y-y_{dec,X})} \nonumber\\
&&-3ic_{\gamma}^{-1}\mu \int_{y_{dec,X}}^y dy' e^{-ic_\gamma^{-1}\mu(y-y')} \Phi^{(0)}_+(y') \,,
\ea
and finally
\ba
D_{X,2}(y > y_{dec,X})
&=& D_{X,0}(y_{dec,X})j_2(c_\gamma^{-1}(y-y_{dec,X})) \\
&&+\, D_{X,1}(y_{dec,X})\left[\frac{6}{5}j_1(c_\gamma^{-1}(y-y_{dec,X})) -\frac{9}{5}j_3(c_{\gamma}^{-1}(y-y_{dec,X}))\right]\nonumber\\
&& +\, c_\gamma^{-1}\int_{y_{dec,X}}^y dy' \left[\frac{6}{5}j_1(c_\gamma^{-1}(y-y')) -\frac{9}{5}j_3(c_{\gamma}^{-1}(y-y'))\right]\Phi^{(0)}_+(y')\nonumber\,.
\ea
As before, this expression can be used to find $\Phi^{(1)}_{-}$, which is now dependent on both $y$ and $y_{dec,X}$,
\be
\label{eq:phiminusone}
\Phi^{(1)}_-(y|y_{dec,X}) = \left\{\begin{array}{cc} 0  &\quad y< y_{dec,X}\\ -\frac{4}{3}\frac{\epsilon_X}{y^2}D_{X,2}(y >y_{dec,X}) & \quad y\ge y_{dec,X}\end{array}\,.\right.
\ee
And finally, we can find the expression for $\Phi_{+}^{(1)}$,
\be
\label{eq:phiplusone}
\Phi^{(1)}_+(y| y_{dec,X}) = 3 \Phi^{(1)}_+(y_{in}|y_{dec,X}) \frac{j_1(y)}{y} + \int_{y_{dec,X}}^y dy' \tilde{S}[\Phi^{(1)}_-(y'|y_{dec,X})] G_{\Phi_+}(y,y')\,.
\ee
We can find $\Phi^{(1)}_+(y_{in}| y_{dec,X})$ by matching to the constant super-horizon solution of $\Phi_{+}$. If the $\Phi_+$ modes enter the horizon prior to decoupling of $X$ (i.e. if $y_{dec,X} > 1$), we have
\be
\Phi^{(1)}_+(y_{in}| y_{dec,X}) = 0 \,.
\ee
If they enter the horizon after decoupling of $X$ ($y_{dec,X} <1$), then we have \cite{Bashinsky:2003tk}
\be
\Phi^{(1)}_+(y_{in}| y_{dec,X}) = \frac{-4}{45}\zeta\epsilon_X \,.
\ee
Hence we can write a general solution 
\be
\Phi^{(1)}_+(y| y_{dec,X}) = \frac{-4}{15}\zeta\epsilon_X \left(1 - \Theta(y_{dec,X}-1)\right)\frac{j_1(y)}{y} + \int_{y_{dec,X}}^y dy' \tilde{S}[\Phi^{(1)}_-(y'|y_{dec,X})] G_{\Phi_+}(y,y')\,,
\ee
which can be used to determine $A^{(1)}(y|y_{dec,X})$, $B^{(1)}(y|y_{dec,X})$, and $\theta^{(1)}(y|y_{dec,X})$. The final expression for the phase shift must be solved numerically.
In the $y\rightarrow \infty$ limit, one finds a phase shift $\theta^{(1)}(y\rightarrow \infty| y_{dec,X})$ that decreases with increasing $y_{dec,X}$. This is shown in the left panel of Figure \ref{fig:analyticphaseshift}. 

The solutions we have here are only valid in the radiation-dominated era. We can nevertheless get a good estimate of the shift in peak locations at CMB multipole $\ell$ by associating anisotropies at $\ell$ with $d_\gamma$ evaluated at
$k = \ell/d_{LSS}$ and $\tau = \tau_{dec}$,
\ba
\label{eq:deltaellestimate}
\delta \ell(\ell |\tau_{dec,X}) &\approx& \delta \ell(y = c_\gamma \ell/d_{LSS}\tau_{dec}|\,y_{dec,X} = c_\gamma \ell/d_{LSS}\tau_{dec,X}) \nonumber\\
& \approx & \theta^{(1)}(y = c_\gamma \ell/d_{LSS}\tau_{dec}|\,y_{dec,X} = c_\gamma \ell/d_{LSS}\tau_{dec,X}) \frac{\Delta\ell}{\pi} \,,
\ea
where $\tau_{dec}$ is the photon decoupling time and $d_{LSS}$ is the distance to
the surface of last scattering. The result for this estimate of $\delta\ell(\ell|\tau_{dec,X})$
for several values of $\tau_{dec,X}$ is shown in the right panel of Figure \ref{fig:analyticphaseshift}.
In the limit of early decoupling (e.g $z_{dec,X}\rightarrow 10^{9}$) we recover
the usual analytic result for the phase shift from the Standard Model neutrinos.
From here on, we use the term ``neutrino-like species" to refer to species
that decouple at $z_{dec,X} \sim 10^9$. For later decoupling times, the amplitude
of the phase shift is smaller and $\delta \ell$ develops a clear peak around
$\ell_{dec,X}\simeq\pi/\theta_{dec,X}\simeq\pi d_{LSS}/c_{\gamma}\tau_{dec,X}$,
which is the multipole corresponding to the angular size of the sound horizon at
$z_{dec,X}$ at the surface of last scattering. To compute $\tau_{dec,X}$ and
$d_{LSS}\simeq\tau_{0}-\tau_{dec}$, we adopt the cosmological parameters in Table
\ref{table:fiducial} and $\Delta N_{eff}=1$, and the results are shown as the
vertical lines in the right panel of Figure \ref{fig:analyticphaseshift}. Note
that the oscillations visible in Figure \ref{fig:analyticphaseshift} here are
artificially large because we have assumed that each multipole $\ell$ is sourced
by $d_\gamma$ at a single Fourier mode $k = \ell/d_{LSS}$.

\section{Numerical Computation of the Effects of Finite $z_{dec,X}$}
\label{sec:numerical}
In the previous section we derived an analytic approximation for the phase shift
due to the decoupled species $X$, which was evaluated for various $z_{dec,X}$,
assuming a purely radiation dominated universe and at the lowest order in $\epsilon_X$.
In this section we shall study the effects of finite $z_{dec,X}$ on the temperature
and polarization power spectra using the Boltzmann solver \texttt{CLASS} \cite{Blas:2011rf}
that will not rely on the approximations of the previous section. We outline the
modification of \texttt{CLASS} to model the decoupled species in Section \ref{sec:modified_class}
and present the results in Section \ref{sec:numerical_result}. Throughout this
Section, we adopt the Planck 2015 cosmological parameters \cite{Ade:2015xua},
which are summarized in Table \ref{table:fiducial}.

\label{sec:deltal}
\begin{table}[t]
\centering
\begin{tabular}{|c|c|c|}
\hline\hline
CMB temperature & $T_{\rm cmb}$ & 2.7255 \\
Baryon density & $\Omega_{b}h^{2}$ & 0.0222 \\
Cold dark matter density & $\Omega_{c}h^{2}$ & 0.1197 \\
Angle of sound horizon at photon decoupling & $\theta_{s}$ & 0.010409 \\
Optical depth & $\tau_{reio}$ & 0.06 \\
Primordial scalar fluctuation amplitude & $A_{s}$ & $2.196\times10^{-9}$ \\
Pivot scale & $k_p$ & $0.05~{\rm Mpc}^{-1}$ \\
Scalar spectral index & $n_{s}$ & 0.9655 \\
Effective number of neutrino species & $N_{eff}$ & 3.046 \\
Helium mass fraction & $Y_{p}$ & 0.24664 \\
\hline\hline
\end{tabular}
\caption{Fiducial cosmological parameters adopted in this paper.}
\label{table:fiducial}
\end{table}

\subsection{Modification of the Boltzmann Code for Decoupled Species}
\label{sec:modified_class}
Let us start by introducing the Boltzmann hierarchy of species $X$
in the synchronous gauge, in which the usual Boltzmann codes such as \texttt{CLASS}
and \texttt{camb} \cite{Lewis:1999bs,Howlett:2012mh} solve the set of
equations. The complete Boltzmann equation depends on the physics of the
specific collision, but the hierarchy for a massless species $X$ can schematically
be written as
\begin{align}
 \:&\dot{F}_{X,0}=-kF_{X,1}-\frac{2}{3}\dot{h} \,, \quad
 \dot{F}_{X,1}=\frac{k}{3}\left(F_{X,0}-2F_{X,2}\right)+\dot{\kappa}_XF_{X,1}C_{X,1} \,, \\
 \:&\dot{F}_{X,2}=\frac{k}{5}\left(2F_{X,1}-3F_{X,3}\right)+\frac{4}{15}\dot{h}+\frac{8}{5}\dot{\eta}+\dot{\kappa}_XF_{X,2}C_{X,2} \,, \\
 \:&\dot{F}_{X,\ell\ge3}=\frac{k}{2\ell+1}\left[\ell F_{X,\ell-1}-(\ell+1)F_{X,\ell+1}\right]+\dot{\kappa}_XF_{X,\ell}C_{X,\ell} \,,
\end{align}
where
\be
 F_X(\tau,\vec{k},\hat{n})\equiv\frac{\int dqq^2q\delta f_X(q)}{\int dqq^2q\bar{f}_X(q)}
 =\sum_{\ell=0}^{\infty}(-i)^{\ell}(2\ell+1)F_{X,\ell}(\vec{k},\tau)P_{\ell}(\mu) \,,
\ee
$h$ and $\eta$ are the metric perturbations in the synchronous gauge \cite{Ma:1995ey},
$\dot{\kappa}_X$ is the opacity, and $C_{X,\ell}$ is the collision term depending
on the exact model of interest. For example, if we consider the species $X$
being dark photons interacting with the dark baryons (as proposed by the $N$naturalness
model which is discussed in detail in Section \ref{sec:Nnaturalness}), then
the collision term would have the same form as the tightly-coupled photon-baryon
plasma but with different constants. Alternatively, if we consider the species
$X$ to be self-interacting, then the collision term would follow the formalism presented
in \cite{Lancaster:2017ksf}. Here we do not specify the exact collision term
to keep the discussion general.

In the early universe before the decoupling of the species $X$, $|\dot{\kappa}_X|\gg\mathcal{H}$ and so
only the first two moments of the Boltzmann hierarchy survive. One can thus
truncate the hierarchy at $\ell=1$ and simplify to the fluid equation as
\begin{equation}
 \dot{\delta}_X=-\frac{4}{3}\theta_X-\frac{2}{3}\dot{h} \,, \quad
 \dot{\theta}_X=\frac{k^2}{4}\delta_X \,,
\label{eq:fluid}
\end{equation}
where $\delta_X=F_{X,0}$ is the density perturbation and $\theta_X=\frac{3}{4}kF_{X,1}$
is the velocity divergence. We refer to $X$ during this time as fluid-like. On the other
hand, once $X$ decouples, after which we refer to it as a free-streaming particle, one
has to solve the complete Boltzmann hierarchy and truncate at some higher moment. Therefore,
the primary difference between solving the Boltzmann equation of the species $X$ before
and after its decoupling is the existence of $F_{X,\ell\ge2}$. We shall use this characteristic
to model species $X$ which is fluid-like and free-streaming before and after $z_{dec,X}$.

We now present the specific changes to \texttt{CLASS} for computing
the CMB peak location changes caused by the decoupling of the species $X$.
We use the $\ncdm$ feature in \texttt{CLASS} to model the decoupled
species $X$. We set the mass in the parameter file to be $10^{-8}$ eV,
hence $X$ evolves effectively as a massless particle. Since $\ncdm$
particles are assumed to be fermions in \texttt{CLASS}, we modify the
corresponding distribution function in \texttt{background.c} to allow
for the calculation of bosons, such as for dark photons.

The $\ncdm$ particles are assumed to be a decoupled species such as neutrinos,
so in principle one has to solve their Boltzmann hierarchy. In order to
accelerate the computation, \texttt{CLASS} makes the $\ncdm$ fluid approximation
\cite{Lesgourgues:2011rh} when $k\tau$ is greater than a given threshold,
quantifying how deep the mode is in horizon. Specifically, the equations
of the $\ncdm$ fluid approximation are given by
\begin{align}
\label{eq:classfluid1}
 \dot{\delta}_\ncdm\:&=-(1+w_\ncdm)\left(\theta_\ncdm+\frac{\dot{h}}{2}\right)
 -3{\mathcal H}(c_\ncdm^2-w_\ncdm)\delta_\ncdm \,, \\
\label{eq:classfluid2}
 \dot{\theta}_\ncdm\:&=-{\mathcal H}\left(1-3c_\ncdm^2\right)\theta_\ncdm
 +\frac{c_\ncdm^2}{1+w_\ncdm}k^2\delta_\ncdm-k^2\sigma_\ncdm \,, \\
\label{eq:classfluid3}
 \dot{\sigma}_\ncdm\:&=-3\left[\frac{1}{\tau}+{\mathcal H}
 \left(\frac{2}{3}-c_\ncdm^2-\frac{1}{3}\frac{{\mathbb P}_\ncdm}{P_\ncdm}\right)\right]\sigma_\ncdm
 +\frac{4}{3}\frac{c_{{\rm vis},\ncdm}^2}{1+w_\ncdm}\left(2\theta_\ncdm+\dot{h}\right) \,,
\end{align}
where $\sigma_\ncdm$ is the shear stress,
${\mathbb P}_\ncdm$ is the pseudo pressure given by
\begin{equation}
 {\mathbb P}_\ncdm\equiv\frac{4\pi}{3}\frac{1}{a^4}\int dq \bar{f}_\ncdm(q)\frac{q^6}{\epsilon_\ncdm^3} \,,
\end{equation}
and $c_{{\rm vis},\ncdm}$ is the viscosity parametrizing the shear stress.
As a result, the code effectively computes only the evolution of the first
three moments of the Boltzmann hierarchy. Note that if the species is fluid-like,
then the absence of the anisotropic stress causes the shear stress to be zero.
Thus, in \cite{Baumann:2015rya} the fluid-like particle is modeled by setting
the viscosity to be zero under the \texttt{CLASS} $\ncdm$ fluid approximation,
i.e. Eq.~(\ref{eq:classfluid1})-Eq.~(\ref{eq:classfluid3}), and the equations
for massless particles (hence $w_\ncdm=c^2_\ncdm=1/3$ and $\epsilon_\ncdm=q$)
reduce to the fluid equations, i.e. Eq.~(\ref{eq:fluid}). This is equivalent
to setting $F_{\ncdm,\ell\ge2}=0$. On the other hand, if the species is free-streaming,
then the shear stress and so the viscosity are non-zero.

For a more precise calculation of the evolution of a species that behaves
as a fluid first and then starts free-streaming at $z_{dec,X}$, we force
\texttt{CLASS} to solve the Boltzmann hierarchy for the $\ncdm$ particles
without adopting the $\ncdm$ fluid approximation by setting $\texttt{ncdm\_fluid\_trigger\_tau\_over\_tau\_k}=10^8$.
The exact calculation depends on the collision terms, but they are only important for a short period of time as $|\dot{\kappa}_X|\gg\mathcal{H}$
for fluid hence effectively only the first two moments of the Boltzmann
hierarchy survive and $|\dot{\kappa}_X|\ll\mathcal{H}$ for free-streaming
particle.
To simplify the calculation, we assume that decoupling at $z_{dec,X}$ can be
modeled by multiplying $F_{\ncdm,\ell\ge2}$ by
\begin{equation}
 f(z)=\frac{1}{2}\left[\tanh\frac{(z_{dec,X}-z)}{\Delta z}+1\right] \,,
\label{eq:fz}
\end{equation}
where $\Delta z$ characterizes the width of the decoupling. Therefore,
in the limit that $z\gg z_{dec,X}$, $f(z)\to0$ and so $F_{\ncdm,\ell\ge2}=0$ 
for the fluid-like species. On the other hand, in the limit $z\ll z_{dec,X}$
$f(z)\to1$, $F_{\ncdm,\ell\ge2}$ evolve according to the dynamics of the
species, which now contain the anisotropic stress as the decoupling happens
at $z_{dec,X}$. Since all moments are coupled via the Boltzmann hierarchy,
this captures the evolution of the free-streaming particles.

\begin{figure}[t]
\centering
\includegraphics[width=0.495\textwidth]{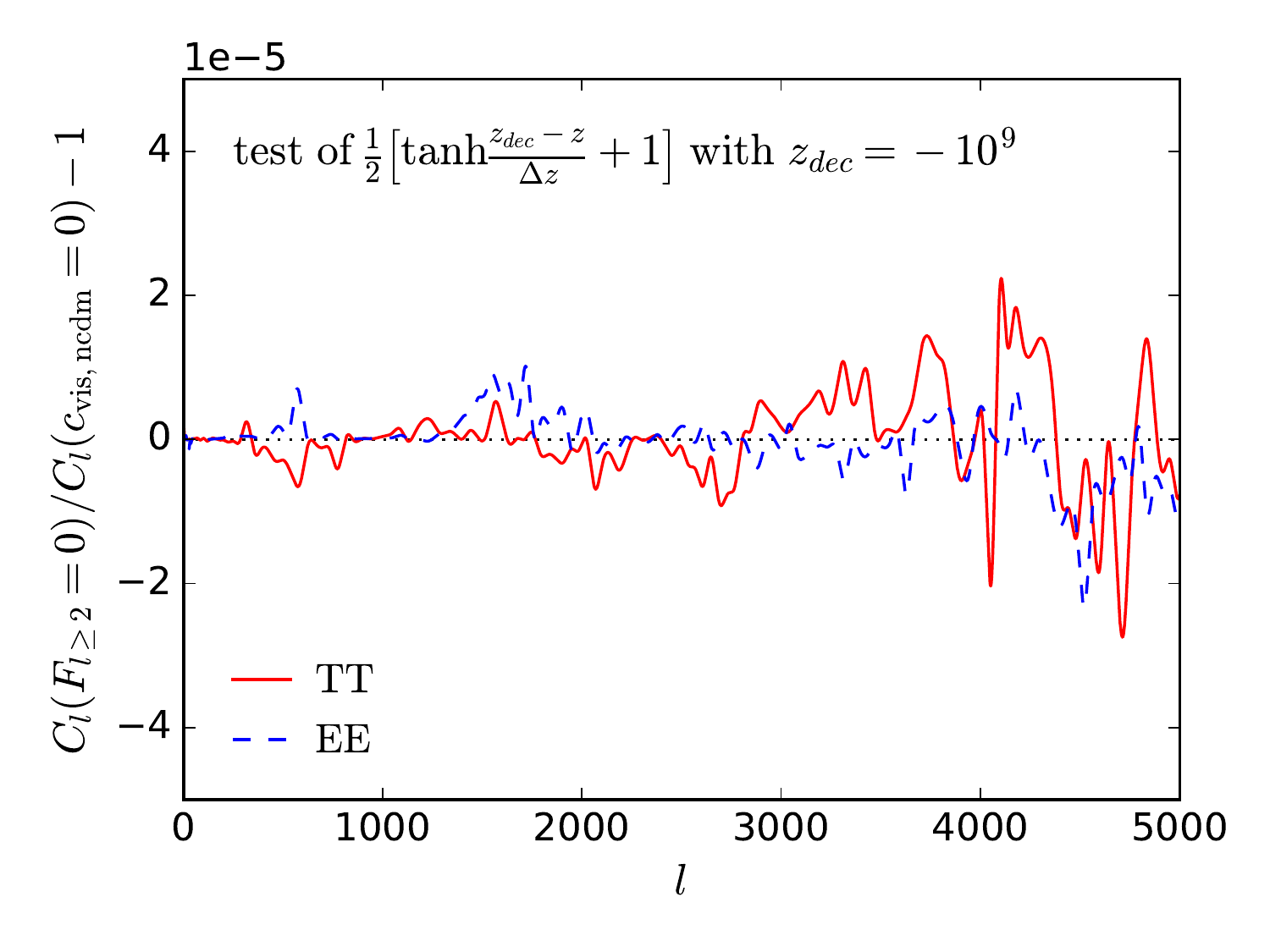}
\includegraphics[width=0.495\textwidth]{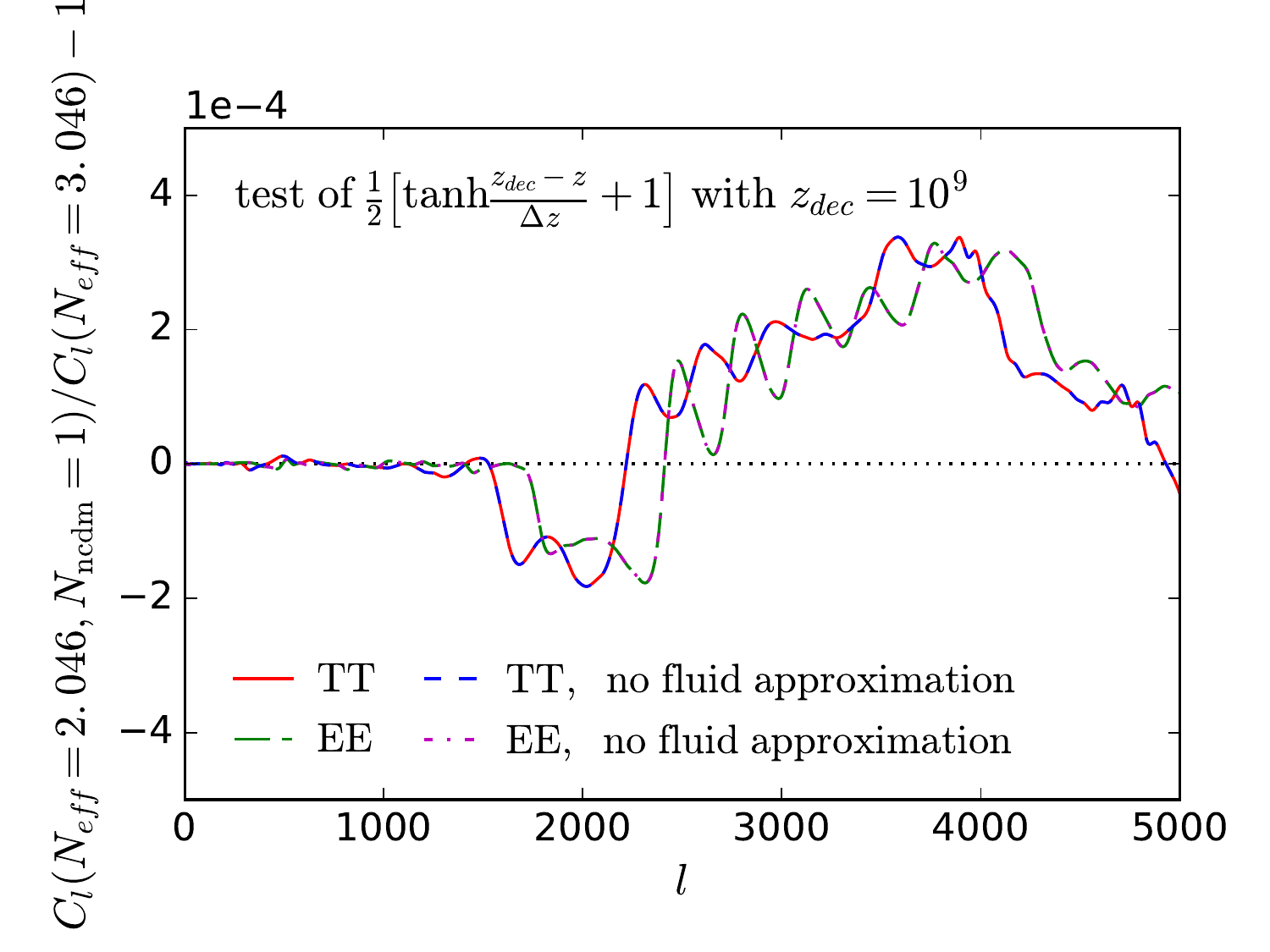}
\caption{Test of our fluid approximation, i.e. Eq.~(\ref{eq:fz}),
in the two limiting cases. (Left) Fractional difference of the unlensed
CMB power spectra in the presence of a fluid-like particle with $N_{eff}=1$
between our fluid approximation with $z_{dec,X}=-10^{9}$ and the one adopted
in \cite{Baumann:2015rya} with $c_{{\rm vis},\ncdm}=0$. The red solid and
blue dashed lines show TT and EE power spectra. (Right) Fractional difference
of the unlensed CMB power spectra of $N_{eff}=2.046$ and a neutrino-like
particle with $N_{eff}=1$ between our fluid approximation and $N_{eff}=3.046$
with \texttt{CLASS} $\ncdm$ fluid approximation. The red solid and green
long dashed lines show TT and EE power spectra of our fluid approximation,
whereas the blue dashed and magenta dot-dashed lines show those of no fluid
approximation, i.e. solving the Boltzmann hierarchy.}
\label{fig:test_fl_fs_1}
\end{figure}

We first test the modification for the fluid-like particle. We adopt the
fiducial cosmological parameters in Table \ref{table:fiducial} with
$N_{eff}=2.046$ for the neutrino-like particle and one $\ncdm$ particle
with the temperature accounting for $N_{eff}=1$. We compare the unlensed
CMB power spectra computed by setting $c_{{\rm vis},\ncdm}=0$ as used in
\cite{Baumann:2015rya} to our fluid approximation of Eq.~(\ref{eq:fz})
with $z_{dec,X}=-10^{9}$, hence effectively $F_{\ncdm,\ell\ge2}=0$ at all time.
The results are shown in the left panel of Figure \ref{fig:test_fl_fs_1},
and we find that the fractional differences for both TT and EE are less than
0.002\% for $\ell\le5000$. We next test the modification for the free-streaming
particle with the fiducial cosmology. We compare the unlensed CMB power spectra
computed by setting $N_{eff}=3.046$ to those with $N_{eff}=2.046$ and one
$\ncdm$ particle that decouples at $z_{dec,X}=10^{9}$ with a temperature
defined to give $N_{eff}=1$. The results are shown as the red solid (TT) and green
long dashed (EE) lines in the right panel of Figure \ref{fig:test_fl_fs_1}.
We find that the fractional differences are less than 0.04\% on all scales,
and the difference is mainly driven by the \texttt{CLASS} $\ncdm$ fluid
approximation. Namely, if we turn off the $\ncdm$ fluid approximation and
compute the complete Boltzmann hierarchy, we find that the results, shown
as the blue dashed and magenta dot-dashed lines, are in perfect agreement
with our fluid approximation. Note that while both tests show the fractional
differences are at most 0.04\%, for a more consistent comparison in the
rest of this paper we shall always use our fluid approximation and change
only the decoupling redshift.

\begin{figure}[t]
\centering
\includegraphics[width=0.495\textwidth]{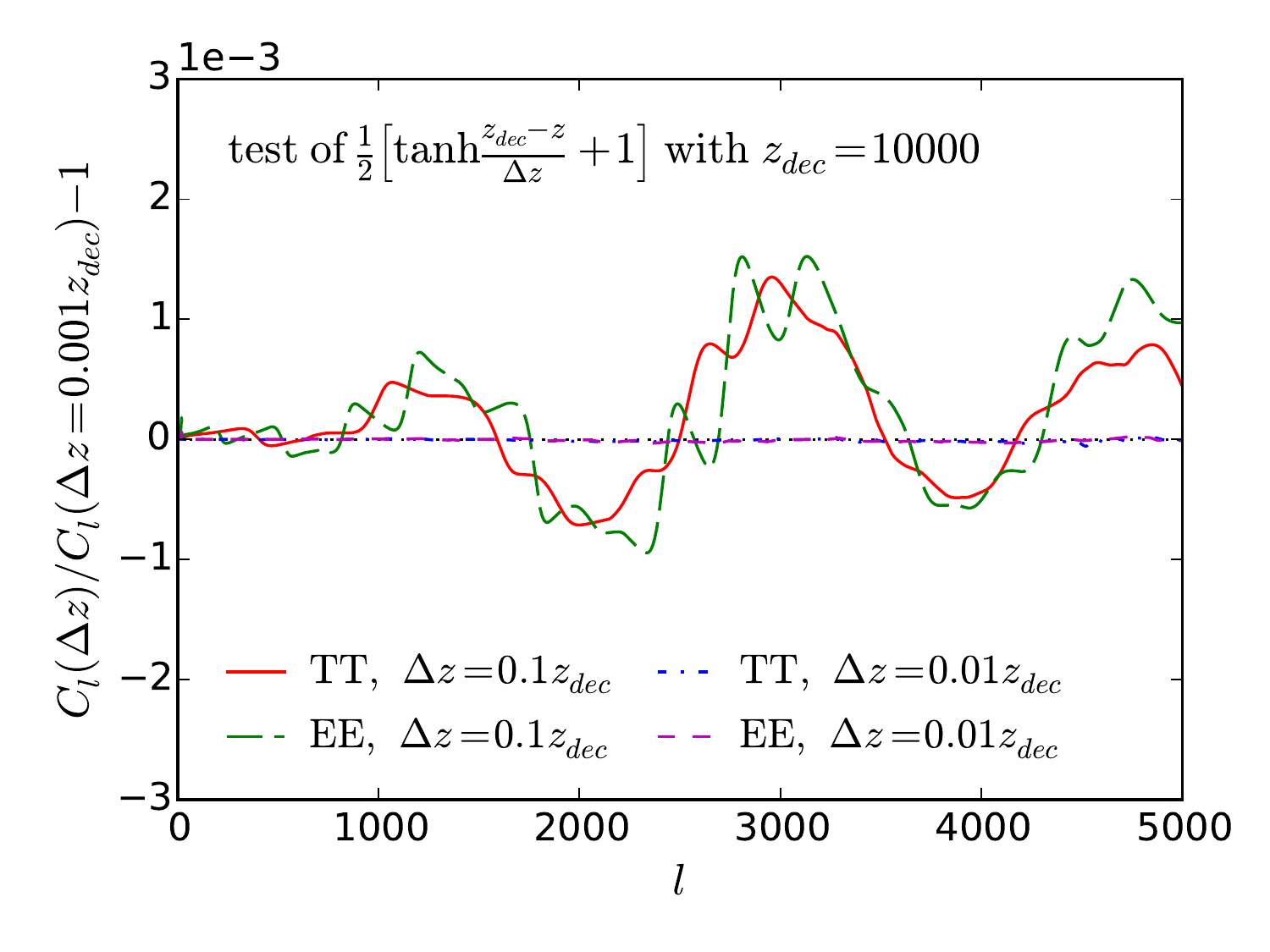}
\includegraphics[width=0.495\textwidth]{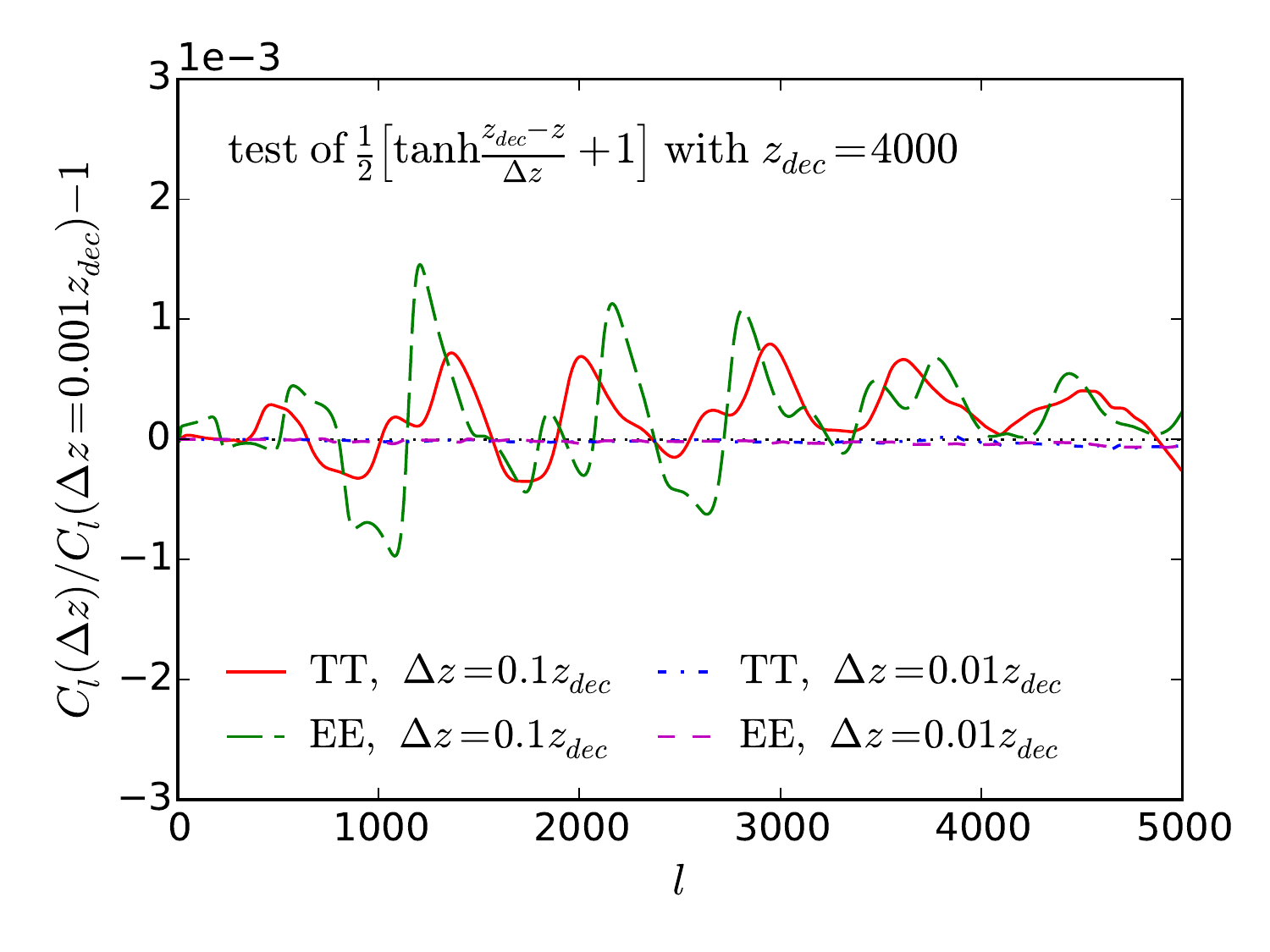}
\caption{Fractional difference of the CMB power spectra computed with our
fluid approximation for one $\ncdm$ particle with $N_{eff}=1$ decouples at
$z_{dec,X}=10000$ (left) and 4000 (right) between various $\Delta z/z_{dec,X}$
and $\Delta z/z_{dec,X}=0.001$. The red solid and green long dashed lines show
$\Delta z/z_{dec,X}=0.1$ for TT and EE, whereas the blue dot-dashed and magenta
dashed lines show $\Delta z/z_{dec,X}=0.01$ for TT and EE.}
\label{fig:test_fl_fs_2}
\end{figure}

We stress that our fluid approximation is a simplified picture because it is
only exact in the two limiting conditions. Specifically, at early times when
$|\dot{\kappa}_X|\gg\mathcal{H}$ only $F_{\ncdm,\ell\le1}$ survive and at late
times when $|\dot{\kappa}_X|\ll\mathcal{H}$ the collision term plays a little
role in the evolution of $F_{\ncdm,\ell}$, but when $|\dot{\kappa}_X|\sim\mathcal{H}$
the evolution depends on the specific model of the collision. However, usually
the opacity $\dot{\kappa}_X$ is a steep function so there is only a narrow
window that the collision term will affect the evolution of $F_{X,\ell}$. To
test the convergence of instantaneous decoupling we compute the CMB power
spectra using our fluid approximation for a neutrino-like particle with $N_{eff}=2.046$
and one $\ncdm$ particle with $N_{eff}=1$ that decouples at $z_{dec,X}$ for various
$\Delta z$. In Figure \ref{fig:test_fl_fs_2}, we compare the results for $z_{dec,X}=10000$
(left) and 4000 (right) with $\Delta z/z_{dec,X}=0.001$, 0.01, and 0.1, and the
fractional differences between $\Delta z/z_{dec,X}=(0.1,0.001)$ and (0.01,0.001)
are less than 0.2\% and 0.01\% for $\ell\le5000$, respectively. We find that the
results converge with smaller $\Delta z$, and to avoid the sensitivity of the
result to $\Delta z$, we will fix $\Delta z=0.01z_{dec,X}$ in the rest of this
paper.

We also test the sensitivity of the CMB power spectra to whether the $\ncdm$
particle is bosonic or fermionic, and we find that as long as the temperature
is set so that the corresponding $N_{eff}$ is the same, the CMB power spectra
are not affected by the nature of the $\ncdm$ particle. In the following we
fix the decoupled species $X$ to be bosonic.

\subsection{Results}
\label{sec:numerical_result}
In this section we use the modified version of \texttt{CLASS} to compute
the unlensed CMB temperature and polarization power spectra in the universe
with an additional radiation component $X$. 
 
\begin{figure}[t]
\centering
\includegraphics[width=0.49\textwidth]{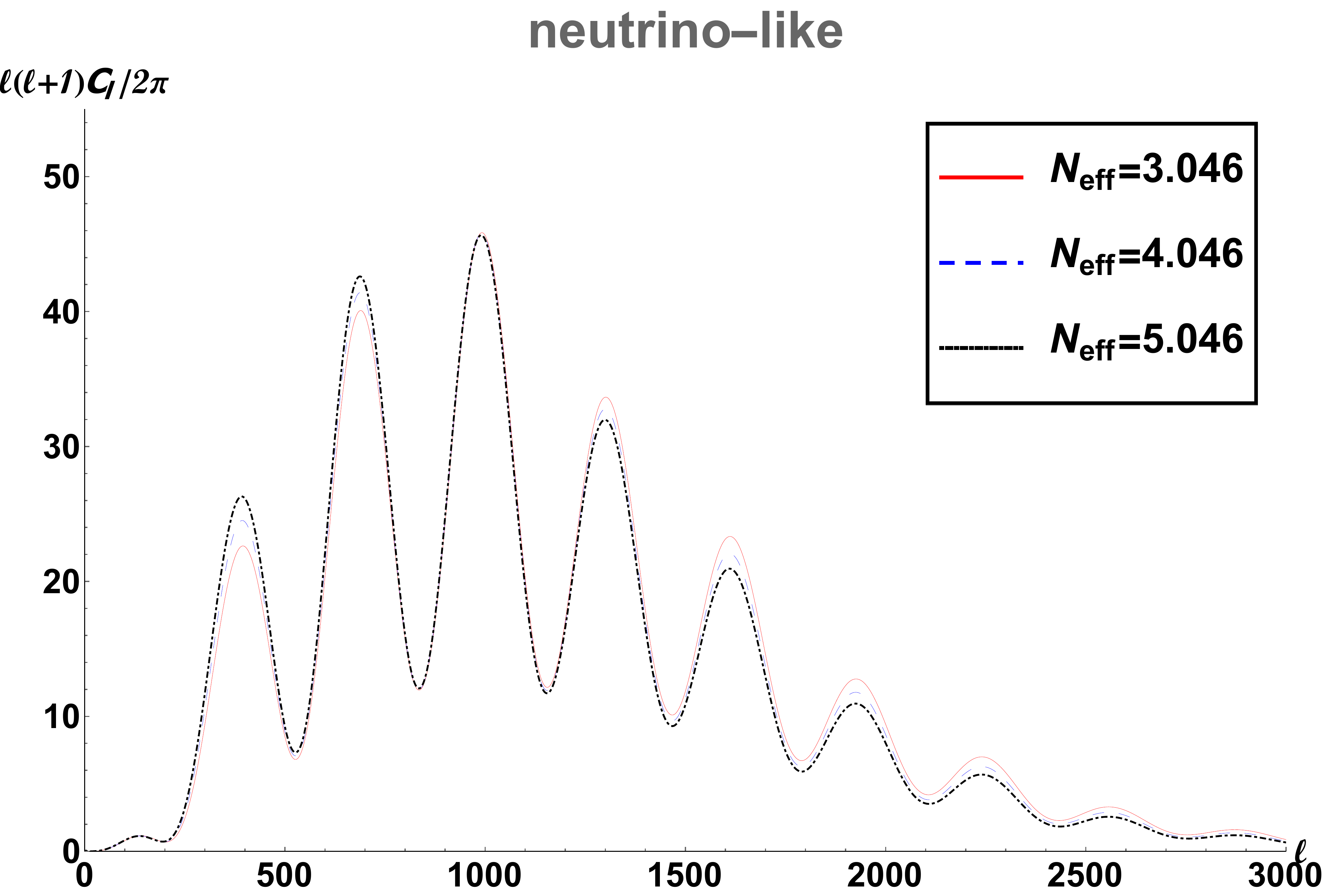}
\includegraphics[width=0.49\textwidth]{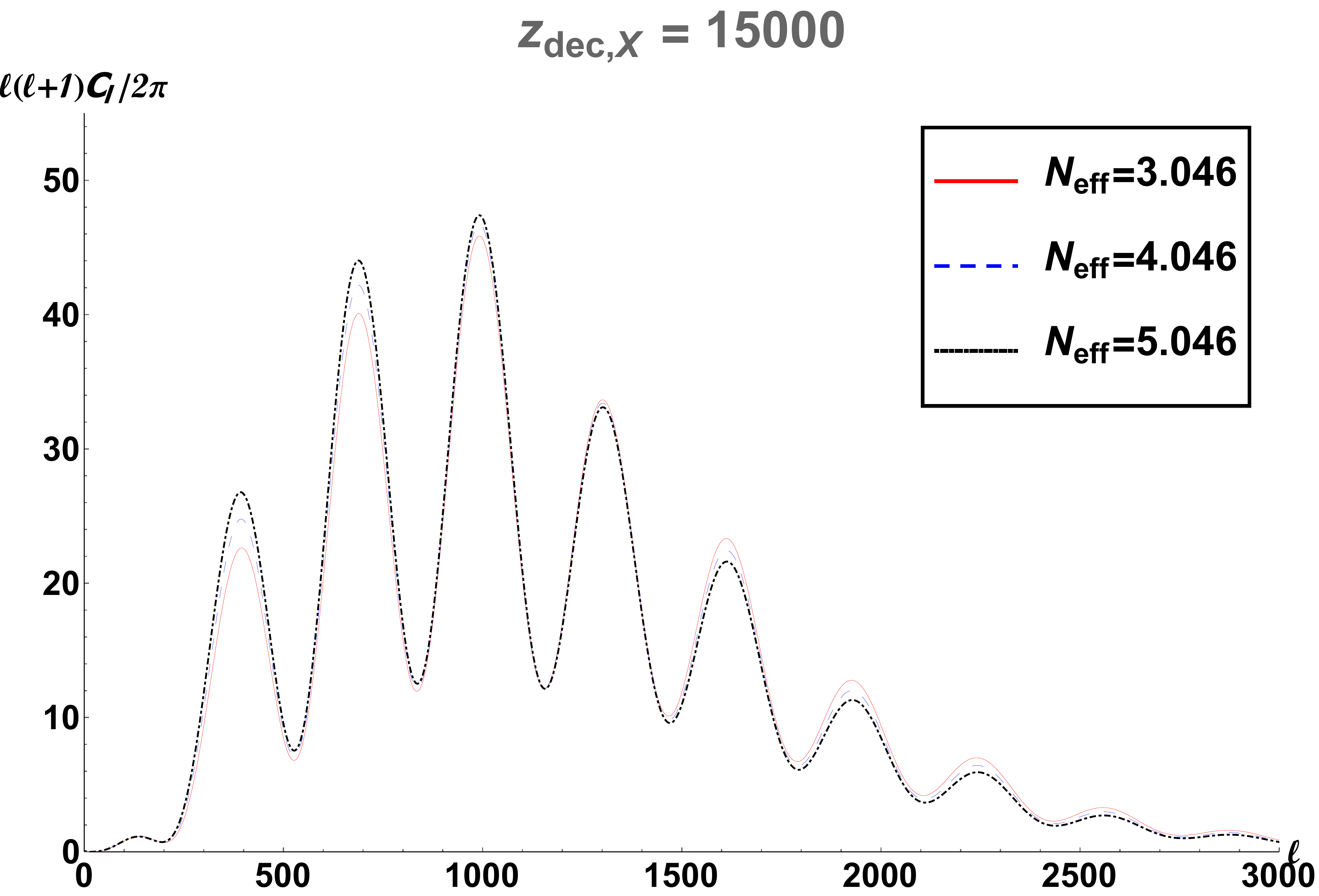}
\includegraphics[width=0.49\textwidth]{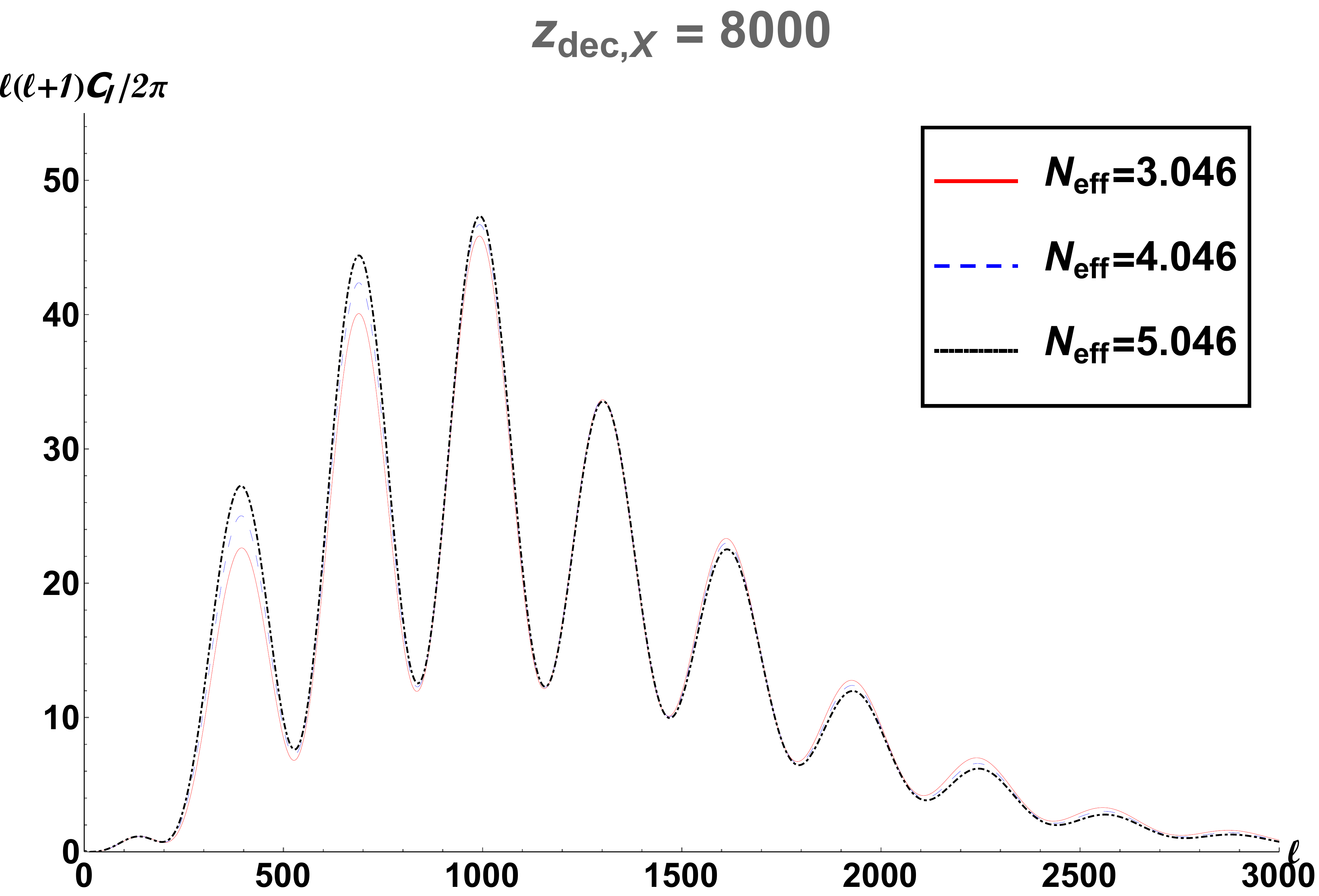}						
\includegraphics[width=0.49\textwidth]{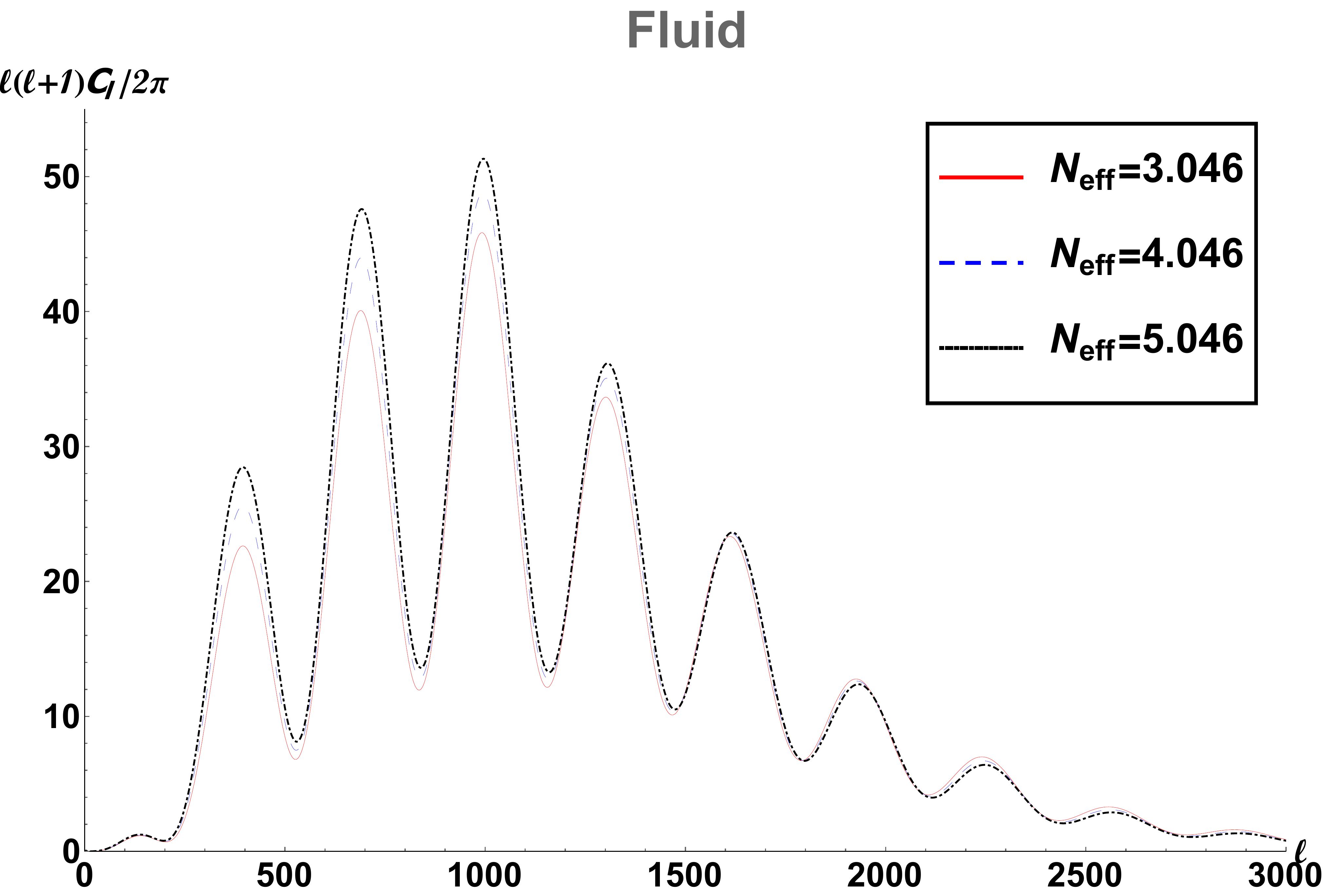}
\caption{The unlensed EE power spectra for several values of $z_{dec,X}$ and $N_{eff}$.
Each panel shows a fixed value of $z_{dec,X}$ and three different $N_{eff}$ values. From top left to bottom right we show a neutrino-like species ($z_{dec,X} =10^{9}$),
$z_{dec,X} =15000$, $z_{dec,X} = 8000$, and a species that never decouples (fluid).
The red solid, blue dashed, and black dot-dashed lines correspond to $N_{eff}$=3.046,
4.046 and 5.046, respectively. The curve with $N_{eff}$=3.046 is without the extra
species $X$ that decouples at $z_{dec,X}$ and is the same in every panel.
The vertical axis $\ell(\ell+1)C_{\ell}/2\pi$ is in units of $\mu {\rm K}^{2}$.}
\label{fig:PWS_zdec}
\end{figure}

In Figure \ref{fig:PWS_zdec}, the EE power spectra are plotted for a range of $z_{dec,X}$ and $\Delta N_{eff}$ values. All other cosmological parameters are held fixed with the values given in Table \ref{table:fiducial}.  As $\Delta N_{eff}$ increases the mean radiation density increases, which changes the matter-radiation equality time, the early integrated Sachs-Wolfe effect, and the sound horizon and damping scales \cite{Bashinsky:2003tk, Hou:2011ec}. These effects from changes to the mean radiation density do not depend on $z_{dec,X}$ and are common to each panel. The differences between the panels with different $z_{dec,X}$ and common $N_{eff}$ values are caused by the differences in the behavior of the perturbations in the relativistic species, whether they are fluid-like, neutrino-like, or transition from one to the other.  As discussed in Section \ref{sec:thdeltal}, once $X$ decouples the amplitude of photon perturbations is suppressed and a phase shift is induced in the power spectrum. In Figure \ref{fig:PWS_zdec}, the dominant effect that is visibly different between panels with different $z_{dec,X}$ values is the decrease in amplitude of the power spectra for earlier decoupling times. 

We next study the phase shift between power spectra with fixed $\Delta N_{eff} =1$ and different $z_{dec,X}$.
We define the shift via changes of the peak locations $\delta\ell\equiv\ell_{\rm fl}-\ell (z_{dec,X})$,
where $\ell_{\rm fl}$ and $\ell(z_{dec,X})$ are the multipoles of the power spectra
peaks for $\Delta N_{eff} = 1$ caused by a fluid-like species and a species decoupling at $z_{dec,X}$, respectively.
To determine the peak locations, we identify the local maxima of the power spectrum
using spline interpolation, and we find eight and thirteen peaks in TT and EE
power spectra, respectively. Note that we study $\delta \ell$ of the unlensed power
spectra rather than the lensed ones because lensing smears the power spectrum,
reducing our ability to locate the peaks \cite{Seljak:1995ve,Baumann:2015rya}.

\begin{figure}[t]
\centering
\includegraphics[width=0.3\textwidth]{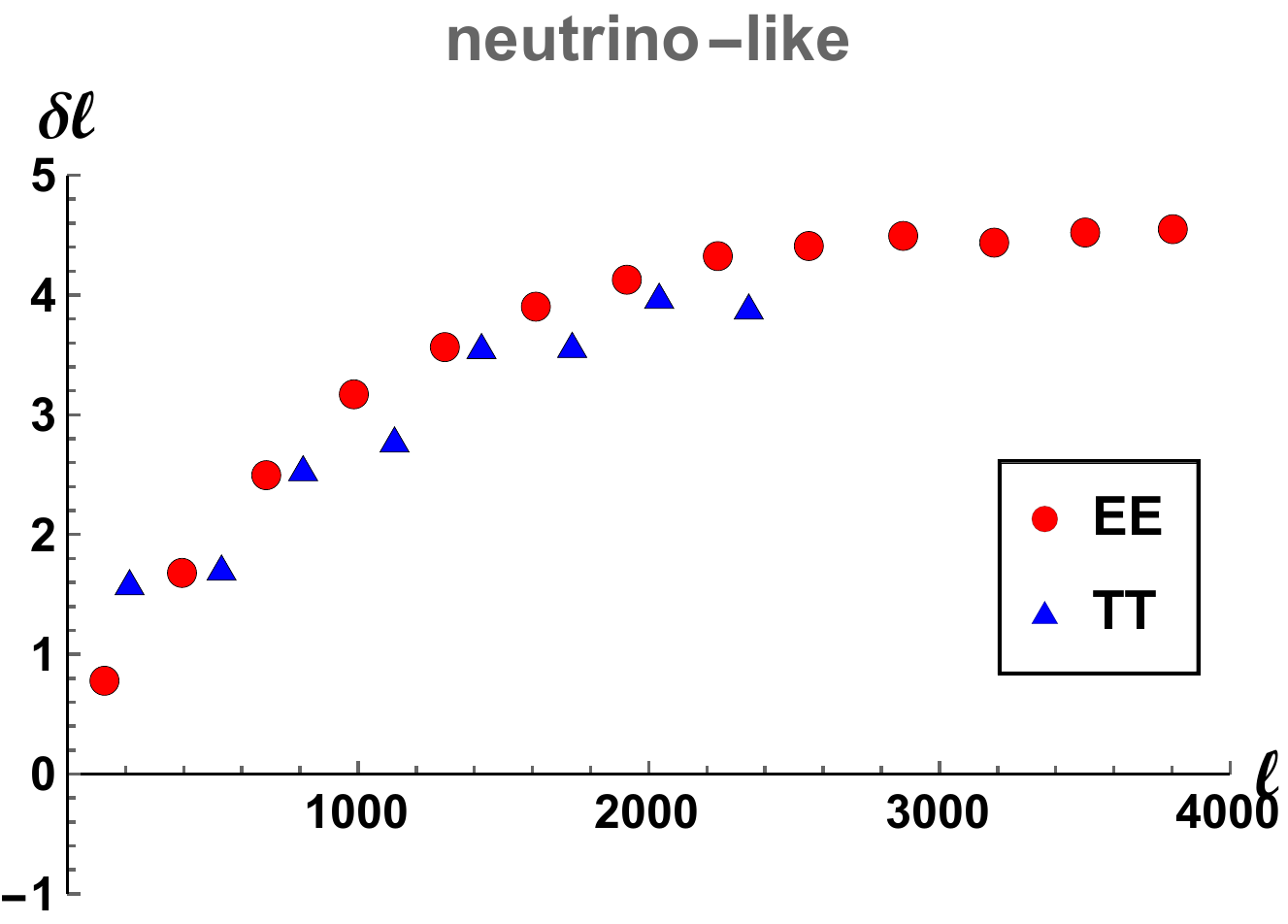}
\includegraphics[width=0.3\textwidth]{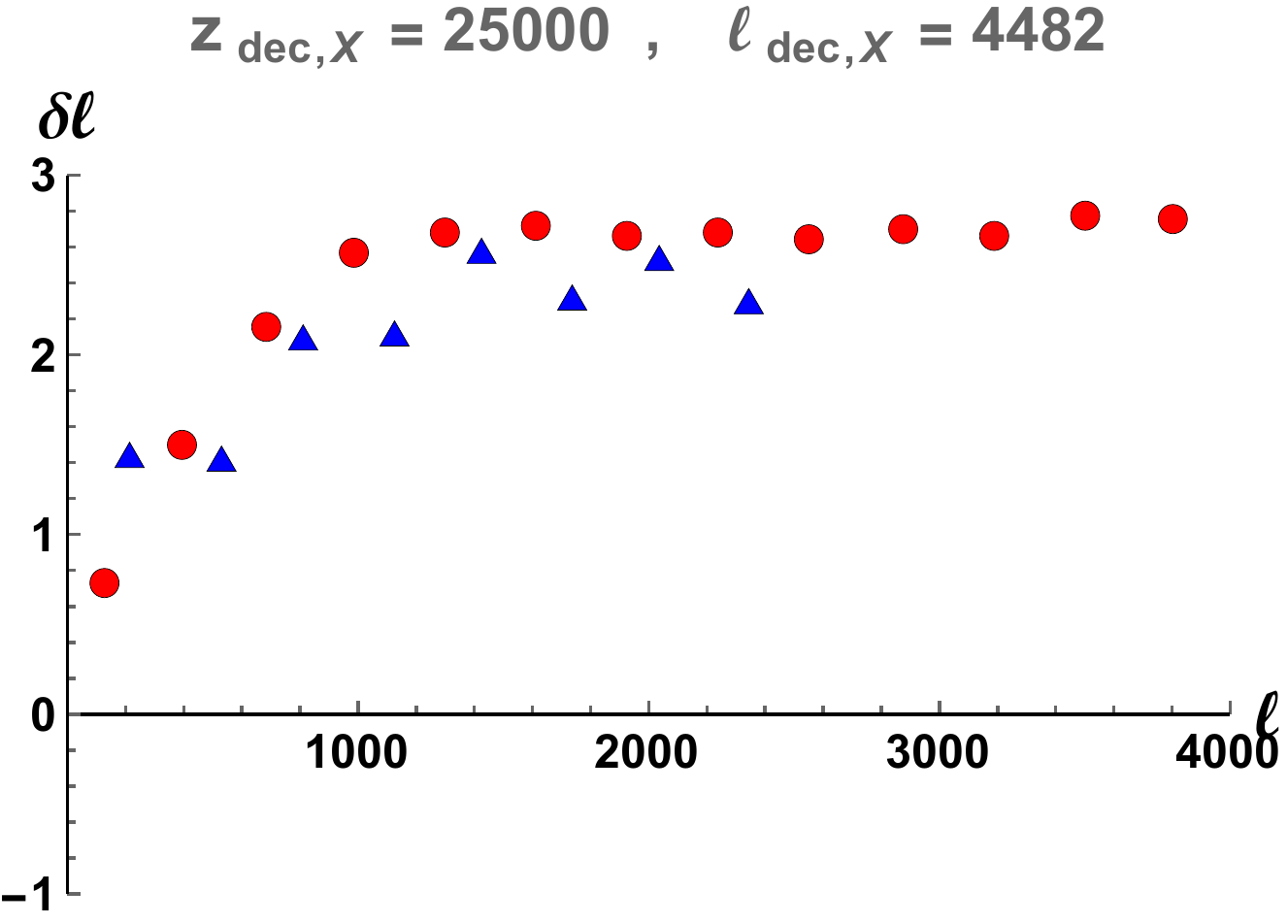}
\includegraphics[width=0.3\textwidth]{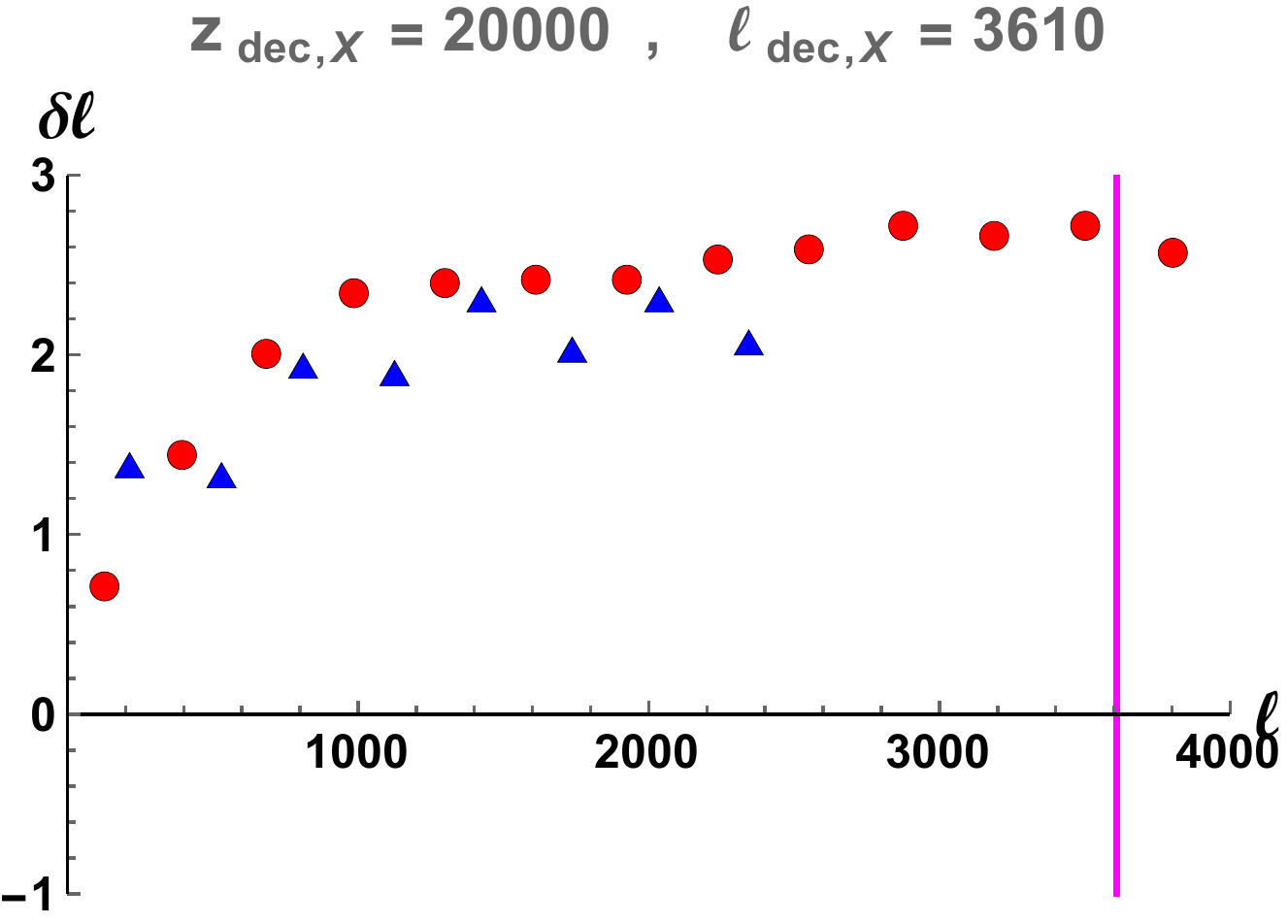}						
\includegraphics[width=0.3\textwidth]{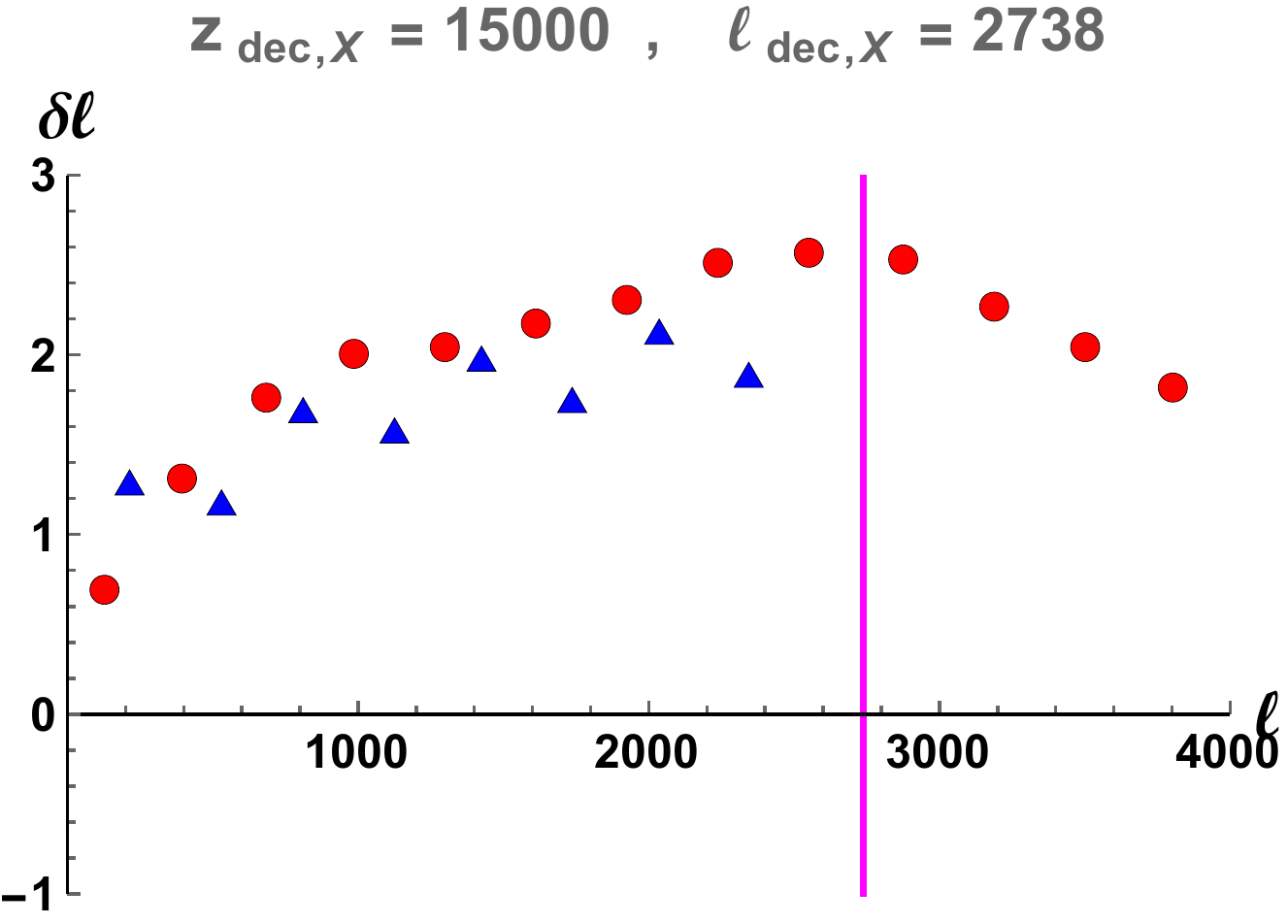}
\includegraphics[width=0.3\textwidth]{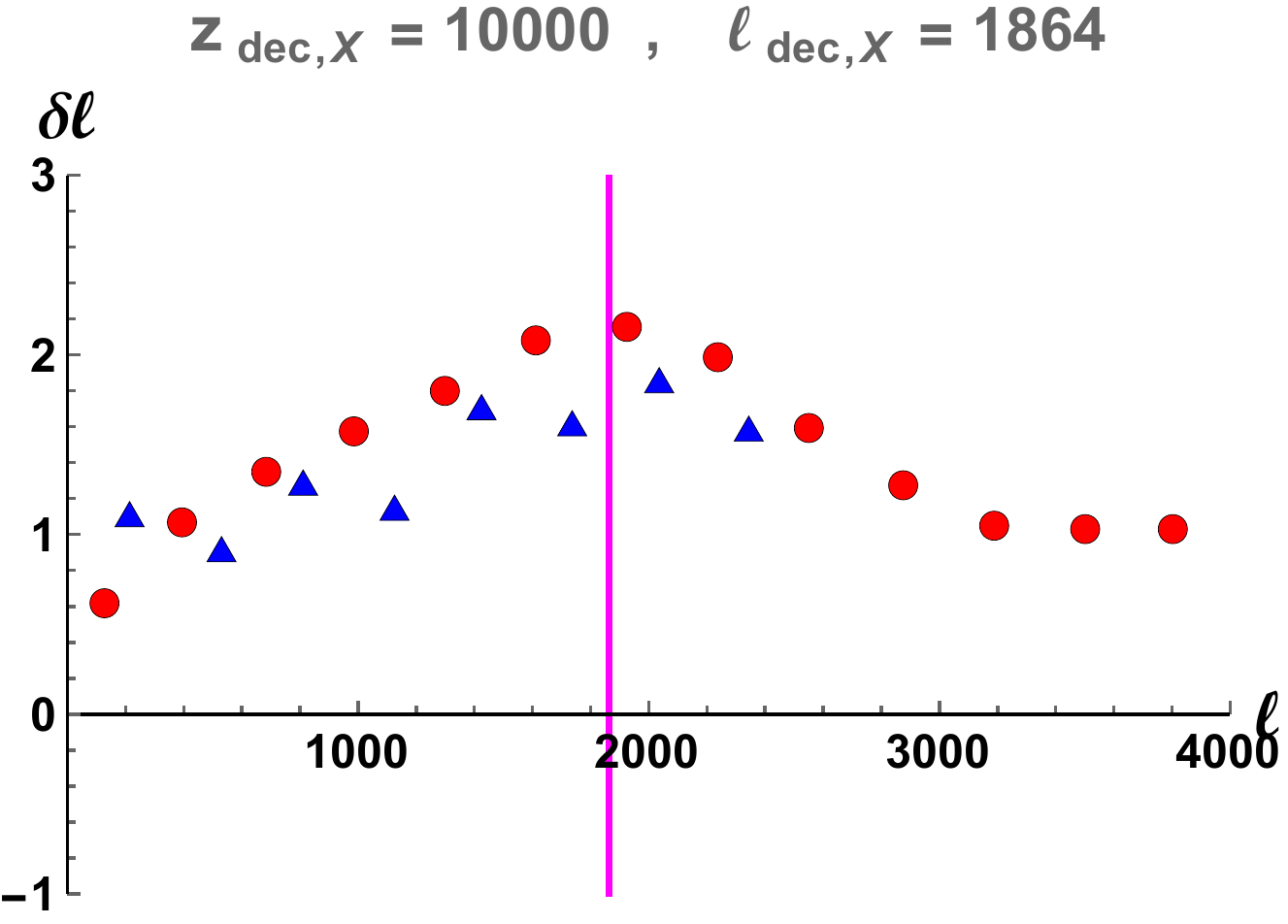}						
\includegraphics[width=0.3\textwidth]{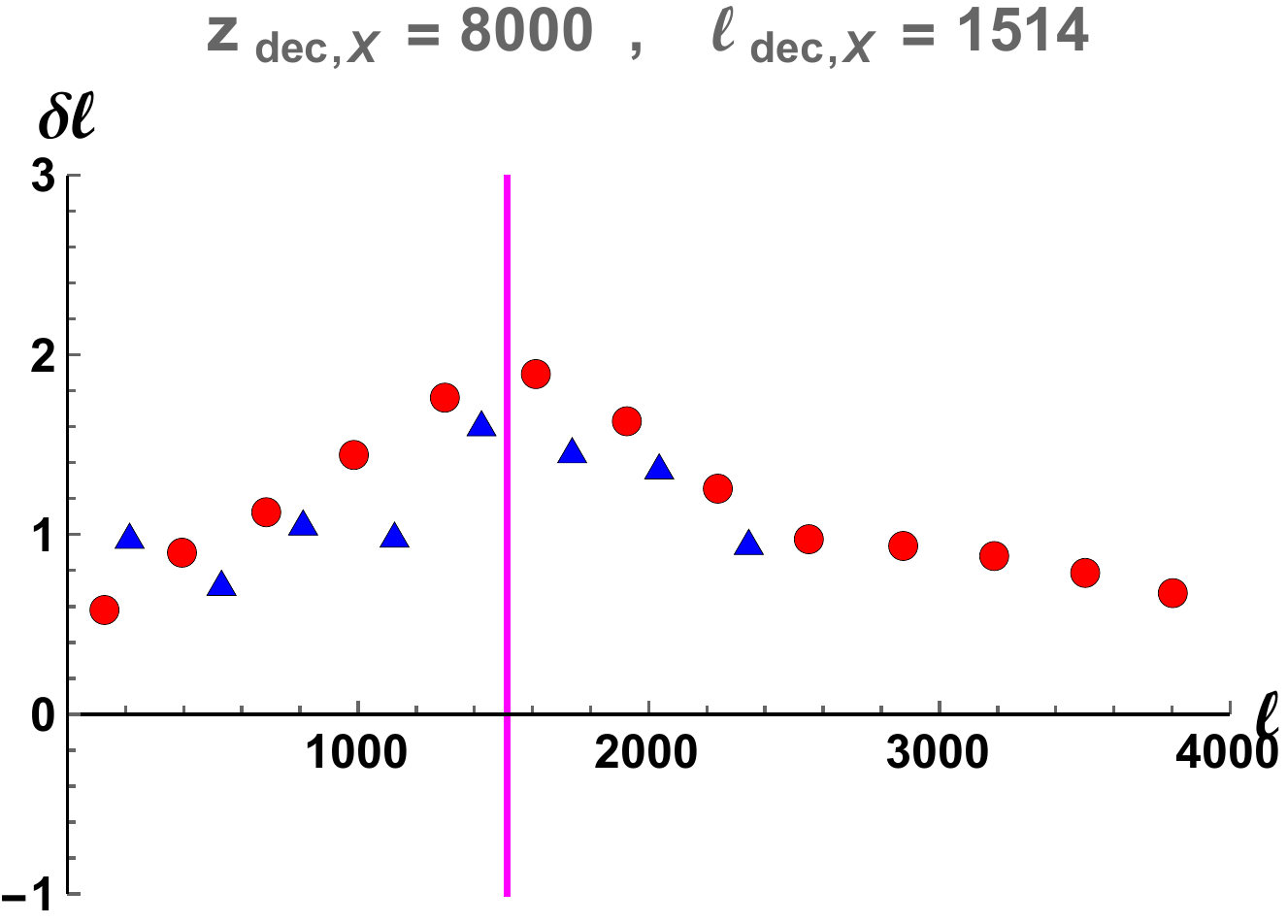}
\includegraphics[width=0.3\textwidth]{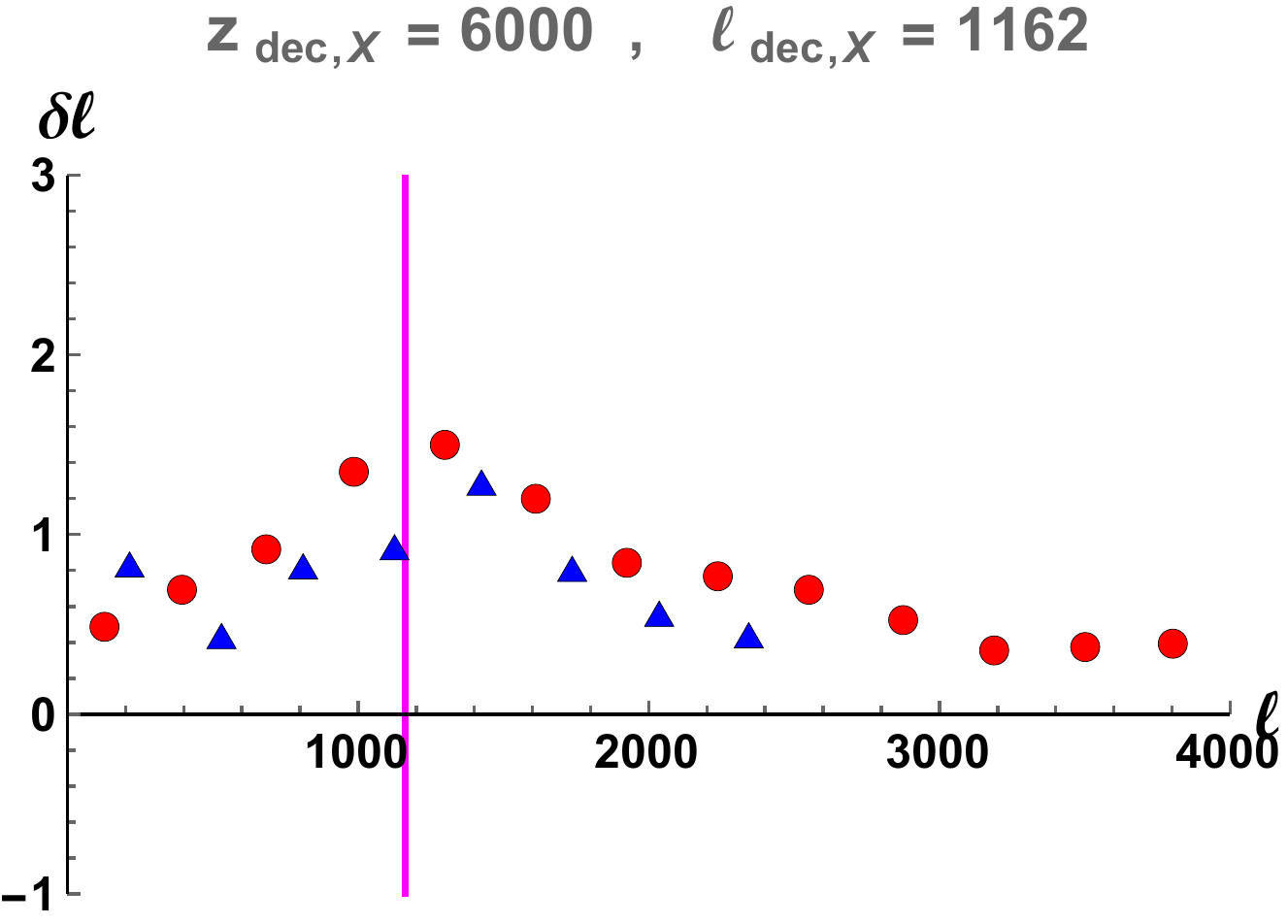}
\includegraphics[width=0.3\textwidth]{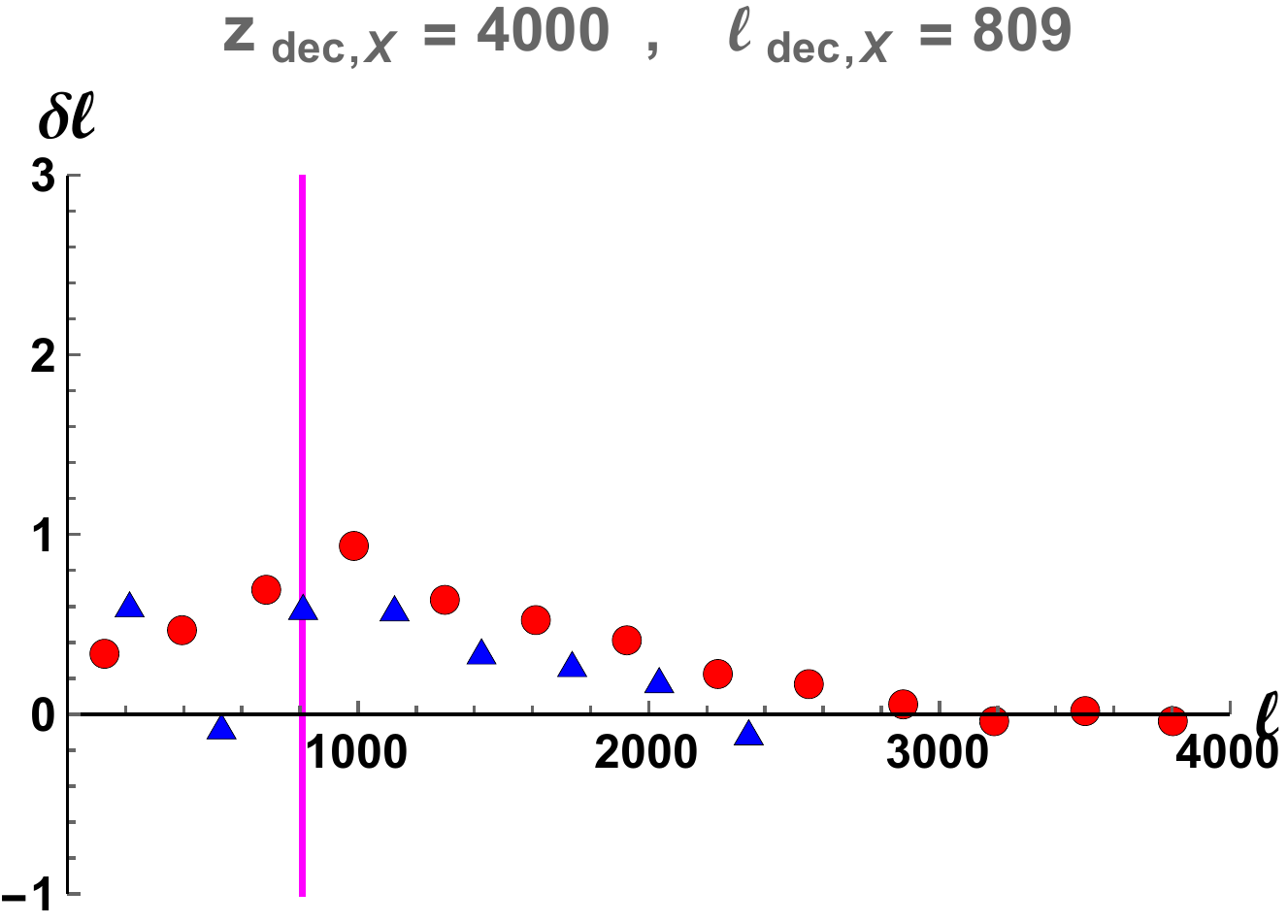}
\includegraphics[width=0.3\textwidth]{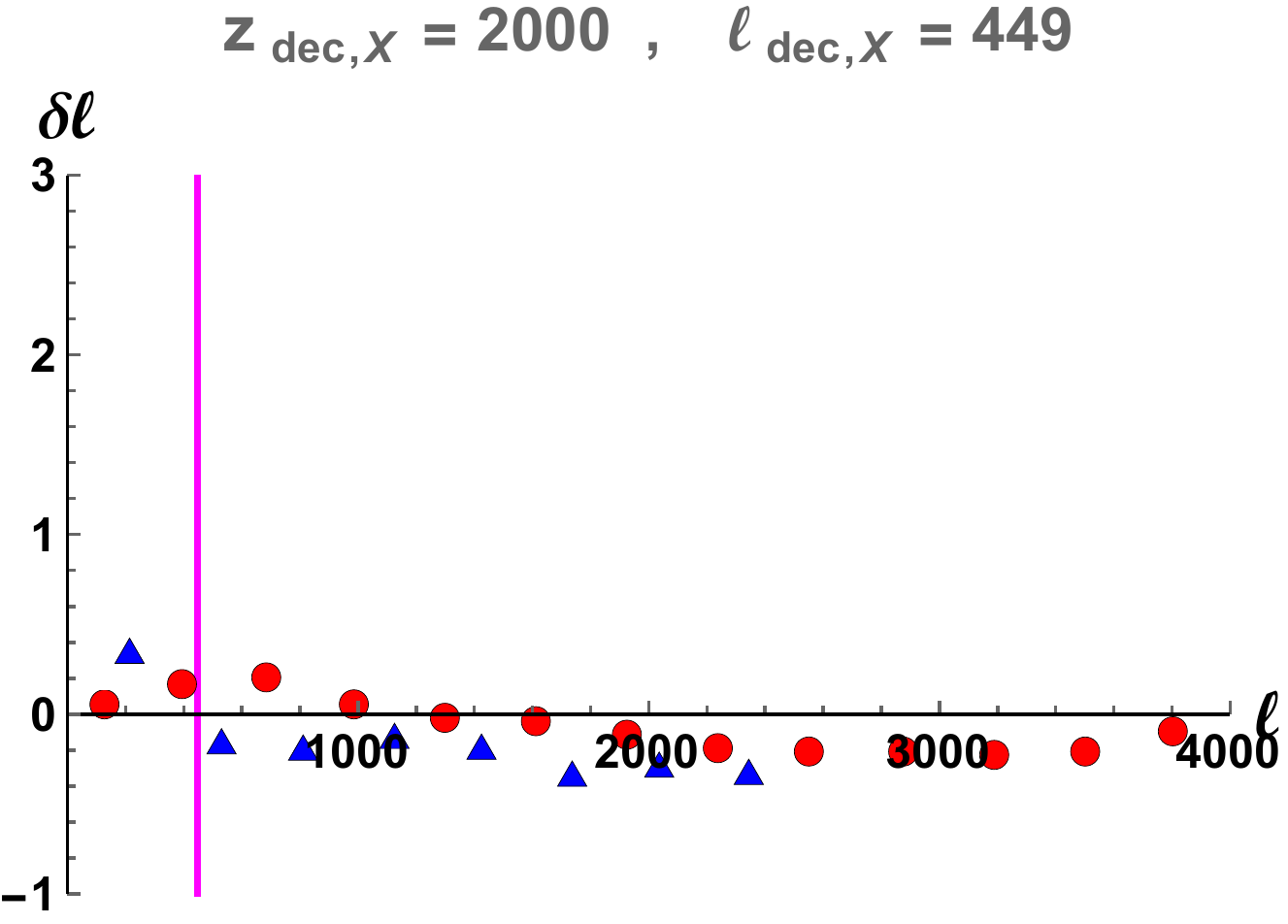}
\caption{The phase shift $\delta\ell$ of the TT (blue triangle) and EE (red circle)
power spectra between cosmologies with an additional fluid-like species and a decoupling species
$X$ with $z_{dec,X}$ decreasing from top left to bottom right panel. We fix the
additional species $\Delta N_{eff}=1$ so the difference in the phase shift is completely caused
by variations in $z_{dec,X}$. The magenta line shows the position of the multipole
$\ell_{dec,X}$ of which the associated comoving scale is equal to that of the
sound horizon at $z_{dec,X}$.}
\label{fig:psjoint}
\end{figure}

Figure \ref{fig:psjoint} shows $\delta\ell$ for the TT (blue triangle) and
EE (red circle) power spectra. The vertical magenta lines represent the multipole moment at which the associated comoving scale
is equal to that of the sound horizon at $z_{dec,X}$, i.e.
$\ell_{dec,X}\simeq\sqrt{3}\pi d_{LSS}/\tau_{dec,X}$. We display various $z_{dec,X}$
with values decreasing from top left to bottom right panels. The neutrino-like case is equivalent to the phase shift for a free-streaming particle that decouples at a very early time, i.e. $z_{dec,X}\rightarrow\mathcal{O}(10^{9})$.

The phase shifts between the TT and EE power spectra are in good agreement,
although their peaks are at different locations. This is because the
multipole moments of the amplitude of polarization anisotropy, $\Theta_{P,l}$,
are proportional to $\dot{d_{\gamma}}$, hence the polarization experiences the
same phase shift as the temperature anisotropy \cite{Zaldarriaga:1995gi,Baumann:2015rya}.
We also find that $\delta\ell$ depends strongly on $z_{dec,X}$. Specifically,
for $z_{dec,X}>20000$ the phase shift is nearly constant at high-$\ell$,
which is in qualitative agreement with the standard result for neutrinos
\cite{Bashinsky:2003tk, Follin:2015hya, Baumann:2015rya}. On the other hand,
for $z_{dec,X}\leq15000$ (corresponding to $\ell_{dec,X}\leq3000$), $\delta\ell$
peaks at around $\ell_{dec,X}$ and decreases for $\ell > \ell_{dec,X}$. This trend is in good agreement with the analytic study shown
in Figure \ref{fig:test_fl_fs_1} and can be understood as the result of
decreasing $\theta^{(1)}(y_{dec}|y_{dec,X})$ in Eq.~(\ref{eq:firstordertheta})
with increasing $y_{dec,X}$. The magnitude of $\delta\ell$ shown in
Figure \ref{fig:psjoint} is, however, smaller than the analytic prediction by about
$40-50\%$ for $z_{dec,X}\leq15000$. Previous
works studying the phase shift from free-streaming neutrinos (e.g. \cite{Bashinsky:2003tk, Follin:2015hya}) 
have also noted that the analytic calculation
in Section \ref{sec:dgammaevolution} and \ref{sec:thdeltal} over-predicts
the phase shift, in that case by $\sim 20\%$. A large part
of the discrepancy is due to the assumption of radiation domination in the analytic calculation. We
attribute the larger discrepancy seen here for lower $z_{dec,X}$ to the fact that the
radiation-dominated approximation is significantly worse as $z_{dec,X}$ gets
closer to $z_{eq}$. In particular, if species $X$ decouples too late, radiation
density does not dominate the gravitational potentials and the anisotropic
stress generated after $z_{dec,X}$ will have small impact on the evolution
of $\Phi_{\pm}$.

The strong dependence of the overall amplitude of the power spectra and the phase shift $\delta\ell$ on $z_{dec,X}$ shown in Figures \ref{fig:PWS_zdec} and \ref{fig:psjoint} demonstrates that 
the CMB has the power to constrain properties of the decoupled species $X$. In principle not only can the energy density of $X$ be probed by the
magnitude of $\delta\ell$, but $z_{dec,X}$ can be determined from the angular
scale of the largest phase shift, i.e. the $\ell$ of the maximum $\delta\ell$, and the relative heights of the peaks seen in Figure  \ref{fig:PWS_zdec}.
Since the decoupling time and temperature are closely related to the nature
and strength of non-gravitational interactions of $X$,
our result is potentially a useful tool for identifying candidates of new radiation from
different BSM models. The potential to constrain $z_{dec,X}$, however, will require that the particle $X$ decouples late enough that the effects of finite $z_{dec,X}$ are visible in the observable $\ell$-range of CMB, but not so late that matter dominates the energy budget of the universe and perturbations in $X$ plays a little role in determining the CMB. 

\section{An Example from $N$naturalness}
\label{sec:Nnaturalness}
In the previous sections we discussed the effects of a species that decouples at intermediate redshifts on the CMB power spectra. In this section we present
one of the possible models, $N$naturalness \cite{Arkani-Hamed:2016rle},
that can produce a light degree of freedom $X$ satisfying the following
three features:
\begin{itemize}
 \item $X$ does not belong to the Standard Model particle contents;
 \item $X$ contributes additional relativistic energy density, $\Delta N_{eff}$;
 \item $X$ becomes a free-streaming particle at $z_{dec,X}\sim\mathcal{O}(10^{3})$
 with the sound speed $c_{X}>c_{\gamma}\approx1/\sqrt{3}$ before photons decouple in our sector.
\end{itemize}
We shall summarize the key physics of $N$naturalness for different sectors with
negative Higgs mass squared in Section \ref{sec:sectors} (see
\cite{Arkani-Hamed:2016rle} for details). In Section \ref{sec:decoupling} we focus on calculating the decoupling
redshifts of photons from these additional sectors, which do not interact non-gravitationally with the
substances in our sector. In Section \ref{sec:exzreci}, 
we provide a numerical calculation for a specific set of parameters in $N$naturalness
as an example. We stress that $N$naturalness is just one possible model for a species that can decouple at $z\sim 10^3-10^4$. 
For another example, see \cite{Cyr-Racine:2013jua, Lancaster:2017ksf} in which self-interacting neutrinos with
decoupling redshifts $\sim\mathcal{O}(10^{3})-\mathcal{O}(10^{4})$ were considered.

\subsection{Properties of Different Sectors}
\label{sec:sectors}
In $N$naturalness, $N$ copies of our Standard Model with the same gauge
groups and elementary particles are introduced to solve the hierarchy problem. Each
sector is labeled with an index $i$ and differentiated only by the different values
of the Higgs mass squared
\begin{equation}
 (m_{H}^{2})_{i}=-\frac{\Lambda^{2}_{H}}{N}(2i+r) \,,
\label{eq:higgsmass} 
\end{equation}
where $\Lambda^{2}_{H}$ is UV cut-off for the diverging Higgs mass correction, and
$r$ is the fine-tuning parameter to satisfy $(m_{H}^{2})_{us}=-(88~{\rm GeV})^2$
in our own sector, which is assigned to $i=0$. Since $(m_{H}^{2})_{i}<0$ holds, we
can expect spontaneous breaking of electroweak symmetry for $i\geq0$ sectors.
The larger sector index implies a greater vacuum expectation value (vev) of the
Higgs field and thus the particle masses increase with the sector index $i\geq0$. While particles
in different sectors do not interact with each other, apart from gravitational interactions, they each contribute
to the matter and radiation energy density of the universe. Moreover, the presence of photons and baryons in other sectors
will cause a part of the radiation energy density from those sectors to be tightly-coupled and fluid-like until photon decoupling occurs in 
those sectors. As we shall see, the leading few sectors with light Higgs vev will affect the evolution of the CMB
in our sector gravitationally. The $\Delta N_{eff}$ visible in the CMB will then have contributions from $i\ge1$.

After inflation ends, the reheaton populates the particle contents in each sector through
its decay. If each sector is equally reheated, then the model is inconsistent with
the current constraint on $\Delta N_{eff}$ since $\Delta N_{eff}\sim N$. Thus, this
requires the model to have a mechanism to dominantly reheat only the first few sectors.
The simplest model for a scalar reheaton $\phi$ to accomplish this is
\begin{equation}
 \mathcal{L}_{\phi}\supset-\tilde{a}\phi\sum\limits_{i}H_{i}^{2}-\frac{1}{2}m_{\phi}^{2}\phi^{2} \,,
\label{eq:reheatonhiggs} 
\end{equation}
where $\tilde{a}$ is a dimensionful coupling and $H_{i}$ is the Higgs field in the
$i^{\rm th}$ sector. Note that we only focus on the scalar reheaton, but the fermionic
reheaton has also been considered in \cite{Arkani-Hamed:2016rle}. For sufficiently
light reheaton $(m_{\phi}\ll |m_{H_{i}}|$ for all $i\geq0$), the most important
reheaton decay operator for sectors with $i\geq0$ is
\begin{equation}
 \mathcal{L}_{\phi}^{\nu\neq0}\sim\tilde{a}y_{f}\frac{\nu_{i}}{m_{h_{i}}^{2}}\phi f_{i}f_{i}^{c} \,,
\label{eq:decayoperator}
\end{equation}
where for the $i^{\rm th}$ sector, $m_{h_{i}}$ is the physical Higgs particle mass
after electroweak symmetry breaking (EWSB), $\nu_{i}$ is the Higgs vev after
EWSB, $f_{i}$ and $f_{i}^c$ are the fermion matter field and its conjugate,
and $y_{f}$ is the common Yukawa coupling associated with the fermion field $f$
(the model assumes the same Yukawa structure for all sectors).
The resulting fermion produced via Eq.~(\ref{eq:decayoperator}) must satisfy
$2m_{f_{i}}\le m_{\phi}$, and we define the largest sector index fulfilling this
condition as $N_{f}$. Therefore, if the reheaton mass is around the electroweak
scale, the reheaton would decay to bottom quarks for the first few sectors with $i\le N_{b}$,
and to charm quarks when $2m_{b,i}>m_{\phi}$ for the rest of the sectors with $i\le N_{c}$
($N_{b}<N_{c}$). Sectors of $i>N_{c}$ with $2m_{c,i}>m_{\phi}$ make a negligible contribution to the cosmological observables
 because the Yukawa couplings are small for up, down, and strange quarks relative to bottom quark,
$(y_{u,d,s}/y_{b})^{2}\leq6.25\times10^{-4}$ so their decay width is also very small.  For this reason, we restrict
our analysis to sectors with $i\leq N_{c}$. 

Following \cite{Arkani-Hamed:2016rle}, we assume that the contribution to energy densities of additional sectors due to scattering with our sector is negligible so that we can assume that energy density of the $i^{\rm th}$ sector is mainly sourced by the reheaton decay. Thus, we can estimate the energy density ratio $\rho_{i}/\rho_{us}$ by comparing the amount of energy deposited to the $i^{\rm th}$ sector to ours, using the decay width from Eq.~(\ref{eq:decayoperator}). Specifically, as
$m_{h_i}\propto\nu_i$ after EWSB, the ratio of energy densities at reheating ($a=a_{RH}$) is
\begin{eqnarray}
 ER(i,r)\equiv\frac{\rho_{i}}{\rho_{us}}(a_{RH})
 &=& \left(\frac{y_{q_{i}}}{y_{q_{us}}}\right)^2\frac{\nu_{us}^{2}}{\nu_{i}^{2}}
 = \left(\frac{y_{q_{i}}}{y_{q_{us}}}\right)^2\frac{(m_{H}^{2})_{us}}{(m_{H}^{2})_{i}} \nonumber\\ 
 &=& \left(\frac{y_{q_{i}}}{y_{q_{us}}}\right)^2\frac{-\frac{\Lambda^{2}_{H}}{N}(r)}{-\frac{\Lambda^{2}_{H}}{N}(2i+r)}
 = \left(\frac{y_{q_{i}}}{y_{q_{us}}}\right)^2\frac{r}{2i+r} \,,
\label{eq:ratio}
\end{eqnarray}
where $y_{q_{i}}$ is the Yukawa coupling of the heaviest quark in the $i^{\rm th}$ sector which satisfies $2m_{q_{i}}\le m_{\phi}$.
With the parameter $r$ we can numerically evaluate the energy density ratio. Note that
$ER(i=0,r)=1$, and $ER(i,r)$ decreases with increasing $i$, so that higher sectors
receive less energy from reheaton decay.

In order to avoid large thermal corrections to the Higgs mass that would change the branching ratios between different sectors, the maximum temperature achieved after inflation must be below the electroweak scale. As in \cite{Arkani-Hamed:2016rle}, we assume that the reheating temperature in our sector is $T_{us}(a_{RH})=T_{RH}\simeq100~{\rm GeV}$ (note, however, that it is possible to have a separation of scales between the maximum temperature after inflation and the reheating temperature \cite{Chung:1998rq}).  
Once the reheaton populates the particle contents in all sectors, each sector evolves with an
initial temperature $T_i(a_{RH})$. 

To determine the radiation density and particle content in our sector at the end of reheating we use the following criteria. If the mass of a particle $m > 6T$, then we assume that the particle state is not populated since $T \sim m/6$ is the threshold at which $80\%$ of the species of mass $m$ will have self-annihilated. This leads to the energy density of our sector at
$a_{RH}$ to be $\rho_{us}(a_{RH})=106.75\times\frac{\pi^2}{30}T_{RH}^4$. Using the energy
density ratio, $ER(i,r)$, we have the energy density of a sector $i$ at $a_{RH}$ to be
\begin{equation}
 \rho_{i}(a_{RH}) = ER(i,r)\rho_{us}(a_{RH}) = ER(i,r)\left[106.75\times\frac{\pi^{2}}{30}T_{RH}^{4}\right] \,,
\label{eq:rhoi}
\end{equation}
and assuming that the sector $i$ is fully radiation dominated we get its temperature
\begin{equation}
 T_{i}(a_{RH}) = \left[ER(i,r)\frac{106.75}{g_{*,i}(a_{RH})}\right]^{1/4}(100~{\rm GeV}) \,,
\label{eq:Ti}
\end{equation}
where $g_{*,i}=\mathcal{N}_{b,i}+\frac{7}{8}\mathcal{N}_{f,i}$. To obtain $g_{*,i}$, we start from $g_{*,i}=106.75$ since all
sectors have the same particle contents as the Standard Model in our sector and compute $T_{i}(a_{RH})$.
If $T_{i}(a_{RH})>m_{t,i}/6$ for the top quark in the $i^{\rm th}$ sector, then $g_{*,i}=106.75$
is the correct counting; otherwise we integrate out the top quark and the heaviest relativistic
particle becomes the physical Higgs particle. Then we have $g_{*,i}=106.75-\frac{7}{8}\times3\times2\times2=96.25$, we can recompute $T_{i}(a_{RH})$ and compare it to $m_{h,i}/6$. Continuing this process, one eventually arrives at $T_{i}(a_{RH})$
greater than $1/6$ of mass of the heaviest relativistic particle for the sector. We have checked that the observable quantities, $\Delta N_{eff,i}$ and $z_{dec,i}$, computed in the following subsection are only weakly sensitive to the particular choice of the temperature threshold, $T> m/6$, used to determine whether a particle of mass $m$ is present at temperature $T$.

\subsection{Photon Decoupling Redshift in Different Sectors}
\label{sec:decoupling}
In the early universe, photons in each sector with light enough Higgs mass
are tightly coupled to their baryons through Thomson scattering as the ordinary
photon-baryon plasma in our sector. While these photons contribute to $\Delta N_{eff}$,
they therefore behave as fluid-like particles since the anisotropic stress is suppressed by
the frequent scattering. As the universe expands, the photon temperature decreases
and the baryon density drops, so at some point photons in sector $i$ would decouple
from their own fermions and baryons and start free-streaming. If this happens earlier than the
last scattering of CMB ($z_{dec,us}\sim1100$) in our sector, then the presence of
the anisotropic stress due to decoupled photons in $i^{\rm th}$ sector induces the
phase and amplitude shift to the acoustic oscillations discussed in Sections \ref{sec:dgammaevolution} and \ref{sec:thdeltal}.

The temperature of photon decoupling for a sector $i$, $T_{dec,i}=T_i(a_{dec,i})$,
can be estimated by the Saha equations as \cite{Dodelson:1282338}
\begin{equation}
 \left[\frac{1-X_{e,i}(a)}{X_{e,i}(a)^{2}}\right]_{\rm equilibrium}
 =\frac{2\zeta(3)}{\pi^{2}}\eta_{i}\left(\frac{2\pi T_i(a)}{m_{e,i}}\right)^{3/2}e^{B_{H,i}/T_i(a)} \,,
\label{eq:sahaeqn}
\end{equation}
where $X_{e,i}$ is the fraction of free electrons, $\eta_{i}$ is the baryon-to-photon ratio,
$m_{e,i}$ is the electron mass, and $B_{H,i}$ is the hydrogen binding energy of the sector $i$. Note that
 the Saha equation actually gives the recombination temperature of
photons, but for simplicity we approximate the temperature of recombination and decoupling
to be the same. We define the decoupling of photons in sector $i$ to be the time when $X_{e,i}(a_{dec,i})=1/2$ is reached.
Using Eq.~(\ref{eq:higgsmass}), we have the electron mass and hydrogen binding
energy in the sector $i$ as
\begin{equation}
 m_{e,i}=\sqrt{\frac{2i+r}{r}}m_{e,us} \,, \quad
 B_{H,i}=\sqrt{\frac{2i+r}{r}}B_{H,us} \,.
\label{eq:BE}
\end{equation}
Defining $A_i\equiv T_{dec,i}[r/(2i+r)]^{1/2}$,
we can simplify Eq.~(\ref{eq:sahaeqn}) to be
\begin{equation}
 2=\frac{2\zeta(3)}{\pi^{2}}\eta_{i}\left(\frac{2\pi}{m_{e,us}}\right)^{3/2}A_i^{3/2}e^{B_{H,us}/A_i} \,.
\label{eq:sahaeqn1}
\end{equation}
For simplicity we assume the same baryon-to-photon ratio $\eta_i=\eta_{us}=5.5\times10^{-10}$
for sectors with $0\leq i\leq N_{c}$ as ours.\footnote{If one assumes a certain source of a lepton asymmetry for each sector within the model and the lepton asymmetry is distributed amongst the sectors as the energy density is distributed, we expect conversion of the lepton asymmetry to the baryon asymmetry to be less efficient than ours for sectors with $i\geq1$. As we will see in Section \ref{sec:exzreci}, both $T_{i}(a_{RH})<T_{RH}$ and $M_{W,i}>M_{W,us}$ hold due to increasing Higgs vev for $i\geq1$. Since the transition rate between vacua of different baryon or lepton numbers induced by the sphaleron is proportional to $\exp(-M_{sph}/T)$ where $M_{sph}$ is the sphaleron's mass and $M_{sph}\sim M_{W}$ \cite{1985PhLB..155...36K}, a greater exponential suppression is expected for the conversion of the lepton asymmetry to the baryon asymmetry for sectors with $i\geq1$ when compared to ours. Nonetheless, we assume $\eta_{i}=\eta_{us}$ for the leading few sectors with $0\leq i\leq N_{c}$ in computing $T_{dec,i}$ in Eq.~(\ref{eq:Tdec}) for simplicity. } 
With this value for $\eta_i$, the numerical solution yields $A_i\approx0.3232~{\rm eV}$, or equivalently
\begin{equation}
 T_{dec,i}=0.3232\sqrt{\frac{2i+r}{r}}~{\rm eV} \,.
\label{eq:Tdec}
\end{equation}
We see that for $\eta_i = \eta_{us}$, photon decoupling in the other sectors generically occurs earlier than in our own sector. Further, we note that the value of $A_i$ and therefore $T_{dec,i}$ is very weakly sensitive to $\eta_i$. For instance, for $\eta_i = 10^{-15}$, $A_i = 0.252$eV, while for $\eta_i = 10^{-6}$, $A_i = 0.4187$eV.  Large changes in $\eta_i$ therefore only change $T_{dec,i}$ by a small amount so that $T_{dec,i}$ remains relatively close to $T_{dec,us}$ for small $i$ and $r\sim \mathcal{O}(0.1)$.

To compute the photon decoupling redshift, we use the conservation of entropy to write
\begin{equation}
 g_{*s,i}(a_{dec,i})T_{dec,i}^{3}a_{dec,i}^{3} = g_{*s,i}(a_{RH})T_{i}(a_{RH})^{3}a_{RH}^{3} \,,
\label{eq:entropy}
\end{equation}
where $g_{*s,i}$ is the effective number of relativistic degrees of freedom for
entropy in a sector and is equal to $g_{*,i}$ when all the relativistic species
are in thermal equilibrium at the same temperature in the sector.  Since neutrinos decouple prior to electron-positron annihilation and
photon recombination\footnote{The neutrino interaction rate of the $i^{\rm th}$ sector is
$\Gamma_i\sim G_{F,i}^{2}T_i^{5}$, where $G_{F,i}\sim\alpha_{W}/M_{W,i}^{2}$ is the
Fermi constant, $M_{W,i}$ is the mass of W boson, and $\alpha_{W}=g^{2}_{EW}/(4\pi)$
is the coupling constant. Assuming that the Hubble expansion is dominated by the radiation
in our sector, $H\sim T_{us}^{2}/M_{pl}$, and neutrinos decouple when $\Gamma_i\approx H$.
Using Eq.~(\ref{eq:ratio}) and Eq.~(\ref{eq:Ti}) and $M_{W,i}=\sqrt{(2i+r)/r}M_{W,us}$,
we have
\begin{equation}
 T_{\nu,i}^{3}=\frac{1}{G_{F,i}^{2}M_{pl}}\left(\frac{T_{us}}{T_{i}}\right)^{2}
 =\frac{M_{W,i}^{4}}{\alpha_{W}^{2}M_{pl}}\frac{y_{q_{us}}}{y_{q_{i}}}\sqrt{\frac{2i+r}{r}}\sqrt{\frac{g_{*,i}}{g_{*,us}}}
 =T_{\nu,us}^{3}\frac{y_{q_{us}}}{y_{q_{i}}}\left(\frac{2i+r}{r}\right)^{5/2}\sqrt{\frac{g_{*,i}}{g_{*,us}}} \,,
\label{eq:dectemp}
\end{equation}
where $T_{\nu,us}^{3}=M_{W,us}^{4}/(\alpha_{W}^{2}M_{pl})\simeq~{\rm (1~MeV)^{3}}$
is the cube of the neutrino decoupling temperature of our sector. From Eq.~(\ref{eq:BE}), we have
\begin{equation}
 \frac{T_{\nu,i}}{m_{e,i}}=\frac{T_{\nu,us}}{m_{e,us}}\left(\frac{y_{q_{us}}}{y_{q_{i}}}\right)^{1/3}
 \left(\frac{2i+r}{r}\right)^{1/3}\left(\frac{g_{*,i}}{g_{*,us}}\right)^{1/6}
 \simeq\left(\frac{y_{q_{us}}}{y_{q_{i}}}\right)^{1/3}
 \left(\frac{2i+r}{r}\right)^{1/3}\left(2^{6}\cdot\frac{g_{*,i}}{g_{*,us}}\right)^{1/6} \,,
\label{eq:Tmratio}
\end{equation}
where $T_{\nu,us}/m_{e,us}\simeq2$. For a radiation-dominated sector before neutrino
decoupling, $g_{*,i}\ge2$ because at least photons are relativistic and $g_{*,i}\le g_{*,us}=106.75$,
so the last term in Eq.~(\ref{eq:Tmratio}) is greater than unity. In addition,
$y_{q_{us}}>y_{q_{i}}$ for $i\geq1$, hence $T_{\nu,i}>m_{e,i}>B_{H,i}$. We find that
for any sectors with $i\ge1$ neutrinos decouple earlier than both electron-positron
annihilation and photon recombination.}, the photon temperature increases relative
to the neutrino temperature after electron-positron annihilation in all sectors with
$i\geq0$. Therefore, we have
$g_{*s,i}(a_{dec,i})=2+\frac{7}{8}\times2\times N_{eff}\times(\frac{4}{11})=3.94$
with $N_{eff}=3.046$. On the other hand, $g_{*s,i}(a_{RH})=g_{*,i}(a_{RH})$ in the
early universe. From Eq.~(\ref{eq:Tdec}) we can determine $a_{dec,us} = 7.26\times 10^{-4}$, which, in combination with Eq.~(\ref{eq:entropy}) and the assumed reheating temperature in our sector $T_{RH, us} = 100$GeV, gives $a_{RH} = 7.82\times 10^{-16}$. We then have
\ba
 a_{dec,i} &=& \frac{1}{1+z_{dec,i}}
 = \left[\frac{g_{*,i}(a_{RH})}{3.94}\right]^{1/3}
 \left[\frac{T_{i}(a_{RH})}{T_{dec,i}}\right]a_{RH}
\label{eq:adec}\\
&=& 4.92 \times 10^{-4}g_{*,i}(a_{RH})^{1/12}\sqrt{\frac{y_{q_{i}}}{y_{q_{us}}}}\left(\frac{r}{2i +r}\right)^{3/4} \,.
\label{eq:compacta_dec}
\ea
Note that the value of $z_{dec}$ in our sector from above is $z_{dec,us} = 1376$, different from the standard value of $z \approx 1100$. This difference is because we are making the approximation that photon decoupling and recombination occur simultaneously, and more importantly because we are using the Saha equation to compute the recombination time, rather than implementing, e.g. the three-level atom or a more complete treatment of the non-equilibrium recombination physics (see, e.g. \cite{Peebles:1968ja,Zeldovich:1969en, Grin:2009ik,AliHaimoud:2010ym}). A more detailed treatment of recombination in other sectors would be interesting, but is beyond the scope of this paper.

\subsection{An Example of $z_{dec,i}$}
\label{sec:exzreci}
Let us now numerically compute the decoupling redshifts for the $i^{\rm th}$ sector
for $(m_{\phi},r)=(10~{\rm GeV},0.5)$ as an example. The results for $i\le3$
are summarized in Table \ref{table:zdeci}.

We first compute the threshold sectors by solving the largest integer satisfying
\begin{equation}
 N_b\le r\left(\frac{m_{\phi}^{2}}{8m_{b,us}^{2}}-\frac{1}{2}\right) \,, \quad
 N_c\le r\left(\frac{m_{\phi}^{2}}{8m_{c,us}^{2}}-\frac{1}{2}\right) \,,
\end{equation}
and this leads to $N_b=0$ and $N_c=3$ for $m_{b,us}\simeq4.18~{\rm GeV}$ and
$m_{c,us}\simeq1.29~{\rm GeV}$. Following the procedure presented at the end
of Section \ref{sec:sectors}, we first find the heaviest relativistic particle
to be the Z boson in the first sector and the bottom quark in sectors with
$i=2, 3$. This is shown in the second column of Table \ref{table:zdeci}.

We next compute the photon properties in the $i^{\rm th}$ sector. We first count
the effective number of relativistic degrees of freedom using the heaviest relativistic
particle in each sector. Assuming $T_{us}(a_{RH})=100~{\rm GeV}$, we compute the
temperature at the $i^{\rm th}$ sector at the reheating time, $T_{i}(a_{RH})$, using Eq.~(\ref{eq:ratio})
and Eq.~(\ref{eq:Ti}) in each sector. Since we assume the Yukawa structure is identical
for all sectors, we have
\begin{equation}
 \frac{y_{q_{0<i\le N_b}}}{y_{q_{us}}}=\frac{y_b}{y_b}=1 \,, \quad
 \frac{y_{q_{N_b<i\le N_c}}}{y_{q_{us}}}=\frac{y_c}{y_b}\approx0.31 \,.
\label{eq:yq}
\end{equation}
The photon decoupling temperature in each sector, $T_{dec,i}$, can be calculated using
Eq.~(\ref{eq:Tdec}). With $T_i(a_{RH})$ and $T_{dec,i}$, we obtain the photon decoupling
redshift $z_{dec,i}$ using Eq.~(\ref{eq:adec}). The values of $z_{dec,i}$ calculated in this way are given in Table \ref{table:zdeci}.\footnote{Recall that we are making the approximation that recombination and decoupling occur at the same time. To confirm that photons in the other sectors remain tightly coupled to baryons until their recombination time we compute the mean free path of photons in the $i^{\rm th}$ sector at that time and compare that to the Hubble radius. The mean free path is given by $\lambda_{\rm mfp}=1/(n_{e}\sigma_{T})$ where $n_{e}$ is the electron number density and $\sigma_{T}=(8\pi\alpha_{em}^{2})/(3m_{e}^{2})$ is Thompson cross section. Using Eq.~(\ref{eq:BE}), Eq.~(\ref{eq:Tdec}), and Eq.~(\ref{eq:compacta_dec}) one finds 
 \begin{equation}
\lambda_{\rm mfp,i}\sim\frac{\eta_{us}}{\eta_{i}}\left(\frac{T_{\gamma,us}}{T_{\gamma,i}}\right)^{3}\left(\frac{2i+r}{r}\right)\lambda_{\rm mfp,us}\sim \frac{\eta_{us}}{\eta_{i}}\left[\frac{106.75}{g_{*,i}(a_{RH})}\right]^{1/4}\left(\frac{y_{q,us}}{y_{q,i}}\right)^{3/2}\left(\frac{2i+r}{r}\right)^{7/4}\lambda_{\rm mfp,us} \,,
\label{eq:mfp}
\end{equation}
where we have assumed the number density of free electrons is approximately the number density of baryons so that $n_e \sim \eta n_\gamma \sim \eta T_\gamma^3$. For $\eta_i\sim \eta_{us}$, the mean free path of photons in other sectors is typically larger than the one in our sector. Using $n_{\gamma,us}|_{a=1}\simeq400~{\rm cm}^{-3}$, $\sigma_{T}\simeq6.65\times10^{-25}~{\rm cm}^{2}$ and $n_{\gamma}\sim a^{-3}$, we find that $\lambda_{\rm mfp,i}|_{a=a_{dec,i}} \sim \mathcal{O}(0.1 c/H|_{a=a_{dec,i}})$ for sectors $i = 1,2,3$. Photons in the extra sectors with $i=1,2,3$ are therefore coupled to baryons until $z_{dec,i}$s in Table \ref{table:zdeci} are reached. For comparison the same calculation for our sector gives $\lambda_{\rm mfp,us}|_{a=a_{dec,us}} \sim 0.01 c/H|_{a=a_{dec,us}}$. We note that this difference between $\lambda_{\rm mfp,us}$ and $\lambda_{\rm mfp,i}$ is more sensitive to $\eta_i$ than $z_{dec,i}$ and the larger mean free path can potentially create additional signatures in the CMB power spectra that we do not explore here.}

Since $T_{i}\propto1/a$ after electron-positron annihilation, we can calculate the
photon temperature of each sector at our last-scattering surface, $T_{i}(a_{dec,us})$,
using $T_{dec,i}$ and $z_{dec,i}$. We then compute $N_{eff,i}$ due to the decoupled
photons from $i^{\rm th}$ sector following Eq.~(\ref{eq:defneff}), and we find that $N_{eff,i}$
adds up to 0.155 for $1\le i\le3$.\footnote{By estimating $T_{\nu,i}(a=1)$ from Eq.~(\ref{eq:Ti}), (\ref{eq:dectemp}) and $a_{RH}\simeq0.782\times10^{-15}$, one can calculate dark neutrino's contribution to $\Delta N_{eff}$, which turns out to be 4\% of dark photon's contribution for our choice of parameters $(m_{\phi},r)=(10~{\rm GeV},0.5)$. The contributions read 0.0031, 0.0019, 0.0013 for $i=1,2$ and 3, respectively. We neglect dark neutrino's contributions for estimating $\Delta N_{eff}$ due to additional sectors here because it is hard for those to make any significant change in shift of phase and amplitude of CMB.} Assuming $\Delta N_{eff}\simeq0.155$, we compute
$\ell_{dec,i}\simeq\sqrt{3}\pi d_{LSS}/\tau_{dec,i}$ with $z_{dec,us}=1376$ and our fiducial
cosmology in Table \ref{table:fiducial}. The third to the eighth columns of Table \ref{table:zdeci}
show the numerical values of $g_{*,i}$, $T_{i}(a_{RH})$ in GeV, $T_{dec,i}$ in eV, $z_{dec,i}$,
$N_{eff,i}$ and $\ell_{dec,i}$.

\begin{table}[t]
\centering\small
\begin{tabular}{|c|c|c|c|c|c|c|c|}
\hline\hline
sector & heaviest relativistic particle & $g_{*,i}$ & $T_{i}(a_{RH})$ [GeV] & $T_{dec,i}$ [eV] & $z_{dec,i}$ & $N_{eff,i}$ & $\ell_{dec,i}$ \\
\hline
0 (us) & top quark    & 106.75 & 100   & 0.3232 & 1376 & N.A. & 325    \\
1      & Z boson      & 95.25 & 38.22 & 0.7227 & 8349 & 0.081 & 1511   \\
2      & bottom quark & 86.25  & 33.83 & 0.9696 & 13083 & 0.044 & 2300  \\
3      & bottom quark & 86.25  & 30.86 & 1.1653 & 17238 & 0.030 & 2991  \\
\hline\hline
\end{tabular}
\caption{The properties of sectors of $i\le3$ for $(m_\phi,r)=(10~{\rm GeV},0.5)$.
The second column shows the heaviest relativistic particle, and the third to
eighth columns show respectively the values of $g_{*,i}$,
$T_{i}(a_{RH})$ in GeV, $T_{dec,i}$ in eV, $z_{dec,i}$, $N_{eff,i}$ and $\ell_{dec,i}$.
The decoupled photons from the sectors with $1\leq i\leq3$ contribute a total of $\Delta N_{eff} = 0.155$. Note that the decoupling redshift for our sector differs from the standard value of $1100$ because we are performing a simpler calculation to estimate the decoupling time (for details see Sec. \ref{sec:decoupling}).}
\label{table:zdeci}
\end{table}

An interesting aspect of the result is presence of sectors that produce
$z_{dec,X}\sim\mathcal{O}(10^{3}-10^{4})$. The decoupled photons from these sectors
will induce the amplitude suppression and the $\ell$-dependent peak location change of the CMB power spectra, and the effect lies in the range that
is observable by the upcoming CMB experiments. Since the size of the effect
is proportional to the energy density of the decoupled photons and $N_{eff,i}$ decreases rapidly
for $i\ge4$, only the first few sectors are relevant for the cosmological observables.
The photons from the listed sectors in Table \ref{table:zdeci} decouple later than $z=18000$,
hence their anisotropic stress will change the acoustic peaks of the CMB power spectrum
in terms of both amplitude and phase, with the largest phase shift appearing at $\ell_{dec,i}$.
We set $(m_\phi,r)=(10~{\rm GeV},0.5)$ to perform the numerical calculation
as an example, but there are other choices in the parameter space consistent with
the $\Delta N_{eff}$ constraint. Different choices will lead to different values
of $z_{dec,i}$ and thus different effects on the CMB power spectrum.
This thus allows a direct test for the existence of such decoupled photons in other sectors and their decoupling time,
and hence a constraints on the parameter space of $N$naturalness model. 

We note that if a fraction of the dark matter is made up of the matter components of additional sectors, then the $N$naturalness model becomes a class of partially interacting dark matter (PIDM) model. In \cite{Cyr-Racine:2013fsa}, the authors studied cosmologies with massless dark photons of $U(1)_{D}$ interacting with dark atoms and noted that these scenarios would produce a ``dark acoustic oscillation" (DAO) feature in the matter power spectrum and correlation function. Applying the logic from \cite{Cyr-Racine:2013fsa} to $N$naturalness, we expect that there must be analogues of the usual BAO feature in each other sector due to the tight coupling between the matter and photons in extra sectors. The characteristic scales of the DAO of additional sectors are smaller than the BAO scale in our sector and decrease with the sector index $i\geq1$ because an earlier decoupling implies a smaller corresponding sound horizon at decoupling. As a result, the DAO bumps in the galaxy correlation function will appear at smaller scales. For parameters $\Sigma_{DAO}\equiv\alpha_{D}(B_{D}/eV)^{-1}(m_{D}/GeV)^{-1/6}\geq10^{-3}$, $\xi\equiv T_{D,dec}/T_{CMB}|_{z=0}\geq0.2$, it was shown that $f_{int}\equiv\rho_{int}/\rho_{DM}$ can be at most 5\% while the constraint becomes weakened greatly for the smaller $\Sigma_{DAO}$ \cite{Cyr-Racine:2013fsa}. Here $\alpha_{D}$ is the dark fine structure constant, $B_{D}$ is the binding energy of dark atom, $m_{D}$ is the mass of dark atom, $T_{D,dec}$ is the decoupling temperature of dark photon and $\rho_{int}$ is the energy density of the PIDM. In the context of the $N$naturalness model, however, it turns out that the additional sectors with $i\geq1$ yields $\Sigma_{DAO}\lesssim\mathcal{O}(10^{-4})$ and this quantity decreases with the index $i$. So current large-scale structure data does not place significant constraints on the $N$naturalness model. 
Additionally, in \cite{Shandera:2018xkn} the authors studied the gravitational waves
produced by the black holes formed by dark atoms. They demonstrated that it is possible
to constrain the dark atom mass, which can be translated into constraints on the $N$naturalness
parameters, by future experiment such as Advanced LIGO and Einstein Telescope.
It would be interesting to explore synergies between future DAO and CMB power spectrum
change due to dark radiation interacting with dark baryons, as well as different observables
generated by the dark sectors.

Lastly, we will comment on dark radiation's effect on primordial gravitational waves. In this paper, we have focused on scalar mode perturbations to the metric as in Eq.~(\ref{eq:scalarpotential}). However, as was pointed out and studied in \cite{Pritchard:2004qp, Xia:2008gm, Boyle:2005se, Weinberg:2003ur, Dicus:2005rh, Watanabe:2006qe, Miao:2007cw, Zhao:2009we}, non-zero anisotropic stresses will also alter the dynamics of tensor perturbations. For a fixed non-zero tensor-to-scalar ratio, they reduce the CMB B-mode polarization power spectrum beyond $\ell\simeq200$ as compared to the case without free-streaming neutrinos. In view of this, we expect dark decoupled photons from extra sectors in Nnaturalness model to also change CMB B-mode polarization power spectrum. If primordial B-modes are observed, future experiments could therefore potentially provide additional information about dark radiation and dark decoupling \cite{Benson:2014qhw,Henderson:2015nzj,Abazajian:2016yjj}.

\section{Forecast for a Stage IV CMB Experiment}
\label{sec:forecast}
Upcoming and future CMB experiments \cite{Benson:2014qhw,Henderson:2015nzj,Abazajian:2016yjj}
will provide unprecedented measurements of the CMB temperature and polarization
power spectra, allowing an accurate characterization of the small-scale acoustic
features. As presented in Section \ref{sec:cosmologicalPT} and Section \ref{sec:deltal},
a species that decouples at $z_{dec,X}\simeq{\mathcal O}(10^3-10^4)$ changes the CMB temperature and polarization power spectra
at $\ell \lesssim 5000$ (see Figures \ref{fig:PWS_zdec} and \ref{fig:psjoint}). Therefore, the high resolution CMB measurements can potentially probe the
existence of the additional light relic beyond the Standard Model and constrain
its decoupling time. In this section we shall explore the detectability of the decoupled
species for a Stage IV CMB (CMB-S4) experiment.

To forecast the expected constraint on $z_{dec,X}$, we use the Fisher matrix as
(see e.g. \cite{Tegmark:1996bz} for a review)
\begin{equation}
 \mathcal{F}_{ij} = \sum\limits_{\ell}\frac{\partial\vec{x}^T_{\ell}}{\partial\theta_{i}}
 {\mathbb C}_{\ell}^{-1}\frac{\partial\vec{x}_{\ell}}{\partial\theta_{j}} \,,
\end{equation}
where $\vec{x}_{\ell}=(C_{\ell}^{TT},C_{\ell}^{EE},C_{\ell}^{TE})$ is the data vector, ${\mathbb C}_{\ell}$
is the covariance of $\vec{x}_{\ell}$, and $\vec{\theta}$ is the parameter vector. Since
the measurement of the amplitude and peak location change relies significantly on the determination
of the acoustic features, we adopt the unlensed CMB power spectrum as our data
vector assuming that the operation of delensing can perfectly recover the unsmeared
peaks  \cite{Hirata:2003ka, Smith:2010gu, Sherwin:2015baa, Simard:2014aqa}.
In the limit of very low noise, forecasts for the lensed CMB, along with lensing
power spectrum including the full lensing-induced non-Gaussian covariance, approach
forecasts that work with only the unlensed CMB \cite{Green:2016cjr}. For CMB-S4 we include $30\le\ell\le3000$
for $C_{\ell}^{TT}$ to account for foregrounds dominating at high $\ell$ and take $30\le\ell\le5000$
for $C_{\ell}^{EE}$ and $C_{\ell}^{TE}$. Assuming that the data covariance is dominated by
the disconnected four-point function, the data covariance is given by
\begin{equation}
 {\mathbb C}\left(C_{\ell}^{\alpha\beta},C_{\ell}^{\gamma\delta}\right)=\frac{1}{(2\ell+1)f_{\rm sky}}
 \left[(C_{\ell}^{\alpha\gamma}+N_{\ell}^{\alpha\gamma})(C_{\ell}^{\beta\delta}+N_{\ell}^{\beta\delta})
 +(C_{\ell}^{\alpha\delta}+N_{\ell}^{\alpha\delta})(C_{\ell}^{\beta\gamma}+N_{\ell}^{\beta\gamma})\right] \,,
\end{equation}
where $(\alpha,\beta,\gamma,\delta)\in\{T,E\}$, $f_{\rm sky}$ is the effective fractional
area of the sky used, and $N_{\ell}^{\alpha\alpha}=\Delta_\alpha^2\exp[{\ell}(\ell+1)\theta_{b}^{2}/(8\ln 2)]$
is the noise spectrum. Following \cite{Abazajian:2016yjj}, we set $f_{\rm sky}=0.4$,
the instrumental noise $\Delta_T=1~\mu$K-arcmin and $\Delta_E=\sqrt{2}\Delta_T$, and
the full-width half-maximum of the beam $\theta_b=1$ arcmin. Following \cite{Abazajian:2016yjj, Baumann:2017gkg},
we impose a low-$\ell$ prior from Planck data by including $C_{\ell}^{TT}$ data for $2\le\ell\le29$
and $f_{\rm sky} = 0.8$ with an additional Gaussian prior on $\tau_{reio}$ with $\sigma_{\tau_{reio}}=0.01$. 

\begin{table}[t]
\centering
\begin{tabular}{|c|c|ccc|}
\hline\hline
Additional decoupled species & $\Delta N_{eff}$ & 1 & 0.334 & 0.084 \\
Helium mass fraction (BBN) & $Y_{p}^{\rm BBN}$ & 0.259 & 0.251 & 0.248 \\
Helium mass fraction (free) & $Y_{p}^{\rm free}$ & 0.192 & 0.228 & 0.242 \\
\hline\hline
\end{tabular}
\caption{The fiducial values of helium fractions for three different $\Delta N_{eff}$ when we fix the helium fraction to agree with the BBN prediction ($Y_{p}^{\rm BBN}$) and or allow it to be a free parameter but keep $\theta_{D}/\theta_{s}$ fixed
($Y_{p}^{\rm free}$). See the text for detailed descriptions.}
\label{table:Yp}
\end{table}

We consider the parameters $\vec{\theta}=(\Omega_{b}h^{2},\Omega_{c}h^{2},\theta_{s},\tau_{reio},A_{s},n_{s},\Delta N_{eff},z_{dec,X},Y_p)$.
The first six components take the fiducial values in Table \ref{table:fiducial},
with the cosmic neutrino background contributing $N_{eff,\nu}=3.046$.
We consider three values for the additional decoupled relativistic species $\Delta N_{eff}$,
ranging from completely consistent with the most recent Planck constraints to
$\sim 4\sigma$ away, shown in Table \ref{table:Yp}. We consider various values of the decoupling redshift $z_{dec,X}\sim\mathcal{O}(10^{3}-10^{4})$ to explore the constraining power
for dark decoupling during the radiation-dominated, matter-radiation equality, and matter-dominated regimes. For
the helium mass fraction, we consider two scenarios. First, we fix $Y_{p}^{\rm BBN}$ to be the BBN prediction
for a given choice of $\Omega_bh^2$ and $N_{eff}$ values \cite{Ade:2015xua}. Second, we allow $Y_p^{\rm free}$ to be a free parameter determined from the data. 
In the second case, we adjust the fiducial values of $Y_{p}^{\rm free}$ for each fiducial value of $\Delta N_{eff}$ 
according to 
\begin{equation}
 Y_{P}^{\rm free}=1-(1-Y_{P})\left[\frac{a_\nu+N_{eff,\nu}+\Delta N_{eff}}{a_\nu+N_{eff,\nu}}\right]^{0.56} \,,
\label{eq:Yp}
\end{equation}
where $Y_{p}$ takes the fiducial value provided in Table \ref{table:fiducial}.
This choice keeps the ratio $\theta_{D}/\theta_{s}$ fixed as we vary the central value of $N_{eff} = 3.046 + \Delta N_{eff}$, where $\theta_D$ is the angle subtended by the Silk damping scale. We make this choice because the ratio is well-determined by current data \cite{Hou:2011ec}.
The values of the two choices of helium fractions are summarized in Table \ref{table:Yp}.
In total there are nine and eight components in $\vec{\theta}$ of the Fisher matrix
for the two scenarios. To compute the derivatives of the data with respect to the
parameters, we adopt the same step sizes as used in \cite{Green:2016cjr} except
for $\theta_s$. For $\theta_s$ we take the step size to be $0.0025\theta_{s}$,
which is half of the step size taken in \cite{Green:2016cjr} because we find a better
convergence. For $z_{dec,X}$, we take $0.025z_{dec,X}$ as the step size. We have checked that for the $z_{dec,X}$ values we consider, our results are not very sensitive to the step size in $z_{dec,X}$ for $\Delta z_{dec,X}\lsim 0.1 z_{dec,X}$.

\begin{figure}[t]
\centering
\includegraphics[width=0.495\textwidth]{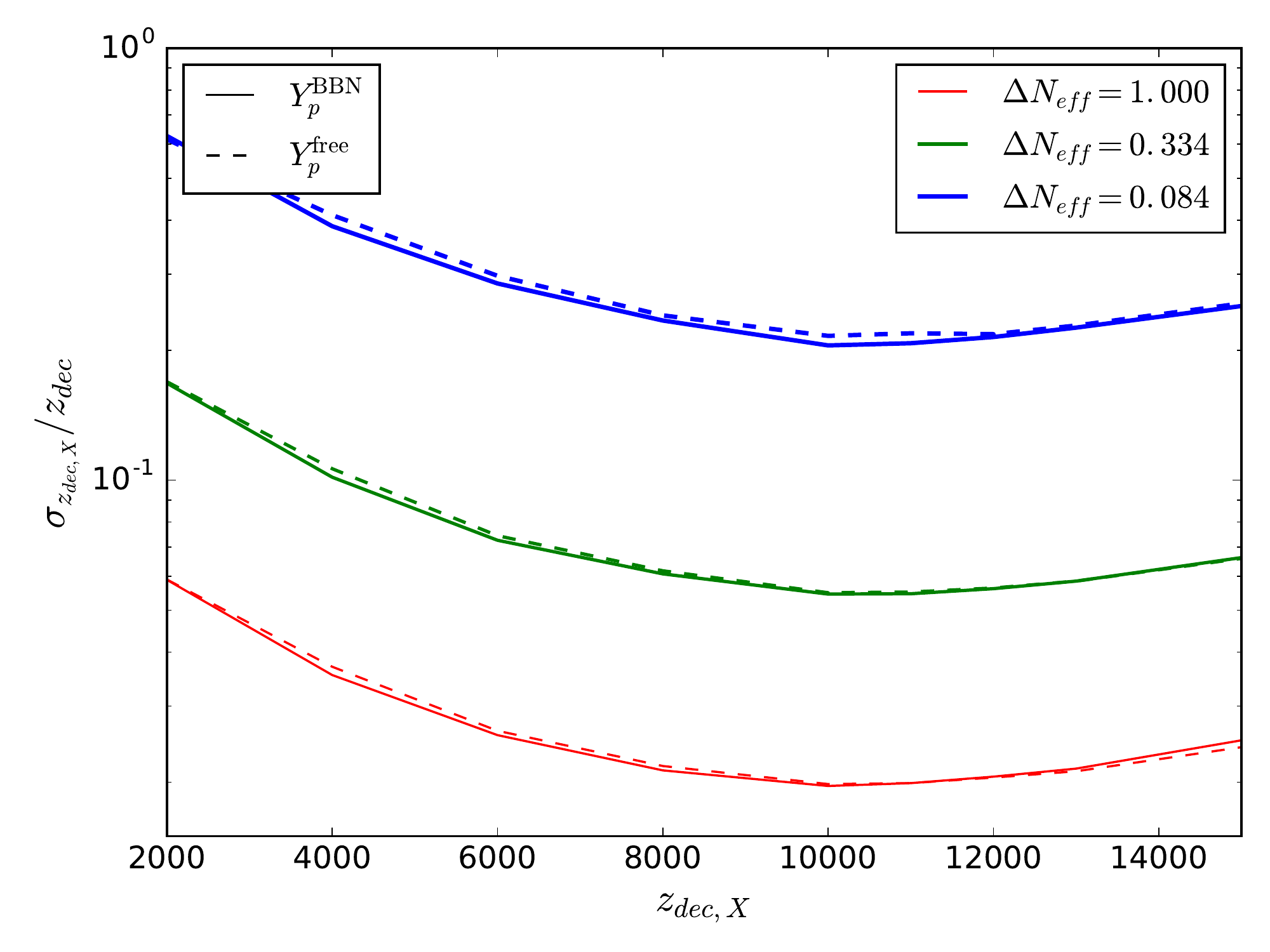}
\includegraphics[width=0.495\textwidth]{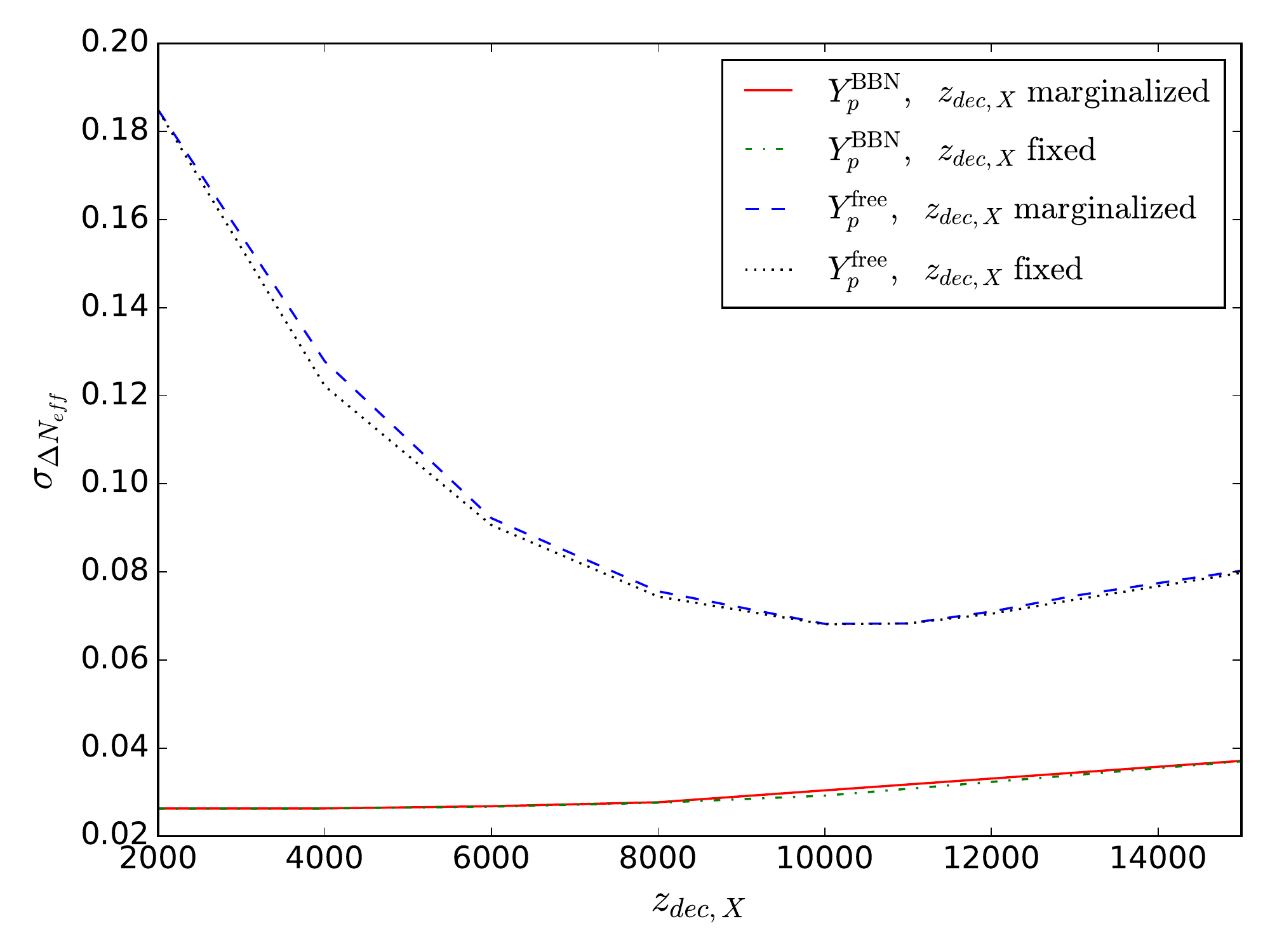}
\caption{(Left) The forecasted fractional error in $z_{dec,X}$, $\sigma_{z_{dec,X}}/z_{dec,X}$,
as a function of $z_{dec,X}$. The solid lines show the constraints when $Y_p^{\rm BBN}$ is fixed by BBN while the dashed lines show the constraints when $Y_p^{\rm free}$ is a free parameter. For a fixed $z_{dec,X}$, the fractional error in $z_{dec,X}$ is larger for smaller
$\Delta N_{eff}$. For a fixed $\Delta N_{eff}$, as $z_{dec,X}$ increases, forecasted
error decreases up to $z_{dec,X}\simeq10000$ and then increases beyond $z_{dec,X}\simeq10000$.
(Right) The forecasted error on $\Delta N_{eff}$ as a function of $z_{dec,X}$ with $\Delta N_{eff}=0.334$.
In the $Y_p^{\rm BBN}$ case, the red solid line shows the constraint with $z_{dec,X}$ marginalized over
while the green dashed-dotted line shows the constraint with $z_{dec,X}$ fixed; in the $Y_p^{\rm free}$
case, the blue dashed line shows the constraint with $z_{dec,X}$ marginalized over while the black dotted
line shows the constraint with $z_{dec,X}$ fixed.
Marginalizing $z_{dec,X}$ has little impact on the constraints on $\Delta N_{eff}$.}
\label{fig:sigmaz}
\end{figure}

The left panel of Figure \ref{fig:sigmaz} shows the forecasted error on $z_{dec,X}$ normalized by $z_{dec,X}$,
$\sigma_{z_{dec,X}}/z_{dec,X}$, as a function of $z_{dec,X}$. For a fixed
$z_{dec,X}$, the fractional error on $z_{dec,X}$ is larger for smaller values $\Delta N_{eff}$.
A lower $\Delta N_{eff}$ means a smaller fractional energy density of
the decoupled species with respect to the total energy density $\epsilon_X$, thus the effects of the decoupled species are smaller and it is more difficult to constrain $z_{dec,X}$. Interestingly, we find a minimum in the fractional error on $z_{dec,X}$ around $z_{dec,X} \sim 10000$. We attribute the existence of a minimum to two reasons. First, if $z_{dec,X}$ is too large then $\ell_{dec,X}$ falls outside the range of $\ell$ probed by CMB-S4 so the $\ell$-dependent effects that provide information about $z_{dec,X}$ are not part of the data. Second, as $z_{dec,X}$ decreases to values near matter radiation equality, the effects of the perturbations in the relativistic species $X$ are increasingly unimportant for determining the CMB power spectra and therefore the physical effects of the decoupling of $X$ are not visible.  

The right panel of Figure \ref{fig:sigmaz} shows the forecasted error on $\Delta N_{eff}$
as a function $z_{dec,X}$ for fiducial $\Delta N_{eff}=0.334$.  In the $Y_p^{\rm BBN}$ case, the constraints on $\Delta N_{eff}$ are relatively
insensitive to the value of $z_{dec,X}$, varying only at the $50\%$ level with
smaller values of $z_{dec,X}$ resulting in tighter constraints on $\Delta N_{eff}$.
In the $Y_p^{\rm free}$ case the constraints on $\Delta N_{eff}$ vary by as much
as a factor of three depending on the value of $z_{dec,X}$. For intermediate values
of $z_{dec,X} \sim \mathcal{O}(10^3-10^4)$, the dominant trend is for the forecasted
constraints on $\Delta N_{eff}$ to increase as $z_{dec,X}$ increases or decreases
relative to the best constrained value $z_{dec,X} \sim 10000$. While not shown in the figure, we find that if all other parameters are held
fixed then $\Delta N_{eff}$ is best constrained when $z_{dec,X} \rightarrow z_{dec, \nu}$
for both $Y_p^{\rm BBN}$ and $Y_p^{\rm free}$ cases. Finally, we note
that for these intermediate values of $z_{dec,X}$ the constraints on $\Delta N_{eff}$
are not very sensitive to whether $z_{dec,X}$ is held fixed or treated as a free
parameter. This demonstrates that there is no loss of constraining power on
$\Delta N_{eff}$ when $z_{dec,X}$ is introduced as a parameter, hence we advocate
of inclusion of $z_{dec,X}$ for future analysis to probe not only the additional
contribution from the light relics but also their properties.

\begin{figure}[t]
\centering
\includegraphics[width=0.495\textwidth]{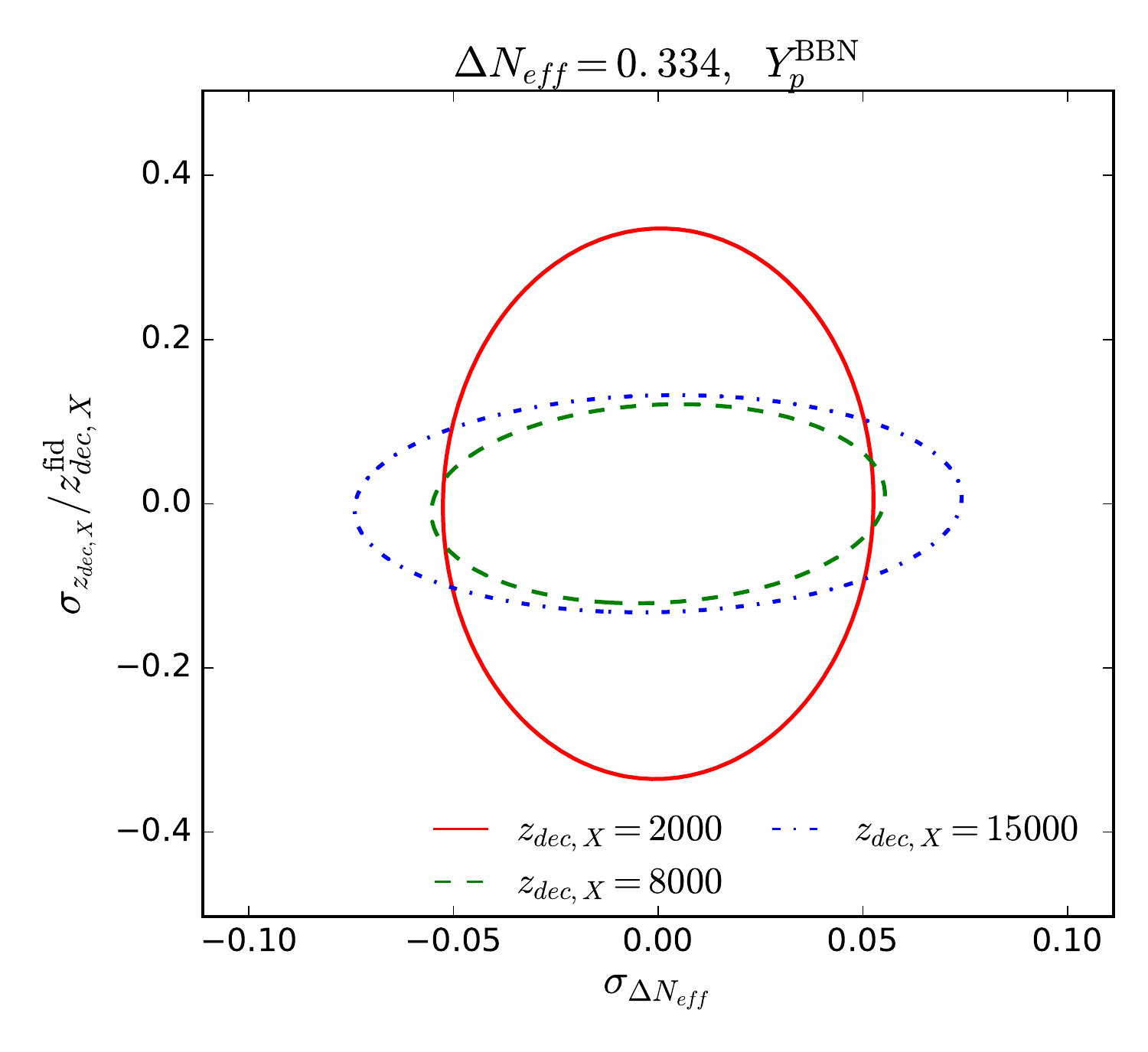}
\includegraphics[width=0.495\textwidth]{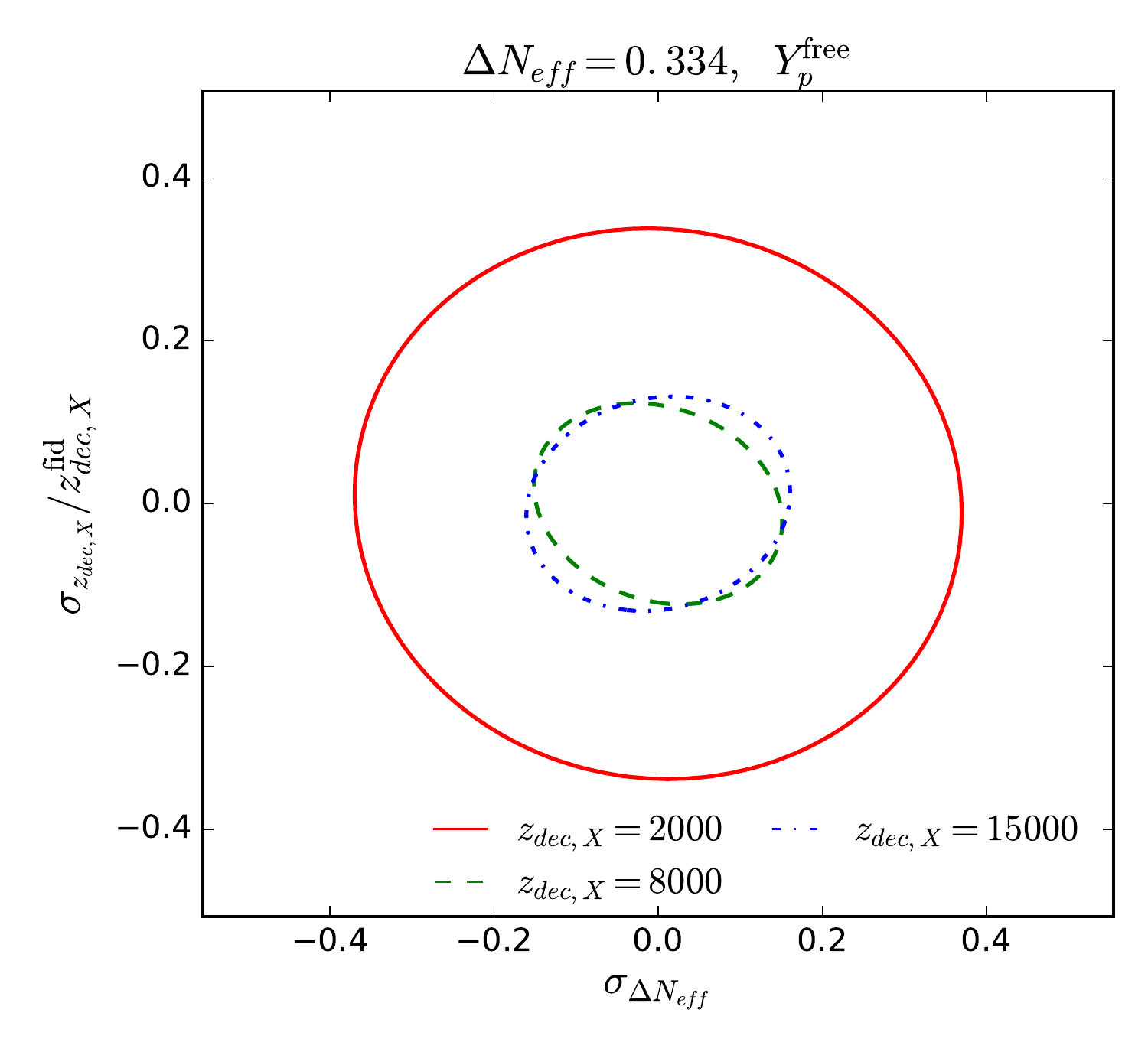}
\caption{Two-dimensional joint constraints on $\Delta N_{eff}$ and $z_{dec,X}$
for $\Delta N_{eff}=0.334$. The left and right panels display
the contours of 95\% C.L. for fixing $Y_p^{\rm BBN}$ to the BBN prediction
and varying $Y_{p}^{\rm free}$ as a free parameter, respectively. The red
solid, green dashed, and blue dot-dashed lines show $z_{dec,X}=2000$, 8000,
and 15000, respectively.}
\label{fig:contour_Ypf}
\end{figure}

Figure \ref{fig:contour_Ypf} shows the two dimensional joint constraints (95\% C.L.)
on $\Delta N_{eff}$ and $z_{dec,X}$ for fiducial $\Delta N_{eff}=0.334$. The left and right panels correspond to fixing the
value of $Y_p^{\rm BBN}$ for consistency with BBN and allowing $Y_p^{\rm free}$
to be an independent parameter, respectively. The red solid, green dashed, and
blue dot-dashed lines display $z_{dec,X}=2000$, 8000, and 15000, respectively.
In all cases we find a relatively weak correlation between $\Delta N_{eff}$ and
$z_{dec,X}$. The weak correlation explains the small difference on the $\Delta N_{eff}$
constraints between fixing and marginalizing over $z_{dec,X}$.

For a CMB-S4 experiment, we find that for $\Delta N_{eff}=0.334$, which is consistent
with the current constraint from the Planck using TT+lowP+BAO at the 68\% confidence level \cite{Ade:2015xua},
the forecasted constraint on $z_{dec,X}$ is $\sigma_{z_{dec,X}} \approx (0.05-0.1)z_{dec,X}$
for $z_{dec,X}\gtrsim 5000$. The constraint degrades significantly for smaller $z_{dec,X}$,
or decoupling during eras when matter dominates the energy budget of the universe.
For a fixed fiducial value of $z_{dec,X}$, the forecasted constraints on $z_{dec,X}$
weaken as $\Delta N_{eff}$ becomes smaller.

The constraints on $\Delta N_{eff}$ and $z_{dec,X}$ will provide constraints on $N$naturalness model parameters. In the implementation we have considered, the value of the reheaton mass $m_\phi$ determines what particles are produced in other sectors and therefore the total value of $\Delta N_{eff}$ and the reheating temperature in the extra sectors. The fine-tuning parameter $r$, sets the spacing between particle masses in our sector and other sectors, and therefore the binding energies and the recombination temperature in other sectors. The combination of $T_{dec,i}$ (set by $r$) and $T_{i}(a_{RH})$ (set by $r$ and $m_\phi$) determine $z_{dec,i}$. From Eq.~(\ref{eq:compacta_dec}), the uncertainty of $r$
can be expressed in terms of that of $z_{dec}$ as
\begin{equation}
 \sigma_{r}=\frac{2r(2i+r)}{3i}\frac{\sigma_{z_{dec,i}}}{z_{dec,i}+1}\,.
\label{eq:constraint_map}
\end{equation}
From Figure \ref{fig:sigmaz}, we see that CMB data are best at constraining decoupling
times near $z_{dec,X} \sim 10000$. Eq.~(\ref{eq:Tdec}) relates the decoupling temperature
in the $i^{\rm th}$ sector to $r$. For the concrete example that we studied in Section \ref{sec:exzreci},
we forecast that $z_{dec,i}$ can be determined to roughly 10\% for the additional sectors
of $i=1, 2$ and $3$, which is converted into $\sigma_{r}\simeq0.07-0.08$ for $(m_{\phi},r)=(10~{\rm GeV},0.5)$
due to Eq.~(\ref{eq:constraint_map}). 

While we focus exclusively on the constraining power of CMB, the
BAO of large-scale structure data provides additional information in two
specific aspects. First, BAO is sensitive to $\omega_{b}$ and $\omega_{c}$, hence
helps break their degeneracies with $N_{eff}$ and improve the constraint on $\Delta N_{eff}$.
However, due to the weak correlation between $\Delta N_{eff}$ and $z_{dec,X}$, the
improvement on the constraint on $z_{dec,X}$ is limited. Second, as presented in
\cite{Baumann:2017lmt, Baumann:2017gkg, Baumann:2018qnt}, the phase shift from additional light
relic species is also imprinted on the BAO location measured from the distribution of galaxies.
While these papers focused on the effect from neutrino-like species, we have
shown in this paper that light relics that decouple at intermediate redshift will
produce a different phase shift in the CMB power spectrum compared to neutrino-like
species. The same effect can therefore potentially be detected from the BAO of
the large-scale structure and it will be sensitive to $z_{dec,X}$, probing directly
the properties of the decoupled species. We leave a consistent treatment of
the phase shift in BAO generated by species with intermediate decoupling times for future work.

\section{Conclusions}
\label{sec:conclusion}
In this paper, we have studied the  imprint left on the CMB anisotropy power spectrum by a light degree of freedom $X$ that decouples from non-gravitational interactions in its own sector during the epoch probed by the CMB. In Section \ref{sec:cosmologicalPT} we presented an analytic approach toward understanding the effects of this decoupling on the amplitude of photon perturbations and on the shift in the acoustic peak locations, $\delta\ell=(\theta/\pi)\Delta\ell$, as a function of the decoupling redshift of the new species $z_{dec,X}$. These calculations showed that the amplitude of photon perturbation decreases with increasing $z_{dec,X}$, while the phase shift increases with increasing $z_{dec,X}$ (see Figure \ref{fig:analyticphaseshift}). In Section \ref{sec:numerical} we computed the CMB power spectra with a decoupling dark species using a modified version of the \texttt{CLASS} code. We demonstrated that the amplitude of the CMB power spectra decreases slightly with increasing $z_{dec,X}$ while the phase shift $\delta\ell$ increases in amplitude with increasing $z_{dec,X}$. Both the phase shift and amplitude change acquire a new $\ell$-dependence with a feature near the angular scale $\ell_{dec,X}$ corresponding to the horizon size at $z_{dec,X}$ (see Figures \ref{fig:PWS_zdec} and  \ref{fig:psjoint}).  For values of $z_{dec,X}\sim \mathcal{O}(10^3-10^4)$, there is a peak in the phase shift at $\ell_{dec,X}$ within the observable range of a Stage IV CMB experiment. The changes to the CMB power spectra potentially enable direct constraints on $z_{dec,X}$ and thus constraints on the types and strengths of non-gravitational interactions responsible for decoupling of the species $X$. 

We consider the $N$naturalness scenario as a concrete example of a model that includes dark radiation with decoupling redshift $z_{dec,X}\sim\mathcal{O}(10^{3}-10^{4})$. In Section \ref{sec:Nnaturalness} we showed how to relate $N$naturalness parameters to the observables $\Delta N_{eff}$ and $z_{dec,X}$ and discussed the sensitivity of this mapping to assumptions about the implementation of the $N$naturalness model. The relevant properties of the additional sectors for one example choice of parameters is given in Table \ref{table:zdeci}. For $N$naturalness parameter choices consistent with current data, a Stage IV experiment can potentially determine the photon decoupling redshifts, $z_{dec,X}$, in the first few sectors to $\sim 30\%$, which can be translated into comparable constraints on the fine-tuning parameter $r$.

In Section \ref{sec:forecast}, we forecasted constraints on the set of parameters $\Delta N_{eff}$, $z_{dec,X}$ from a Stage IV CMB experiment. The full results of this forecast are shown in Figures \ref{fig:sigmaz} and \ref{fig:contour_Ypf}.
To summarize, we find that for currently allowed values of $\Delta N_{eff}$ (e.g. $\Delta N_{eff} \lsim 0.334$) a Stage IV CMB experiment could determine $z_{dec,X}$ at the tens-of-percent level. For larger values of $\Delta N_{eff}$ the decoupling redshift can be determined even more precisely. The constraints on $\Delta N_{eff}$ and $z_{dec,X}$ are sensitive to the fiducial value of $z_{dec,X}$.  If the primordial helium abundance is fixed by consistency with BBN, constraints on $\Delta N_{eff}$ vary at the $50\%$ level with $z_{dec,X}$ and we find that smaller $z_{dec,X}$ values result in somewhat tighter constraints. On the other hand, if $Y_p$ is allowed to vary, the constraints on $\Delta N_{eff}$ are much more dependent on the choice of $z_{dec,X}$. We find that the forecasted constraints on $\Delta N_{eff}$ are generally tighter when $\Delta N_{eff}$ is generated by the species with earlier decoupling times, but are strongest for a new species that decouples at $z_{dec,X} \sim 10000$ corresponding to $\ell_{dec,X}\sim 1800$. In both cases adding the parameter $z_{dec,X}$ to the forecast does not degrade the constraints on $\Delta N_{eff}$ so long as $z_{dec,X} \sim 10^3-10^4$. \\

\noindent {\bf Acknowledgments}\\
We are grateful to Joel Meyers, Daniel Green, and Benjamin Wallisch for helpful discussions, code comparisons, and for sharing their modified version of the \texttt{CLASS} code from \cite{Baumann:2015rya}. We thank Patrick Meade and Eiichiro Komatsu for helpful discussions and comments on this draft. GC also thanks David Pinner, Leonardo Senatore, Robert Shrock, Suzanne Staggs and Raffaele Tito D$^\prime$Agnolo for fruitful discussions. Results in this paper were obtained using the high-performance computing system at the Institute for Advanced Computational Science at Stony Brook University.
GC, CC and ML are supported by grant NSF PHY-1620628. ML is also supported by DOE DE-SC0017848.\\

\bibliographystyle{JHEP}
\bibliography{main}

\providecommand{\href}[2]{#2}\begingroup\raggedright\begin{thebibliography}{10}

\bibitem{Ade:2015xua}
{\scshape Planck} collaboration, P.~A.~R. Ade et~al., \emph{{Planck 2015
  results. XIII. Cosmological parameters}},
  \href{https://doi.org/10.1051/0004-6361/201525830}{\emph{Astron. Astrophys.}
  {\bfseries 594} (2016) A13},
  [\href{https://arxiv.org/abs/1502.01589}{{\ttfamily 1502.01589}}].

\bibitem{Mangano:2001iu}
G.~Mangano, G.~Miele, S.~Pastor and M.~Peloso, \emph{{A Precision calculation
  of the effective number of cosmological neutrinos}},
  \href{https://doi.org/10.1016/S0370-2693(02)01622-2}{\emph{Phys. Lett.}
  {\bfseries B534} (2002) 8--16},
  [\href{https://arxiv.org/abs/astro-ph/0111408}{{\ttfamily
  astro-ph/0111408}}].

\bibitem{Mangano:2005cc}
G.~Mangano, G.~Miele, S.~Pastor, T.~Pinto, O.~Pisanti and P.~D. Serpico,
  \emph{{Relic neutrino decoupling including flavor oscillations}},
  \href{https://doi.org/10.1016/j.nuclphysb.2005.09.041}{\emph{Nucl. Phys.}
  {\bfseries B729} (2005) 221--234},
  [\href{https://arxiv.org/abs/hep-ph/0506164}{{\ttfamily hep-ph/0506164}}].

\bibitem{Gnedin:1997vn}
N.~Y. Gnedin and O.~Y. Gnedin, \emph{{Cosmological neutrino background
  revisited}}, \href{https://doi.org/10.1086/306469}{\emph{Astrophys. J.}
  {\bfseries 509} (1998) 11--15},
  [\href{https://arxiv.org/abs/astro-ph/9712199}{{\ttfamily
  astro-ph/9712199}}].

\bibitem{Hannestad:1995rs}
S.~Hannestad and J.~Madsen, \emph{{Neutrino decoupling in the early universe}},
  \href{https://doi.org/10.1103/PhysRevD.52.1764}{\emph{Phys. Rev.} {\bfseries
  D52} (1995) 1764--1769},
  [\href{https://arxiv.org/abs/astro-ph/9506015}{{\ttfamily
  astro-ph/9506015}}].

\bibitem{Heckler:1994tv}
A.~F. Heckler, \emph{{Astrophysical applications of quantum corrections to the
  equation of state of a plasma}},
  \href{https://doi.org/10.1103/PhysRevD.49.611}{\emph{Phys. Rev.} {\bfseries
  D49} (1994) 611--617}.

\bibitem{Dolgov:1992qg}
A.~D. Dolgov and M.~Fukugita, \emph{{Nonequilibrium effect of the neutrino
  distribution on primordial helium synthesis}},
  \href{https://doi.org/10.1103/PhysRevD.46.5378}{\emph{Phys. Rev.} {\bfseries
  D46} (1992) 5378--5382}.

\bibitem{Dolgov:2002wy}
A.~D. Dolgov, \emph{{Neutrinos in cosmology}},
  \href{https://doi.org/10.1016/S0370-1573(02)00139-4}{\emph{Phys. Rept.}
  {\bfseries 370} (2002) 333--535},
  [\href{https://arxiv.org/abs/hep-ph/0202122}{{\ttfamily hep-ph/0202122}}].

\bibitem{Dolgov:1997mb}
A.~D. Dolgov, S.~H. Hansen and D.~V. Semikoz, \emph{{Nonequilibrium corrections
  to the spectra of massless neutrinos in the early universe}},
  \href{https://doi.org/10.1016/S0550-3213(97)00479-3}{\emph{Nucl. Phys.}
  {\bfseries B503} (1997) 426--444},
  [\href{https://arxiv.org/abs/hep-ph/9703315}{{\ttfamily hep-ph/9703315}}].

\bibitem{Dodelson:1992km}
S.~Dodelson and M.~S. Turner, \emph{{Nonequilibrium neutrino statistical
  mechanics in the expanding universe}},
  \href{https://doi.org/10.1103/PhysRevD.46.3372}{\emph{Phys. Rev.} {\bfseries
  D46} (1992) 3372--3387}.

\bibitem{Rana:1991xk}
N.~C. Rana and B.~M. Seifert, \emph{{Effect of neutrino heating in the early
  universe on neutrino decoupling temperatures and nucleosynthesis}},
  \href{https://doi.org/10.1103/PhysRevD.44.393}{\emph{Phys. Rev.} {\bfseries
  D44} (1991) 393--397}.

\bibitem{Esposito:2000hi}
S.~Esposito, G.~Miele, S.~Pastor, M.~Peloso and O.~Pisanti,
  \emph{{Nonequilibrium spectra of degenerate relic neutrinos}},
  \href{https://doi.org/10.1016/S0550-3213(00)00554-X}{\emph{Nucl. Phys.}
  {\bfseries B590} (2000) 539--561},
  [\href{https://arxiv.org/abs/astro-ph/0005573}{{\ttfamily
  astro-ph/0005573}}].

\bibitem{deSalas:2016ztq}
P.~F. de~Salas and S.~Pastor, \emph{{Relic neutrino decoupling with flavour
  oscillations revisited}},
  \href{https://doi.org/10.1088/1475-7516/2016/07/051}{\emph{JCAP} {\bfseries
  1607} (2016) 051}, [\href{https://arxiv.org/abs/1606.06986}{{\ttfamily
  1606.06986}}].

\bibitem{Abazajian:2016yjj}
{\scshape CMB-S4} collaboration, K.~N. Abazajian et~al., \emph{{CMB-S4 Science
  Book, First Edition}},  \href{https://arxiv.org/abs/1610.02743}{{\ttfamily
  1610.02743}}.

\bibitem{2011PhLB..701..296M}
G.~{Mangano} and P.~D. {Serpico}, \emph{{A robust upper limit on N$_{}$ from
  BBN, circa 2011}},
  \href{https://doi.org/10.1016/j.physletb.2011.05.075}{\emph{Physics Letters
  B} {\bfseries 701} (July, 2011) 296--299},
  [\href{https://arxiv.org/abs/1103.1261}{{\ttfamily 1103.1261}}].

\bibitem{Benson:2014qhw}
{\scshape SPT-3G} collaboration, B.~A. Benson et~al., \emph{{SPT-3G: A
  Next-Generation Cosmic Microwave Background Polarization Experiment on the
  South Pole Telescope}}, \href{https://doi.org/10.1117/12.2057305}{\emph{Proc.
  SPIE Int. Soc. Opt. Eng.} {\bfseries 9153} (2014) 91531P},
  [\href{https://arxiv.org/abs/1407.2973}{{\ttfamily 1407.2973}}].

\bibitem{Henderson:2015nzj}
S.~W. Henderson et~al., \emph{{Advanced ACTPol Cryogenic Detector Arrays and
  Readout}}, \href{https://doi.org/10.1007/s10909-016-1575-z}{\emph{J. Low.
  Temp. Phys.} {\bfseries 184} (2016) 772--779},
  [\href{https://arxiv.org/abs/1510.02809}{{\ttfamily 1510.02809}}].

\bibitem{Jungman:1995bz}
G.~Jungman, M.~Kamionkowski, A.~Kosowsky and D.~N. Spergel, \emph{{Cosmological
  parameter determination with microwave background maps}},
  \href{https://doi.org/10.1103/PhysRevD.54.1332}{\emph{Phys. Rev.} {\bfseries
  D54} (1996) 1332--1344},
  [\href{https://arxiv.org/abs/astro-ph/9512139}{{\ttfamily
  astro-ph/9512139}}].

\bibitem{Kojima:2009gw}
K.~Kojima, T.~Kajino and G.~J. Mathews, \emph{{Generation of Curvature
  Perturbations with Extra Anisotropic Stress}},
  \href{https://doi.org/10.1088/1475-7516/2010/02/018}{\emph{JCAP} {\bfseries
  1002} (2010) 018}, [\href{https://arxiv.org/abs/0910.1976}{{\ttfamily
  0910.1976}}].

\bibitem{Cadamuro:2010cz}
D.~Cadamuro, S.~Hannestad, G.~Raffelt and J.~Redondo, \emph{{Cosmological
  bounds on sub-MeV mass axions}},
  \href{https://doi.org/10.1088/1475-7516/2011/02/003}{\emph{JCAP} {\bfseries
  1102} (2011) 003}, [\href{https://arxiv.org/abs/1011.3694}{{\ttfamily
  1011.3694}}].

\bibitem{Menestrina:2011mz}
J.~L. Menestrina and R.~J. Scherrer, \emph{{Dark Radiation from Particle Decays
  during Big Bang Nucleosynthesis}},
  \href{https://doi.org/10.1103/PhysRevD.85.047301}{\emph{Phys. Rev.}
  {\bfseries D85} (2012) 047301},
  [\href{https://arxiv.org/abs/1111.0605}{{\ttfamily 1111.0605}}].

\bibitem{Boehm:2012gr}
C.~Boehm, M.~J. Dolan and C.~McCabe, \emph{{Increasing Neff with particles in
  thermal equilibrium with neutrinos}},
  \href{https://doi.org/10.1088/1475-7516/2012/12/027}{\emph{JCAP} {\bfseries
  1212} (2012) 027}, [\href{https://arxiv.org/abs/1207.0497}{{\ttfamily
  1207.0497}}].

\bibitem{Brust:2013xpv}
C.~Brust, D.~E. Kaplan and M.~T. Walters, \emph{{New Light Species and the
  CMB}}, \href{https://doi.org/10.1007/JHEP12(2013)058}{\emph{JHEP} {\bfseries
  12} (2013) 058}, [\href{https://arxiv.org/abs/1303.5379}{{\ttfamily
  1303.5379}}].

\bibitem{Weinberg:2013kea}
S.~Weinberg, \emph{{Goldstone Bosons as Fractional Cosmic Neutrinos}},
  \href{https://doi.org/10.1103/PhysRevLett.110.241301}{\emph{Phys. Rev. Lett.}
  {\bfseries 110} (2013) 241301},
  [\href{https://arxiv.org/abs/1305.1971}{{\ttfamily 1305.1971}}].

\bibitem{Cyr-Racine:2013fsa}
F.-Y. Cyr-Racine, R.~de~Putter, A.~Raccanelli and K.~Sigurdson,
  \emph{{Constraints on Large-Scale Dark Acoustic Oscillations from
  Cosmology}}, \href{https://doi.org/10.1103/PhysRevD.89.063517}{\emph{Phys.
  Rev.} {\bfseries D89} (2014) 063517},
  [\href{https://arxiv.org/abs/1310.3278}{{\ttfamily 1310.3278}}].

\bibitem{Vogel:2013raa}
H.~Vogel and J.~Redondo, \emph{{Dark Radiation constraints on minicharged
  particles in models with a hidden photon}},
  \href{https://doi.org/10.1088/1475-7516/2014/02/029}{\emph{JCAP} {\bfseries
  1402} (2014) 029}, [\href{https://arxiv.org/abs/1311.2600}{{\ttfamily
  1311.2600}}].

\bibitem{Millea:2015qra}
M.~Millea, L.~Knox and B.~Fields, \emph{{New Bounds for Axions and Axion-Like
  Particles with keV-GeV Masses}},
  \href{https://doi.org/10.1103/PhysRevD.92.023010}{\emph{Phys. Rev.}
  {\bfseries D92} (2015) 023010},
  [\href{https://arxiv.org/abs/1501.04097}{{\ttfamily 1501.04097}}].

\bibitem{Chacko:2015noa}
Z.~Chacko, Y.~Cui, S.~Hong and T.~Okui, \emph{{Hidden dark matter sector, dark
  radiation, and the CMB}},
  \href{https://doi.org/10.1103/PhysRevD.92.055033}{\emph{Phys. Rev.}
  {\bfseries D92} (2015) 055033},
  [\href{https://arxiv.org/abs/1505.04192}{{\ttfamily 1505.04192}}].

\bibitem{Lancaster:2017ksf}
L.~Lancaster, F.-Y. Cyr-Racine, L.~Knox and Z.~Pan, \emph{{A tale of two modes:
  Neutrino free-streaming in the early universe}},
  \href{https://doi.org/10.1088/1475-7516/2017/07/033}{\emph{JCAP} {\bfseries
  1707} (2017) 033}, [\href{https://arxiv.org/abs/1704.06657}{{\ttfamily
  1704.06657}}].

\bibitem{Arkani-Hamed:2016rle}
N.~Arkani-Hamed, T.~Cohen, R.~T. D'Agnolo, A.~Hook, H.~D. Kim and D.~Pinner,
  \emph{{Solving the Hierarchy Problem at Reheating with a Large Number of
  Degrees of Freedom}},
  \href{https://doi.org/10.1103/PhysRevLett.117.251801}{\emph{Phys. Rev. Lett.}
  {\bfseries 117} (2016) 251801},
  [\href{https://arxiv.org/abs/1607.06821}{{\ttfamily 1607.06821}}].

\bibitem{Buen-Abad:2015ova}
M.~A. Buen-Abad, G.~Marques-Tavares and M.~Schmaltz, \emph{{Non-Abelian dark
  matter and dark radiation}},
  \href{https://doi.org/10.1103/PhysRevD.92.023531}{\emph{Phys. Rev.}
  {\bfseries D92} (2015) 023531},
  [\href{https://arxiv.org/abs/1505.03542}{{\ttfamily 1505.03542}}].

\bibitem{Salvio:2013iaa}
A.~Salvio, A.~Strumia and W.~Xue, \emph{{Thermal axion production}},
  \href{https://doi.org/10.1088/1475-7516/2014/01/011}{\emph{JCAP} {\bfseries
  1401} (2014) 011}, [\href{https://arxiv.org/abs/1310.6982}{{\ttfamily
  1310.6982}}].

\bibitem{Kawasaki:2015ofa}
M.~Kawasaki, M.~Yamada and T.~T. Yanagida, \emph{{Observable dark radiation
  from a cosmologically safe QCD axion}},
  \href{https://doi.org/10.1103/PhysRevD.91.125018}{\emph{Phys. Rev.}
  {\bfseries D91} (2015) 125018},
  [\href{https://arxiv.org/abs/1504.04126}{{\ttfamily 1504.04126}}].

\bibitem{Baumann:2016wac}
D.~Baumann, D.~Green and B.~Wallisch, \emph{{New Target for Cosmic Axion
  Searches}}, \href{https://doi.org/10.1103/PhysRevLett.117.171301}{\emph{Phys.
  Rev. Lett.} {\bfseries 117} (2016) 171301},
  [\href{https://arxiv.org/abs/1604.08614}{{\ttfamily 1604.08614}}].

\bibitem{Abazajian:2001nj}
K.~Abazajian, G.~M. Fuller and M.~Patel, \emph{{Sterile neutrino hot, warm, and
  cold dark matter}},
  \href{https://doi.org/10.1103/PhysRevD.64.023501}{\emph{Phys. Rev.}
  {\bfseries D64} (2001) 023501},
  [\href{https://arxiv.org/abs/astro-ph/0101524}{{\ttfamily
  astro-ph/0101524}}].

\bibitem{Strumia:2006db}
A.~Strumia and F.~Vissani, \emph{{Neutrino masses and mixings and...}},
  \href{https://arxiv.org/abs/hep-ph/0606054}{{\ttfamily hep-ph/0606054}}.

\bibitem{Boyarsky:2009ix}
A.~Boyarsky, O.~Ruchayskiy and M.~Shaposhnikov, \emph{{The Role of sterile
  neutrinos in cosmology and astrophysics}},
  \href{https://doi.org/10.1146/annurev.nucl.010909.083654}{\emph{Ann. Rev.
  Nucl. Part. Sci.} {\bfseries 59} (2009) 191--214},
  [\href{https://arxiv.org/abs/0901.0011}{{\ttfamily 0901.0011}}].

\bibitem{Boyle:2007zx}
L.~A. Boyle and A.~Buonanno, \emph{{Relating gravitational wave constraints
  from primordial nucleosynthesis, pulsar timing, laser interferometers, and
  the CMB: Implications for the early Universe}},
  \href{https://doi.org/10.1103/PhysRevD.78.043531}{\emph{Phys. Rev.}
  {\bfseries D78} (2008) 043531},
  [\href{https://arxiv.org/abs/0708.2279}{{\ttfamily 0708.2279}}].

\bibitem{Stewart:2007fu}
A.~Stewart and R.~Brandenberger, \emph{{Observational Constraints on Theories
  with a Blue Spectrum of Tensor Modes}},
  \href{https://doi.org/10.1088/1475-7516/2008/08/012}{\emph{JCAP} {\bfseries
  0808} (2008) 012}, [\href{https://arxiv.org/abs/0711.4602}{{\ttfamily
  0711.4602}}].

\bibitem{Meerburg:2015zua}
P.~D. Meerburg, R.~Hlo??ek, B.~Hadzhiyska and J.~Meyers, \emph{{Multiwavelength
  constraints on the inflationary consistency relation}},
  \href{https://doi.org/10.1103/PhysRevD.91.103505}{\emph{Phys. Rev.}
  {\bfseries D91} (2015) 103505},
  [\href{https://arxiv.org/abs/1502.00302}{{\ttfamily 1502.00302}}].

\bibitem{Kaplan:2011yj}
D.~E. Kaplan, G.~Z. Krnjaic, K.~R. Rehermann and C.~M. Wells, \emph{{Dark
  Atoms: Asymmetry and Direct Detection}},
  \href{https://doi.org/10.1088/1475-7516/2011/10/011}{\emph{JCAP} {\bfseries
  1110} (2011) 011}, [\href{https://arxiv.org/abs/1105.2073}{{\ttfamily
  1105.2073}}].

\bibitem{Ackerman:mha}
L.~Ackerman, M.~R. Buckley, S.~M. Carroll and M.~Kamionkowski, \emph{{Dark
  Matter and Dark Radiation}},
  \href{https://doi.org/10.1103/PhysRevD.79.023519,
  10.1142/9789814293792_0021}{\emph{Phys. Rev.} {\bfseries D79} (2009) 023519},
  [\href{https://arxiv.org/abs/0810.5126}{{\ttfamily 0810.5126}}].

\bibitem{CyrRacine:2012fz}
F.-Y. Cyr-Racine and K.~Sigurdson, \emph{{Cosmology of atomic dark matter}},
  \href{https://doi.org/10.1103/PhysRevD.87.103515}{\emph{Phys. Rev.}
  {\bfseries D87} (2013) 103515},
  [\href{https://arxiv.org/abs/1209.5752}{{\ttfamily 1209.5752}}].

\bibitem{Steigman:2013yua}
G.~Steigman, \emph{{Equivalent Neutrinos, Light WIMPs, and the Chimera of Dark
  Radiation}}, \href{https://doi.org/10.1103/PhysRevD.87.103517}{\emph{Phys.
  Rev.} {\bfseries D87} (2013) 103517},
  [\href{https://arxiv.org/abs/1303.0049}{{\ttfamily 1303.0049}}].

\bibitem{Boehm:2013jpa}
C.~Boehm, M.~J. Dolan and C.~McCabe, \emph{{A Lower Bound on the Mass of Cold
  Thermal Dark Matter from Planck}},
  \href{https://doi.org/10.1088/1475-7516/2013/08/041}{\emph{JCAP} {\bfseries
  1308} (2013) 041}, [\href{https://arxiv.org/abs/1303.6270}{{\ttfamily
  1303.6270}}].

\bibitem{Zaldarriaga:1995gi}
M.~Zaldarriaga and D.~D. Harari, \emph{{Analytic approach to the polarization
  of the cosmic microwave background in flat and open universes}},
  \href{https://doi.org/10.1103/PhysRevD.52.3276}{\emph{Phys. Rev.} {\bfseries
  D52} (1995) 3276--3287},
  [\href{https://arxiv.org/abs/astro-ph/9504085}{{\ttfamily
  astro-ph/9504085}}].

\bibitem{Bashinsky:2003tk}
S.~Bashinsky and U.~Seljak, \emph{{Neutrino perturbations in CMB anisotropy and
  matter clustering}},
  \href{https://doi.org/10.1103/PhysRevD.69.083002}{\emph{Phys. Rev.}
  {\bfseries D69} (2004) 083002},
  [\href{https://arxiv.org/abs/astro-ph/0310198}{{\ttfamily
  astro-ph/0310198}}].

\bibitem{Baumann:2015rya}
D.~Baumann, D.~Green, J.~Meyers and B.~Wallisch, \emph{{Phases of New Physics
  in the CMB}},
  \href{https://doi.org/10.1088/1475-7516/2016/01/007}{\emph{JCAP} {\bfseries
  1601} (2016) 007}, [\href{https://arxiv.org/abs/1508.06342}{{\ttfamily
  1508.06342}}].

\bibitem{Follin:2015hya}
B.~Follin, L.~Knox, M.~Millea and Z.~Pan, \emph{{First Detection of the
  Acoustic Oscillation Phase Shift Expected from the Cosmic Neutrino
  Background}},
  \href{https://doi.org/10.1103/PhysRevLett.115.091301}{\emph{Phys. Rev. Lett.}
  {\bfseries 115} (2015) 091301},
  [\href{https://arxiv.org/abs/1503.07863}{{\ttfamily 1503.07863}}].

\bibitem{Baumann:2017lmt}
D.~Baumann, D.~Green and M.~Zaldarriaga, \emph{{Phases of New Physics in the
  BAO Spectrum}},
  \href{https://doi.org/10.1088/1475-7516/2017/11/007}{\emph{JCAP} {\bfseries
  1711} (2017) 007}, [\href{https://arxiv.org/abs/1703.00894}{{\ttfamily
  1703.00894}}].

\bibitem{Baumann:2017gkg}
D.~Baumann, D.~Green and B.~Wallisch, \emph{{Searching for Light Relics with
  Large-Scale Structure}},  \href{https://arxiv.org/abs/1712.08067}{{\ttfamily
  1712.08067}}.

\bibitem{Baumann:2018qnt}
D.~Baumann, F.~Beutler, R.~Flauger, D.~Green, M.~Vargas-Maga?a, A.~Slosar
  et~al., \emph{{First Measurement of Neutrinos in the BAO Spectrum}},
  \href{https://arxiv.org/abs/1803.10741}{{\ttfamily 1803.10741}}.

\bibitem{Archidiacono:2017slj}
M.~Archidiacono, S.~Bohr, S.~Hannestad, J.~H. Jørgensen and J.~Lesgourgues,
  \emph{{Linear scale bounds on dark matter--dark radiation interactions and
  connection with the small scale crisis of cold dark matter}},
  \href{https://doi.org/10.1088/1475-7516/2017/11/010}{\emph{JCAP} {\bfseries
  1711} (2017) 010}, [\href{https://arxiv.org/abs/1706.06870}{{\ttfamily
  1706.06870}}].

\bibitem{Cyr-Racine:2013jua}
F.-Y. Cyr-Racine and K.~Sigurdson, \emph{{Limits on Neutrino-Neutrino
  Scattering in the Early Universe}},
  \href{https://doi.org/10.1103/PhysRevD.90.123533}{\emph{Phys. Rev.}
  {\bfseries D90} (2014) 123533},
  [\href{https://arxiv.org/abs/1306.1536}{{\ttfamily 1306.1536}}].

\bibitem{Cui:2018imi}
Y.~Cui and R.~Huo, \emph{{Visualizing Invisible Dark Matter Annihilation with
  the CMB and Matter Power Spectrum}},
  \href{https://arxiv.org/abs/1805.06451}{{\ttfamily 1805.06451}}.

\bibitem{Blas:2011rf}
D.~Blas, J.~Lesgourgues and T.~Tram, \emph{{The Cosmic Linear Anisotropy
  Solving System (CLASS) II: Approximation schemes}},
  \href{https://doi.org/10.1088/1475-7516/2011/07/034}{\emph{JCAP} {\bfseries
  1107} (2011) 034}, [\href{https://arxiv.org/abs/1104.2933}{{\ttfamily
  1104.2933}}].

\bibitem{Ma:1995ey}
C.-P. Ma and E.~Bertschinger, \emph{{Cosmological perturbation theory in the
  synchronous and conformal Newtonian gauges}},
  \href{https://doi.org/10.1086/176550}{\emph{Astrophys. J.} {\bfseries 455}
  (1995) 7--25}, [\href{https://arxiv.org/abs/astro-ph/9506072}{{\ttfamily
  astro-ph/9506072}}].

\bibitem{0067-0049-148-1-233}
L.~Page, M.~R. Nolta, C.~Barnes, C.~L. Bennett, M.~Halpern, G.~Hinshaw et~al.,
  \emph{First-year wilkinson microwave anisotropy probe (wmap) observations:
  Interpretation of the tt and te angular power spectrum peaks}, {\emph{The
  Astrophysical Journal Supplement Series} {\bfseries 148} (2003) 233}.

\bibitem{Lewis:1999bs}
A.~Lewis, A.~Challinor and A.~Lasenby, \emph{{Efficient computation of CMB
  anisotropies in closed FRW models}},
  \href{https://doi.org/10.1086/309179}{\emph{Astrophys. J.} {\bfseries 538}
  (2000) 473--476}, [\href{https://arxiv.org/abs/astro-ph/9911177}{{\ttfamily
  astro-ph/9911177}}].

\bibitem{Howlett:2012mh}
C.~Howlett, A.~Lewis, A.~Hall and A.~Challinor, \emph{{CMB power spectrum
  parameter degeneracies in the era of precision cosmology}},
  \href{https://doi.org/10.1088/1475-7516/2012/04/027}{\emph{JCAP} {\bfseries
  1204} (2012) 027}, [\href{https://arxiv.org/abs/1201.3654}{{\ttfamily
  1201.3654}}].

\bibitem{Lesgourgues:2011rh}
J.~Lesgourgues and T.~Tram, \emph{{The Cosmic Linear Anisotropy Solving System
  (CLASS) IV: efficient implementation of non-cold relics}},
  \href{https://doi.org/10.1088/1475-7516/2011/09/032}{\emph{JCAP} {\bfseries
  1109} (2011) 032}, [\href{https://arxiv.org/abs/1104.2935}{{\ttfamily
  1104.2935}}].

\bibitem{Hou:2011ec}
Z.~Hou, R.~Keisler, L.~Knox, M.~Millea and C.~Reichardt, \emph{{How Massless
  Neutrinos Affect the Cosmic Microwave Background Damping Tail}},
  \href{https://doi.org/10.1103/PhysRevD.87.083008}{\emph{Phys. Rev.}
  {\bfseries D87} (2013) 083008},
  [\href{https://arxiv.org/abs/1104.2333}{{\ttfamily 1104.2333}}].

\bibitem{Seljak:1995ve}
U.~Seljak, \emph{{Gravitational lensing effect on cosmic microwave background
  anisotropies: A Power spectrum approach}},
  \href{https://doi.org/10.1086/177218}{\emph{Astrophys. J.} {\bfseries 463}
  (1996) 1}, [\href{https://arxiv.org/abs/astro-ph/9505109}{{\ttfamily
  astro-ph/9505109}}].

\bibitem{Chung:1998rq}
D.~J.~H. Chung, E.~W. Kolb and A.~Riotto, \emph{{Production of massive
  particles during reheating}},
  \href{https://doi.org/10.1103/PhysRevD.60.063504}{\emph{Phys. Rev.}
  {\bfseries D60} (1999) 063504},
  [\href{https://arxiv.org/abs/hep-ph/9809453}{{\ttfamily hep-ph/9809453}}].

\bibitem{Dodelson:1282338}
S.~Dodelson, \emph{{Modern cosmology}}.
\newblock Academic Press, San Diego, CA, 2003.

\bibitem{1985PhLB..155...36K}
V.~A. {Kuzmin}, V.~A. {Rubakov} and M.~E. {Shaposhnikov}, \emph{{On anomalous
  electroweak baryon-number non-conservation in the early universe}},
  \href{https://doi.org/10.1016/0370-2693(85)91028-7}{\emph{Physics Letters B}
  {\bfseries 155} (May, 1985) 36--42}.

\bibitem{Peebles:1968ja}
P.~J.~E. Peebles, \emph{{Recombination of the Primeval Plasma}},
  \href{https://doi.org/10.1086/149628}{\emph{Astrophys. J.} {\bfseries 153}
  (1968) 1}.

\bibitem{Zeldovich:1969en}
{\relax Ya}.~B. Zeldovich, V.~G. Kurt and R.~A. Sunyaev, \emph{{Recombination
  of hydrogen in the hot model of the universe}}, {\emph{Sov. Phys. JETP}
  {\bfseries 28} (1969) 146}.

\bibitem{Grin:2009ik}
D.~Grin and C.~M. Hirata, \emph{{Cosmological hydrogen recombination: The
  effect of extremely high-n states}},
  \href{https://doi.org/10.1103/PhysRevD.81.083005}{\emph{Phys. Rev.}
  {\bfseries D81} (2010) 083005},
  [\href{https://arxiv.org/abs/0911.1359}{{\ttfamily 0911.1359}}].

\bibitem{AliHaimoud:2010ym}
Y.~Ali-Haimoud, D.~Grin and C.~M. Hirata, \emph{{Radiative transfer effects in
  primordial hydrogen recombination}},
  \href{https://doi.org/10.1103/PhysRevD.82.123502}{\emph{Phys. Rev.}
  {\bfseries D82} (2010) 123502},
  [\href{https://arxiv.org/abs/1009.4697}{{\ttfamily 1009.4697}}].

\bibitem{Shandera:2018xkn}
S.~Shandera, D.~Jeong and H.~S.~G. Gebhardt, \emph{{Gravitational Waves from
  Binary Mergers of Sub-solar Mass Dark Black Holes}},
  \href{https://arxiv.org/abs/1802.08206}{{\ttfamily 1802.08206}}.

\bibitem{Pritchard:2004qp}
J.~R. Pritchard and M.~Kamionkowski, \emph{{Cosmic microwave background
  fluctuations from gravitational waves: An Analytic approach}},
  \href{https://doi.org/10.1016/j.aop.2005.03.005}{\emph{Annals Phys.}
  {\bfseries 318} (2005) 2--36},
  [\href{https://arxiv.org/abs/astro-ph/0412581}{{\ttfamily
  astro-ph/0412581}}].

\bibitem{Xia:2008gm}
T.~Y. Xia and Y.~Zhang, \emph{{Analytic Spectra of CMB Anisotropies and
  Polarization Generated by Relic Gravitational Waves with Modification due to
  Neutrino Free-Streaming}},
  \href{https://doi.org/10.1103/PhysRevD.78.123005}{\emph{Phys. Rev.}
  {\bfseries D78} (2008) 123005},
  [\href{https://arxiv.org/abs/0811.4008}{{\ttfamily 0811.4008}}].

\bibitem{Boyle:2005se}
L.~A. Boyle and P.~J. Steinhardt, \emph{{Probing the early universe with
  inflationary gravitational waves}},
  \href{https://doi.org/10.1103/PhysRevD.77.063504}{\emph{Phys. Rev.}
  {\bfseries D77} (2008) 063504},
  [\href{https://arxiv.org/abs/astro-ph/0512014}{{\ttfamily
  astro-ph/0512014}}].

\bibitem{Weinberg:2003ur}
S.~Weinberg, \emph{{Damping of tensor modes in cosmology}},
  \href{https://doi.org/10.1103/PhysRevD.69.023503}{\emph{Phys. Rev.}
  {\bfseries D69} (2004) 023503},
  [\href{https://arxiv.org/abs/astro-ph/0306304}{{\ttfamily
  astro-ph/0306304}}].

\bibitem{Dicus:2005rh}
D.~A. Dicus and W.~W. Repko, \emph{{Comment on damping of tensor modes in
  cosmology}}, \href{https://doi.org/10.1103/PhysRevD.72.088302}{\emph{Phys.
  Rev.} {\bfseries D72} (2005) 088302},
  [\href{https://arxiv.org/abs/astro-ph/0509096}{{\ttfamily
  astro-ph/0509096}}].

\bibitem{Watanabe:2006qe}
Y.~Watanabe and E.~Komatsu, \emph{{Improved Calculation of the Primordial
  Gravitational Wave Spectrum in the Standard Model}},
  \href{https://doi.org/10.1103/PhysRevD.73.123515}{\emph{Phys. Rev.}
  {\bfseries D73} (2006) 123515},
  [\href{https://arxiv.org/abs/astro-ph/0604176}{{\ttfamily
  astro-ph/0604176}}].

\bibitem{Miao:2007cw}
H.~X. Miao and Y.~Zhang, \emph{{Analytic spectrum of relic gravitational waves
  modified by neutrino free streaming and dark energy}},
  \href{https://doi.org/10.1103/PhysRevD.75.104009}{\emph{Phys. Rev.}
  {\bfseries D75} (2007) 104009},
  [\href{https://arxiv.org/abs/astro-ph/0703602}{{\ttfamily
  astro-ph/0703602}}].

\bibitem{Zhao:2009we}
W.~Zhao, Y.~Zhang and T.~Xia, \emph{{New method to constrain the relativistic
  free-streaming gas in the Universe}},
  \href{https://doi.org/10.1016/j.physletb.2009.05.046}{\emph{Phys. Lett.}
  {\bfseries B677} (2009) 235--238},
  [\href{https://arxiv.org/abs/0905.3223}{{\ttfamily 0905.3223}}].

\bibitem{Tegmark:1996bz}
M.~Tegmark, A.~Taylor and A.~Heavens, \emph{{Karhunen-Loeve eigenvalue problems
  in cosmology: How should we tackle large data sets?}},
  \href{https://doi.org/10.1086/303939}{\emph{Astrophys. J.} {\bfseries 480}
  (1997) 22}, [\href{https://arxiv.org/abs/astro-ph/9603021}{{\ttfamily
  astro-ph/9603021}}].

\bibitem{Hirata:2003ka}
C.~M. Hirata and U.~Seljak, \emph{{Reconstruction of lensing from the cosmic
  microwave background polarization}},
  \href{https://doi.org/10.1103/PhysRevD.68.083002}{\emph{Phys. Rev.}
  {\bfseries D68} (2003) 083002},
  [\href{https://arxiv.org/abs/astro-ph/0306354}{{\ttfamily
  astro-ph/0306354}}].

\bibitem{Smith:2010gu}
K.~M. Smith, D.~Hanson, M.~LoVerde, C.~M. Hirata and O.~Zahn, \emph{{Delensing
  CMB Polarization with External Datasets}},
  \href{https://doi.org/10.1088/1475-7516/2012/06/014}{\emph{JCAP} {\bfseries
  1206} (2012) 014}, [\href{https://arxiv.org/abs/1010.0048}{{\ttfamily
  1010.0048}}].

\bibitem{Sherwin:2015baa}
B.~D. Sherwin and M.~Schmittfull, \emph{{Delensing the CMB with the Cosmic
  Infrared Background}},
  \href{https://doi.org/10.1103/PhysRevD.92.043005}{\emph{Phys. Rev.}
  {\bfseries D92} (2015) 043005},
  [\href{https://arxiv.org/abs/1502.05356}{{\ttfamily 1502.05356}}].

\bibitem{Simard:2014aqa}
G.~Simard, D.~Hanson and G.~Holder, \emph{{Prospects for Delensing the Cosmic
  Microwave Background for Studying Inflation}},
  \href{https://doi.org/10.1088/0004-637X/807/2/166}{\emph{Astrophys. J.}
  {\bfseries 807} (2015) 166},
  [\href{https://arxiv.org/abs/1410.0691}{{\ttfamily 1410.0691}}].

\bibitem{Green:2016cjr}
D.~Green, J.~Meyers and A.~van Engelen, \emph{{CMB Delensing Beyond the B
  Modes}},  \href{https://arxiv.org/abs/1609.08143}{{\ttfamily 1609.08143}}.

\end{thebibliography}\endgroup
\end{document}